\documentclass[onecolumn,nofootinbib,superscriptaddress,notitlepage,amsmath,amssymb,fleqn,aps,prd,preprintnumbers,10pt]{revtex4-1}
\usepackage{xcolor,braket,graphicx,xspace,hyperref}
\usepackage[all]{hypcap}
\newcommand{\eps}{\varepsilon}

\hypersetup{hidelinks,
  pdfauthor={John Campbell,Stefan Hoeche,Max Knobbe,Christian Preuss,Daniel Reichelt},
  pdftitle={QCD splitting functions beyond kinematical limits},
  pdfkeywords={QCD, Perturbation Theory}
}

\begin{document}
\preprint{FERMILAB-PUB-25-0276-T, CERN-TH-2025-091, MCNET-25-09}
\title{QCD splitting functions beyond kinematical limits}
\author{John M.~Campbell}
\affiliation{Fermi National Accelerator Laboratory, Batavia, IL, 60510}
\author{Stefan H{\"o}che}
\affiliation{Fermi National Accelerator Laboratory, Batavia, IL, 60510}
\author{Max Knobbe}
\affiliation{Fermi National Accelerator Laboratory, Batavia, IL, 60510}
\author{Christian T.~Preuss}
\affiliation{Institut für Theoretische Physik, Georg-August-Universit{\"a}t G{\"o}ttingen, 37077 G{\"o}ttingen, Germany}
\author{Daniel Reichelt}
\affiliation{Theoretical Physics Department, CERN, CH-1211 Geneva, Switzerland}

\begin{abstract}
    We present a systematic decomposition of QCD splitting functions into
    scalar dipole radiators and pure splitting remainders up to second order
    in the strong coupling. The individual components contain terms that are
    formally sub-leading in soft or collinear scaling parameters, but well
    understood and universal due to their origin in scalar QCD.
    The multipole radiator functions which we derive share essential
    features of the known double-soft and one-loop soft gluon currents,
    and are not based on kinematical approximations.
\end{abstract}

\maketitle

\section{Introduction}
For more than half a century, experiments at particle colliders have shaped our understanding
of the building blocks of matter. They continue to be at the forefront of fundamental science~\cite{EuropeanStrategyforParticlePhysicsPreparatoryGroup:2019qin,Butler:2023glv}.
The detailed quantitative description of particle production in the collisions was made possible
by the development of QCD as a gauge theory of the strong interactions~\cite{Gross:2022hyw}.
High-energy collider experiments often measure QCD dynamics in the region where partons
are asymptotically free and form jets. The production and evolution of these jets is one
of the key features of the strong interactions at high energy and remains of greatest interest,
both because of its importance as an often irreducible background to new physics searches,
and as a precision test of QCD dynamics.

Being a non-abelian, asymptotically free gauge theory, QCD presents many obstacles to the practitioner.
Perturbative calculations are typically hampered both by the number and the complexity of the
Feynman diagrams associated with a particular partonic final
state~\cite{Heinrich:2020ybq,FebresCordero:2022psq,Campbell:2022qmc,Huss:2025nlt}.
In addition, at higher orders in the perturbative expansion, scattering matrix elements
exhibit infrared singularities that cancel to all orders between real and virtual
corrections~\cite{Bloch:1937pw,Yennie:1961ad,Kinoshita:1962ur,Lee:1964is}, often leading
to enormous complications for the numerical evaluation of observables with the help of
Monte-Carlo integration methods. Generic techniques to address this problem at
next-to-leading order (NLO) in the perturbative expansion were introduced long
ago~\cite{Frixione:1995ms,Catani:1996vz,Catani:2002hc}, and have been fully automated
in various computer codes. A similar automation at the next-to-next-to leading order (NNLO)
seems within reach~\cite{Huss:2025nlt}.

The two sources of infrared singularities are soft gluon radiation and the
collinear decay of massless partons. However, when multiple particles become collinear
one or more of them can still be soft, leading to an overlap between the two types
of divergence. Removing this overlap is key to identifying all singular
regions of scattering matrix elements, and to providing the infrared
subtraction counterterms that enable numerical computations.
The goal of this work is therefore to construct process-independent NNLO
double-real and real-virtual splitting functions that have no remaining overlaps.
This is achieved as follows. First, the universal component of all splitting functions
involving final-state gluons is extracted by making the {\em scalar} part of the theory
manifest. This part generates the pole structure in the soft limit and is related to
the semi-classical limit~\cite{Nambu:1968rr,Boulware:1968zz,Gell-Mann:1954wra,Brown:1968dzy}.
For the purpose of infrared subtraction, it can be computed in the dipole 
approximation~\cite{Catani:1998bh,Sterman:2002qn}. Consistently going beyond the
leading power soft approximation by using scalar dipoles allows us to capture universal
contributions which appear in the splitting functions in a well-understood manner.
Second, we identify the factorizable part of the higher-order splitting functions
based on diagrammatic considerations, which results in an improved understanding
of sub-leading singular terms, especially those involving azimuthal correlations.
Third, the scalar radiators are formulated differentially in color space, in order
to make them maximally useful for practical calculations. Similarly, the spin-dependent
remainders are differential in spin.
Our algorithm makes use of the techniques introduced in~\cite{Catani:1999ss,Catani:2000pi},
in particular the choice of an axial gauge to obtain a physical interpretation of the
gluon polarization tensor~\cite{Frenkel:1976bia,Dokshitzer:1978hw,
  Amati:1978wx,Amati:1978by,Ellis:1978sf,Ellis:1978ty,Kalinowski:1980ju,Kalinowski:1980wea}.
We provide the scalar extension of the double-soft radiators in both axial and
Feynman gauge, allowing for an unambiguous matching to the splitting functions.
Using the background field method~\cite{DeWitt:1967ub,Honerkamp:1972fd,
  Kluberg-Stern:1974nmx,Abbott:1980hw,Abbott:1981ke,Abbott:1983zw,Meissner:1986tr},
we compute scalar dipole radiator functions at one-loop level, which are matched to
the known one-loop splitting amplitudes.

The outline of this manuscript is as follows. Section~\ref{sec:motivation} gives
a summary of the state of QCD results for higher-order splitting functions with the
goal of contextualizing this work. In Section~\ref{sec:spin_decomposition} we introduce
the basic ideas and illustrate the spin decomposition at the vertex level.
Section~\ref{sec:tree-level_splitting_functions} describes the methodology for the
computation of tree-level splitting functions, reviews the results of~\cite{Catani:1999ss}
and presents their spin decomposition. Section~\ref{sec:loop_mes} includes
our results for the one-loop scalar radiator functions and one-loop
splitting functions and discusses their spin decomposition.
Section~\ref{sec:outlook} contains some concluding remarks and an outlook.

\section{Features of existing splitting functions and motivation for this work}
\label{sec:motivation}
The most frequently used techniques to isolate, regularize and cancel infrared singularities
in perturbative QCD calculations at next-to-leading order are the so-called infrared subtraction
algorithms~\cite{Heinrich:2020ybq,FebresCordero:2022psq,Campbell:2022qmc,Huss:2025nlt}.
They rely on the fact that the factorization of higher-order QCD matrix elements
into lower-order matrix elements and a finite set of splitting functions~\cite{
  Gribov:1972ri,Lipatov:1974qm,Dokshitzer:1977sg,Altarelli:1977zs} and soft insertion
operators~\cite{Bassetto:1984ik} is universal. Splitting functions and soft insertions
can be combined in different ways in order to obtain valid infrared subtraction schemes.
One possibility is a phase-space sectorization, which separates the regions where soft
gluon singularities can occur from the regions where only collinear singularities appear.
This is the basis of the Frixione-Kunszt-Signer technique~\cite{Frixione:1995ms}.
In the soft region, one then approximates the matrix element by the soft insertion
operators, while in the collinear region one makes use of the splitting functions.
Another possibility is to perform a matching of the soft insertion operators to the
splitting functions, and to use the matched expressions across the entire phase space.
This is the essence of the Catani-Seymour dipole subtraction method~\cite{Frixione:1995ms,
  Catani:1996vz}. In both cases, one needs to understand the overlap between the
soft insertion operators and the splitting functions. This overlap depends on the precise
limit in which the splitting function or the soft insertion operator was computed.
As an example, consider the spin-averaged splitting function for quark to quark gluon.
It is given by the following expression:
\begin{equation*}
    \langle P_{q\to q}(z)\rangle=C_F\left[\frac{2z}{1-z}+(1-\varepsilon)(1-z)\right]\,.
\end{equation*}
Its soft component is given by the $z\to 1$ limit and reads
\begin{equation*}
    \langle P_{q\to q}(z)\rangle \overset{z\to 1}{\longrightarrow}\frac{2C_F}{1-z}\;.
\end{equation*}
On the other hand, the soft eikonal is given by the square of the soft-gluon current,
${\bf J}^\mu$, and for a situation involving a quark in the final state, it gives,
in the quark-gluon collinear limit
\begin{equation*}
    {\bf J}^\mu {\bf J}_\mu\overset{q\parallel g}{\longrightarrow}\,\propto\frac{2C_F\,z}{1-z}\;.
\end{equation*}
This differs from the $z\to 1$ limit of the quark-to-quark splitting function.
Clearly, the soft and collinear limits do not commute, not even at the lowest non-trivial
order in QCD perturbation theory. The reason for this is quite simply that sub-leading
power contributions in one limit become leading power contributions in the other.
At the next higher order in the strong coupling expansion, this ambiguity obscures
the decomposition of the splitting functions into lower-order results.

The simplest example of this type has been discussed in Ref.~\cite{Gellersen:2021eci}
and will be re-examined in Sec.~\ref{sec:one_to_three_splittings_quark}. The azimuthal
angle dependence of the tree-level splitting function for a $q\to qq'\bar{q}'$ splitting
(derived in~\cite{Campbell:1997hg,Catani:1999ss}) should be identical to that of the
two-loop soft insertion operator for the production of a soft $q'\bar{q}'$ pair,
because the $q'\bar{q}'$ pair can only be created through a gluon splitting.
However, the angular dependence differs between the two, again because the double soft
insertion operator (typically taken from~\cite{Catani:1999ss}) is determined at
leading power in the soft scaling parameter. This obscures the fact that the soft
insertion operator could be used to capture the complete azimuthal angle dependence
of the $q\to qq'\bar{q}'$ splitting function, and in fact, one could use the
\textit{factorized} expression from which the insertion operator is derived
(see Eq.~(92) in Ref.~\cite{Catani:1999ss}).

We aim to solve this problem, which plagues all existing subtraction schemes that
are based on the results and techniques in~\cite{Campbell:1997hg,Catani:1999ss}.
This requires a re-computation of the splitting functions in a form that retains
the sub-leading power terms. However, rather than focus on the classification of
leading and sub-leading powers according to a scaling parameter, we organize
the calculation in a fashion that will allow us not to take any limits, neither
the soft, nor the collinear ones. For this it is important to realize that the
singularities in gauge theory amplitudes arise as a consequence of degenerate
asymptotic states~\cite{Coleman:1965xm,Sterman:1995fz}. The relevant propagators,
together with the gauge couplings, define conserved scalar currents that capture
the singularity structure in a physical rather than a purely mathematical fashion.
Our calculations will be guided by this principle.

The manuscript is quite long and presents all necessary information on this novel
concept, as well as formulae needed for the final assembly of splitting functions
from the individual building blocks.
The main results can be understood based on Secs.~\ref{sec:spin_decomposition}
and~\ref{sec:two-parton_tree-level}, with explicit NNLO results at tree level
listed in Sec.~\ref{sec:tree-level_splitting_decomposition}, and results at
one loop listed in Sec.~\ref{sec:one-loop_splitting_decomposition}.
The main formulae for splitting functions are also referenced in
Tab.~\ref{tab:limits} (end of Sec.~\ref{sec:tree-level_splitting_functions})
and Tab.~\ref{tab:limits_one-loop} (end of Sec.~\ref{sec:loop_mes}).
The scalar dipole approximation, which forms the backbone of the method,
is explained in detail in Secs.~\ref{sec:tree_level_scalar_multipoles} 
and~\ref{sec:one-loop_scalar_multipoles}.
We will begin by explaining the concept of the scalar currents.

\section{Spin decomposition of QCD amplitudes}
\label{sec:spin_decomposition}
QCD matrix elements exhibit well-understood soft-gluon
singularities~\cite{Mueller:1981ex,Ermolaev:1981cm,Dokshitzer:1982fh,
  Dokshitzer:1982xr,Bassetto:1982ma,Bassetto:1984ik}.
They originate in QCD interactions in the eikonal limit, which can be derived 
by making kinematical approximations to fixed-order scattering matrix elements.
However, their physical origin is better understood by investigating the
minimal coupling of a vector potential to a classical,
accelerated charge~\cite{Jackson:1998nia,Peskin:1995ev}.
In QED, the corresponding current in momentum space reads   
\begin{equation}\label{eq:classical_current}
  j_{ik}^\mu(q)=igQ\left(\frac{p_k^\mu}{p_kq}-\frac{p_i^\mu}{p_iq}\right)\;,
\end{equation}
where $g$ is the coupling constant, $Q$ is the charge of the particle,
and $p_i$ and $p_k$ are the momenta of the radiating charge dipole formed
by the particle before and after an instantaneous impact. 
Once the radiation field, $A_\mu$, sourced by the dipole is quantized,
radiative effects can be computed in perturbation theory using the 
interaction Hamiltonian density $j^\mu(x)A_\mu(x)$.
It can be shown that singularities in the infrared that are induced by this
interaction cancel between real-emission corrections and virtual 
corrections~\cite{Bloch:1937pw,Kinoshita:1962ur,Lee:1964is}.

Equation~\eqref{eq:classical_current} does not account for the quantum nature
of the charged particle. However, a minimal change to it yields the expression
for scalar QED with massless radiators, and thus provides an extension
to a full quantum field theory without kinematical constraints:
\begin{equation}\label{eq:scalar_dicurrent}
  \begin{split}
    j_{ik}^\mu(q)=&\;igQ\Big(S^{\mu}(p_k,q)-S^{\mu}(p_i,q)\Big)\;,
    \qquad\text{where}\qquad
    S^{\mu}(p,q)=\frac{(2p+q)^\mu}{(p+q)^2}\;.
  \end{split}
\end{equation}
For the emission of a single on-shell vector boson, this expression
results in the exact same scattering matrix elements as the eikonal current
in Eq.~\eqref{eq:classical_current}. It has in fact been shown in the case
of scalar QED that both the leading and sub-leading contributions to the
squared amplitudes in the soft-photon limit can be obtained from
classical calculations~\cite{Gell-Mann:1954wra,Brown:1968dzy}.
We make this observation the basis for using scalar currents
when constructing multipole radiator functions.
In the soft or double-soft limit, these radiators yield the known
leading soft or double-soft behavior of the theory, but they also
contain additional sub-leading contributions which are important for the
correct matching to collinear splitting functions away from the soft region.

The QCD equivalent of the QED current in Eq.~\eqref{eq:scalar_dicurrent}
can be obtained by using the techniques of~\cite{Bassetto:1984ik,Catani:1996vz}:
\begin{equation}\label{eq:scalar_current_sum}
  \begin{split}
    {\bf J}^\mu(q)=&\;ig_s\sum_i\hat{\bf T}_i\,S^{\mu}(p_i,q)\;.
  \end{split}
\end{equation}
Here, $\hat{\bf T}$, are the charge operators, which are defined as
$(\hat{\bf T}^c_i)_{ab}=T^c_{ab}$ for quarks, $(\hat{\bf T}^c_i)_{ab}=-T^c_{ba}$
for anti-quarks, and $(\hat{\bf T}^c_i)_{ab}=if^{acb}$
for gluons~\cite{Bassetto:1984ik,Catani:1996vz}. Charge conservation
in the QCD multipole implies $\sum_i\hat{\bf T}_i=0$. 
For the emission of a single on-shell vector boson, Eq.~\eqref{eq:scalar_current_sum}
results in exactly the same scattering matrix elements as the QCD eikonal current.
Moreover, the numerators in Eq.~\eqref{eq:scalar_current_sum}
obey elementary Ward identities, such that the single-gluon current is
conserved, as long as the radiating particles remain on their mass shell.

\begin{figure}[t]
  \centerline{\includegraphics[scale=0.36]{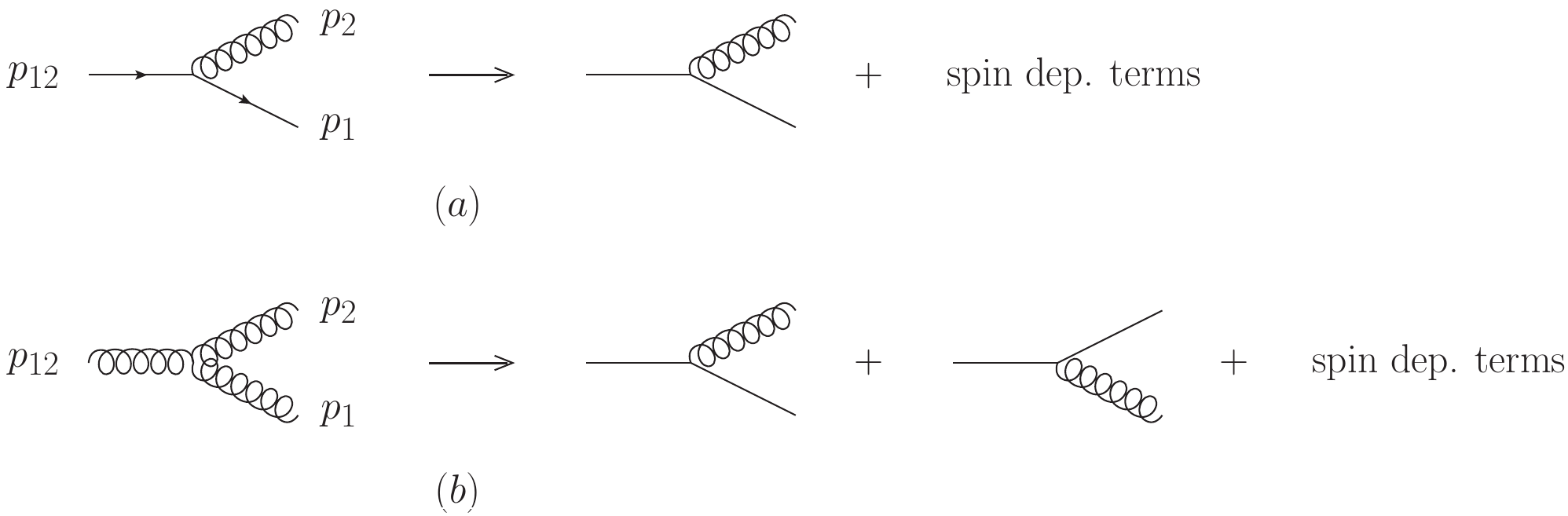}}
  \caption{Examples of one-to-two splitting processes. Figure~(a) shows the splitting of a quark into a quark and a gluon, Fig.~(b) the branching of a gluon into two gluons. The right-hand side sketches the decomposition of the full vertices into scalar and spin-dependent components.
  \label{fig:examples}}
\end{figure}

In Sec.~\ref{sec:tree_level_scalar_multipoles} we will use Eq.~\eqref{eq:scalar_current_sum}
and its extension to the two-emission case to derive the scalar component
of matrix elements that form the basis of the higher-order tree-level
splitting functions. As the complete calculation is quite cumbersome,
we highlight the basic correspondence using two simple examples at the
amplitude level. The first is the quark-to-quark-gluon splitting process
shown in Fig.~\ref{fig:examples}(a) left.
The scattering amplitude is proportional to the following combination of
the coupling-stripped quark-gluon vertex and the quark propagator:
\begin{equation}\label{eq:sc_decomposition_fermion}
  \begin{split}
    \frac{\slash\!\!\!p+\slash\!\!\!q}{(p+q)^2}\,T^a_{ij}\gamma^\mu
    =&\;T^a_{ij}\bigg[\,S^\mu(p,q)+\frac{i\sigma^{\nu\mu}q_\nu}{(p+q)^2}
    -\frac{\gamma^\mu\slash\!\!\!p}{(p+q)^2}\,\bigg]\;.
  \end{split}
\end{equation}
A decomposition of this current has been provided for the first time
in~\cite{Gordon:1928aa}, and an extension to massless fermions is given
in~\cite{Stone:2015bia}. The first term in the square bracket
is the scalar current of Eq.~\eqref{eq:scalar_dicurrent}. The second term,
proportional to $\sigma^{\nu\mu}=\frac{i}{2}[\gamma^\nu,\gamma^\mu]$, describes
the magnetic interaction due to the fermion's spin. This contribution is
sub-leading in the soft and collinear limits. The contribution proportional
to $\gamma^\mu\slash\!\!\!p$ vanishes in the squared amplitude for one-to-two
splittings due to the equations of motion.
Using an axial gauge for the final-state gluon, after squaring the amplitude
we obtain the standard quark-to-quark splitting function, cf.\ 
Sec.~\ref{sec:two-parton_tree-level}. We emphasize that this is achieved
without any kinematical approximations. Additionally, the squared scalar
interaction term can be identified at the amplitude squared level,
where it emerges as a nontrivial combination of the square of the scalar
contribution in Eq.~\eqref{eq:sc_decomposition_fermion} and the interference
of the scalar and the magnetic term. This is a consequence of the fact that
a magnetic interaction is needed to make the chiral kinetic theory
Lorentz invariant~\cite{Son:2012zy,Chen:2014cla}.
The vertex decomposition is sketched in Fig.~\ref{fig:examples}(a) right,
and discussed in detail in App.~\ref{sec:recursion}.

Our second example will be the gluon-to-gluon-gluon splitting,
shown in Fig.~\ref{fig:examples}(b) left. The gluon propagator
times the coupling stripped triple gluon vertex can be written
as a sum of two components in two different permutations
\begin{equation}\label{eq:rearranged_gluon_vertex}
  \begin{split}
  \frac{d^\mu_{\,\sigma}(p_{12})}{p_{12}^2}\,f^{abc}\Gamma^{\sigma\nu\rho}(p_1,p_2)
  =&\;id^\mu_{\,\sigma}(p_{12})\bigg\{{\rm Tr}(T^c[T^b,T^a])
  \bigg[\,S^{\nu}(p_1,p_2)g^{\rho\sigma}-\frac{(p_1-p_2)^\rho}{2p_{12}^2}\,g^{\sigma\nu}\,\bigg]
  +\bigg(\begin{array}{c}
       a\leftrightarrow b  \\[-1mm]
       \sigma\leftrightarrow \nu \\[-1mm]
       1\leftrightarrow 2
  \end{array}\bigg)\bigg\}\;,
  \end{split}
\end{equation}
where $\Gamma^{\mu\nu\rho}(p,q)=g^{\mu\nu}(p-q)^\rho
  +g^{\nu\rho}(2q+p)^\mu-g^{\rho\mu}(2p+q)^\nu$.
Here, $d^{\mu\sigma}(p_{12})$ is the gluon polarization tensor.
The two orderings of the color operators in the commutator appear
because either the color charge or the anti-color charge of the
incoming gluon can be the source of the gluon field. The first, 
scalar term in the square brackets is proportional to a metric 
tensor connecting the polarization vectors of the external particle 
to the incoming gluon. This agrees with the requirement that 
the semi-classical interaction described by $S^\mu(p,k)$ must be both 
spin-independent and helicity preserving. In the limit where the
emitted gluon becomes soft, the second term in the square brackets
does not scale due to the transversality of the Born amplitude.
In the collinear limit, it scales as the inverse of the transverse
momentum. We have again achieved a separation into different
types of splitting vertices, without having to perform any 
kinematical approximations. The one-to-two splitting function
derived in Sec.~\ref{sec:two-parton_tree-level} has a
similar structure. The vertex decomposition is sketched in
Fig.~\ref{fig:examples}(b) right and discussed in detail in
App.~\ref{sec:recursion}. 

Before discussing the tree-level splitting functions in 
Sec.~\ref{sec:tree-level_splitting_functions}, we will comment
on the gauge choices for our calculations. The tree-level expressions
are simplest to obtain in an axial gauge. Axial gauges benefit from being
ghost free~\cite{Arnowitt:1962cv,Konetschny:1975he,Konetschny:1976im,
  Frenkel:1976zk,Konetschny:1979mw}, because they encode only the physical
degrees of freedom of the gluon field~\cite{Frenkel:1976bia,Dokshitzer:1978hw,
  Amati:1978wx,Amati:1978by,Ellis:1978sf,Ellis:1978ty,
  Kalinowski:1980ju,Kalinowski:1980wea,Humpert:1980te}, 
see also the later discussion of Eq.~\eqref{eq:axial_gauge_onshell}
in Section~\ref{sec:tree-level_splitting_functions}.
The corresponding polarization tensor,
\begin{equation}\label{eq:axial_gauge}
    d^{\mu\nu}(p,n)
    =-g^{\mu\nu}+\frac{p^\mu n^\nu+p^\nu n^\mu}{pn}
    -\frac{n^2\,p^\mu p^\nu}{(pn)^2}\;,
\end{equation}
satisfies the physical requirements for on-shell gluons, namely
$-d^\mu_{\;\;\mu}(p,n)=D-2$ and $p_\mu d^{\mu\nu}(p,n)=0$, where $D=4-2\eps$
is the number of space-time dimensions. We discuss its relation to the
polarization sum in the helicity formalism in App.~\ref{sec:polarization_tensor}.
The vector $n^\mu$ is an auxiliary gauge vector. 
When used in the computation of a splitting amplitude, the axial gauge
mimicks the effects of coherent gluon emission off particles that are
omitted from the explicit computation, but whose physical presence
can be inferred from color charge conservation. As a simple example,
consider a process with two charged scalars. The leading-order matrix
element for the production of these scalars is just an overall constant.
The real-emission matrix element is given by the constant times the eikonal
in Eq.~\eqref{eq:classical_current}, contracted with the polarization vector
of the external vector boson. The key observation is that one can
choose the auxiliary vector, $n^\mu$, in the polarization sum such that
one term in the current, Eq.~\eqref{eq:classical_current}, is eliminated.
In this manner, one moves contributions needed to describe coherent vector boson
radiation from one Feynman diagram to another~\cite{Field:1989uq,Dixon:1996wi}.
In the computation of collinear splitting functions to leading power,
this technique makes the result independent of the hard process.

At one-loop order, we will use the axial gauge to compute the
splitting functions using the methods of~\cite{Kosower:1999rx,Bern:2004cz}.
In order to determine their scalar components, we would ideally determine 
the scalar radiators in the same gauge. As this calculation is very
cumbersome we use a different approach. The scalar radiators at one-loop
level are an extension of the one-loop soft current~\cite{Bern:1998sc,
  Bern:1999ry,Catani:2000pi} to the full scalar theory. They are determined
from the same set of diagrams and acquire a physical meaning when computed
in the background field method~\cite{DeWitt:1967ub,Honerkamp:1972fd,
  Kluberg-Stern:1974nmx,Abbott:1980hw,Abbott:1981ke,Abbott:1983zw,Meissner:1986tr,
  Denner:1994nn,Hashimoto:1994ct,Pilaftsis:1996fh}.
We can therefore match the collinear limit of the scalar radiators in the
background field method to the one-loop splitting amplitudes computed
in axial gauge. By means of this technique, the one-loop integrals needed
for the calculation can be limited to a small set which is known to all orders
in the dimensional regularization parameter.

\section{Tree-level expressions}
\label{sec:tree-level_splitting_functions}
In this section, we discuss the structure of the tree-level splitting functions
and the dipole approximation to coherent gluon radiation at second order in the
strong coupling. References~\cite{Campbell:1997hg,Catani:1999ss} first presented
the corresponding results, and Ref.~\cite{Catani:1999ss} gave the first
expressions for the two-gluon current in the double-soft limit which were
fully differential in color space. As discussed in Sec.~\ref{sec:motivation},
these expressions have a non-trivial interplay. It is intuitively clear that there
must be several components of the triple-collinear splitting functions that can be
expressed in terms of the double-soft radiators. However, as we will show in
subsequent sections, determining this overlap by using the double-soft expressions
alone generates remainder functions that contain residual sub-leading soft singularities
which should not be present. As a consequence, in the prevailing subtraction schemes
based on triple collinear and double soft limits, the remainder functions are often
sectorized in order to cleanly extract their singularities~\cite{Czakon:2010td}.
The goal of this section is to overcome this problem for tree-level splittings.
The final results and their scaling behavior in the various soft and collinear limits
will be summarized in Tab.~\ref{tab:limits}. These expressions are obtained by first
separating out the components which correspond to the semi-classical limit.
They include both the leading and all sub-leading soft singular terms that are
also present in the splitting functions, and are computed in
Sec.~\ref{sec:tree_level_scalar_multipoles}. We note that these expressions
have not yet been presented elsewhere, and that the color decompositions
in Sec.~\ref{sec:tree_level_scalar_multipoles} are new results, extending
the expressions of the double-soft limit in Ref.~\cite{Catani:1999ss}.
After separating out the scalar components of the splitting functions,
we remove all remaining terms which can still be expressed in terms of
lower-order results. This decomposition is performed in
Sec.~\ref{sec:tree-level_splitting_decomposition}. The basic methodology 
and the main results of this section can be understood with the help of
Sec.~\ref{sec:two-parton_tree-level} and~\ref{sec:one-gluon_scalar_radiators},
and the summary table, Tab.~\ref{tab:limits}.

\subsection{General form of the splitting functions}
\label{sec:form_of_splitting_functions}
It was shown in Ref.~\cite{Catani:1999ss}, that the $m$-particle quark and gluon
splitting functions can be determined as a ratio of the $m$-particle to $1$-particle
quark and gluon currents after projection onto auxiliary wave functions
conserving color charge and spin. This reflects the fact that the
splitting functions are reduced matrix elements, which are computed with physical
external states, despite their off-shellness. 
The expressions for quarks and gluons are given as follows
\begin{equation}\label{eq:collinear_splittings_recursive}
  \begin{split}
    P_{q}^{ss'}(1,\ldots,m)=&\;\delta^{ss'}
    \left(\frac{s_{1\ldots m}}{8\pi\alpha_s\mu^{2\eps}}\right)^{m-1}
    \frac{{\rm Tr}\big[\,\slash\!\!\!\bar{n}
    \Psi(\{p_1,\ldots,p_m\})
    \bar{\Psi}(\{p_1,\ldots,p_m\})\,\big]}{
    {\rm Tr}\big[\,\slash\!\!\!\bar{n}
    \Psi(\{\bar{p}_{1\ldots m}\})
    \bar{\Psi}(\{\bar{p}_{1\ldots m}\})\,\big]}\;,\\
    P_{g}^{\mu\nu}(1,\ldots,m)=&\;\frac{D-2}{2}
    \left(\frac{s_{1\ldots m}}{8\pi\alpha_s\mu^{2\eps}}\right)^{m-1}
    \frac{d^{\mu\rho}(p_{1\ldots m},\bar{n})
    J_\rho(\{p_1,\ldots,p_m\})
    J_\sigma^{\dagger}(\{p_1,\ldots,p_m\})
    d^{\sigma\nu}(p_{1\ldots m},\bar{n})}{
    d^{\kappa\lambda}(\bar{p}_{1\ldots m},\bar{n})
    J_\lambda(\{\bar{p}_{1\ldots m}\})
    J_\tau^{\dagger}(\{\bar{p}_{1\ldots m}\})
    d^{\tau}_{\;\;\kappa}(\bar{p}_{1\ldots m},\bar{n})}\;.\\
  \end{split}
\end{equation}
The quantities $\Psi$ and $J$ in this context are tree-level currents
that can be determined using recursive techniques such as the
Berends-Giele method~\cite{Berends:1987me,Berends:1988yn,Berends:1990ax,Duhr:2006iq}
\begin{equation}\label{eq:quark_current_new}
  \begin{split}
    &\Psi_i(p_\alpha)=
    \sum_{\substack{\{\beta,\gamma\}\in\\P(\alpha,2)}}
    g_sT^a_{ij}\,
    \frac{i\sigma^{\mu\nu}}{p_{\alpha}^2}\,p_{\gamma,\nu}\,
    J_\mu^a(p_\gamma,n)\Psi_j(p_\beta)\\
    &\;\quad+\sum_{\substack{\{\beta,\gamma\}\in\\P(\alpha,2)}}
    \bigg[\,g_sT^a_{ij}S^\mu(p_\beta,p_\gamma) J_\mu^a(p_\gamma,n)
    -\!\!\!\sum_{\substack{\{\delta,\epsilon\}\in\\ OP(\gamma,2)}}
    \frac{g_s^2}{p_{\alpha}^2}\,\big\{T^a,T^b\big\}_{ij}\,
    J^{\mu,a}(p_\delta,n)J_\mu^b(p_\epsilon,n)\,\bigg]\Psi_j(p_\beta)\;,
  \end{split}
\end{equation}
\begin{equation}\label{eq:gluon_current_new}
  \begin{split}
    &J_\mu^a(p_\alpha,n)=
    \sum_{\substack{\{\beta,\gamma\}\in\\ P(\alpha,2)}}
    \bigg[\,g_s\frac{F^a_{bc}}{2}\,D_\mu(p_\beta,p_\gamma)
    J^{\rho,b}(p_\beta,n)J_\rho^c(p_\gamma,n)
    +g_sT^a_{ij}\,\bar{\Psi}_i(p_\gamma)
    d^{\mu\nu}(p_\alpha,n)\,\frac{\gamma_\nu}{p_\alpha^2}
    \Psi_j(p_\beta)\,\bigg]\\
    &\;\quad-\sum_{\substack{\{\beta,\gamma\}\in\\ P(\alpha,2)}}
    \bigg[\,g_sF^c_{ab}\,S^\sigma(p_\beta,p_\gamma) J_\sigma^c(p_\gamma,n)
    -\!\!\!\sum_{\substack{\{\delta,\epsilon\}\in\\ OP(\gamma,2)}}
    \frac{g_s^2}{p_{\alpha}^2}\big\{F^c,F^d\big\}_{ab}\,J^{\sigma,c}(p_\delta,n)
    J_\sigma^d(p_\epsilon,n)\,\bigg]\,
    d_\mu^{\;\;\nu}(p_\alpha,n)J_\nu^b(p_\beta,n)\;.
  \end{split}
\end{equation}
Details on the above notation can also be found in App.~\ref{sec:recursion}.
The two most important building blocks of the recursion are the scalar production
and decay vertices, which are given by
\begin{equation}\label{eq:tree_level_building_blocks}
  \begin{split}
    S^\mu(p_i,p_j)=&\;\frac{(2p_i+p_j)^\mu}{p_{ij}^2}\;,
    \qquad\qquad
    &D^\mu(p_i,p_j,n)=&\;\frac{d^\mu_{\;\nu}(p_{ij},n)}{p_{ij}^2}\,(p_i-p_j)^\nu\;.
  \end{split}
\end{equation}
They will appear in most of the higher-order results which we derive in
the later sections of this manuscript.

For a discussion of the scaling behavior of the splitting functions, it is
convenient to parametrize the final-state momenta in terms of a forward,
transverse, and backward component according to~\cite{Sudakov:1954sw,Catani:1999ss}
\begin{equation}\label{eq:sudakov_decomposition_cg}
  p_i^\mu=z_i\bar{p}_{1..m}^\mu+\tilde{k}_i^\mu
  -\frac{\tilde{k}_i^2}{z_i}\frac{\bar{n}^\mu}{2\bar{p}_{1..m}\bar{n}}\;,
  \qquad\text{where}\qquad
  \bar{p}_{1..m}^\mu=p_{1..m}^\mu-p_{1..m}^2\,\frac{\bar{n}^\mu}{2p_{1..m}\bar{n}}\;.
\end{equation}
This allows us to express all results in terms of the forward light-cone
momentum fractions, $z_i$, and transverse momenta, $\tilde{k}_i$.
Local four-momentum conservation takes the form
$z_1+z_2+\ldots+z_m=1$ and $\tilde{k}_1+\tilde{k}_2+\ldots+\tilde{k}_m=0$.
To simplify the computation, we have changed the so far arbitrary vector $n^\mu$
defining the axial gauge to a light-like vector, which we denote as $\bar{n}^\mu$.

\subsection{One-to-two splittings}
\label{sec:two-parton_tree-level}
\begin{figure}
    \centerline{\includegraphics[scale=0.36]{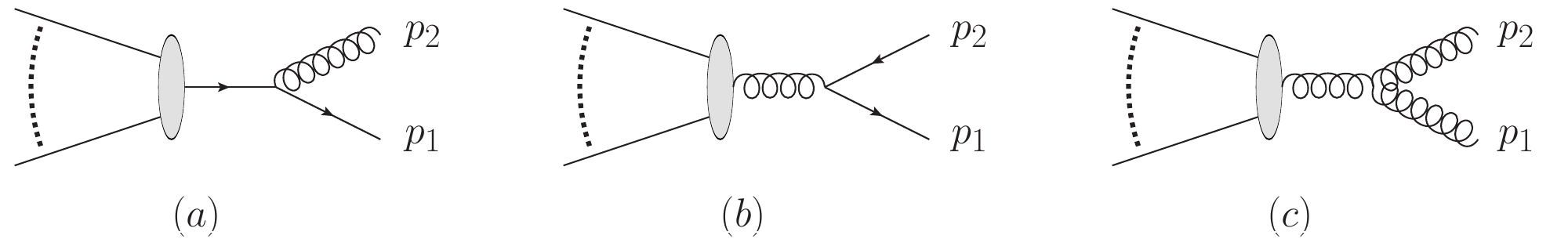}}
    \caption{Feynman diagrams leading to the $1\to 2$ parton splitting functions
    discussed in Sec.~\ref{sec:two-parton_tree-level}. The shaded blob and lines
    to the left represent the hard process with its associated external partons.
    See the main text for details.
    \label{fig:one-to-two_tree-diagrams}}
\end{figure}
In this subsection we re-derive the well-known two-parton splitting functions
and systematically extend them to the off-shell region, which is needed to derive
factorized expressions for the three-parton final states
in Sec.~\ref{sec:three-parton_tree-level}.
The novel aspect of this computation is that the extension is achieved through
a shift in the momentum defining the external wave functions, while leaving
the purely kinematical information in the propagator terms unaffected. 
This idea is in line with the general discussion in Sec.~\ref{sec:motivation},
in particular the fact that singularities arise only for degenerate asymptotic
states, which must have a well-defined polarization. We will limit the discussion
to the minimal information needed for the subsequent calculations and present
details relevant to the case of the two-gluon splitting tensor
in App.~\ref{sec:gluon_splitting_asymmetric}.

\subsubsection{Quark initial state}
\label{sec:two-parton_tree-level_quark}
The $1\to 2$ quark splitting function can be computed
from Eq.~\eqref{eq:collinear_splittings_recursive} using the
2-particle fermion current~\cite{Catani:1999ss}. The corresponding
Feynman diagram is shown in Fig.~\ref{fig:one-to-two_tree-diagrams}(a).
We find
\begin{equation}\label{eq:coll_q_to_qg}
  \begin{split}
    P^{ss'}_{q\to q}(p_1,p_2)
    =&\;\frac{\delta^{ss'}\,s_{12}}{8\pi\alpha_s\mu^{2\eps}}\,
    \frac{{\rm Tr}\big[\,\slash\!\!\!\bar{n}
    \Psi(\{p_1,p_2\})\bar{\Psi}(\{p_1,p_2\})\,\big]}{
    {\rm Tr}\big[\,\slash\!\!\!\bar{n}\Psi(\{\bar{p}_{12}\})
    \bar{\Psi}(\{\bar{p}_{12}\})\,\big]}
    =\delta^{ss'}P_{\tilde{q}\to\tilde{q}}(p_1,p_2)
    +P_{q\to q}^{ss'\,\rm(f)}(p_1,p_2)\;.
  \end{split}
\end{equation}
Following Ref.~\cite{Catani:1999ss}, we denote spin-averaged
quark splitting functions by $\langle P_{q\to X}(1,\ldots,m)\rangle
  =\delta_{ss'} P_{q\to X}^{ss'}(1,\ldots,m)/2$. As the quark splitting
tensors are diagonal in spin space, we will refrain from listing
both their spin-dependent and spin-averaged version.
The individual components of the one-to-two quark splitting function
are given by the $1\to 2$ scalar splitting function and the purely
fermionic term
\begin{equation}\label{eq:coll_q_to_qg_components}
  \begin{split}
    P_{\tilde{q}\to\tilde{q}}(p_1,p_2)=&\;C_F\frac{2z_1}{z_2}
    \bigg(1-\frac{p_1^2}{p_{12}^2}\frac{z_{12}}{z_1}
    -\frac{p_2^2}{p_{12}^2}\frac{z_{12}}{z_2}\bigg)\;,\\
    \langle P_{q\to q}^{\rm(f)}(p_1,p_2)\rangle
    =&\;C_F\,(1-\eps)\bigg(\frac{z_2}{z_{12}}
    -\frac{z_2}{z_1}\frac{p_1^2}{p_{12}^2}
      -\frac{p_2^2}{p_{12}^2}\bigg)\;,
  \end{split}
\end{equation}
where $p_1^2\to 0$ and $p_2^2 \to 0$ in the standard collinear factorization
approach. In order to extend the massless result to the off-shell region,
we have replaced $d^{\mu\nu}(p_2,\bar{n})\to d^{\mu\nu}(\bar{p}_2,\bar{n})$,
see the discussion of Eq.~\eqref{eq:tc_factorization}. In complete analogy,
we have replaced the spinors $u(p_1)$ by $u(\bar{p}_1)$. Both these shifts
are needed to achieve the expected scaling behavior of the remainder functions
computed in Sec.~\ref{sec:tree-level_splitting_decomposition}.
The gluon-spin dependent quark-to-quark splitting tensor is given by
\begin{equation}\label{eq:pqq_munu_sc}
  \begin{split}
    \langle P_{q\to q}^{\mu\nu}(p_1,p_2)\rangle
    =&\;P_{\tilde{q}\to\tilde{q}}^{\mu\nu}(p_1,p_2)
    +\langle P^{{\rm(f)}\mu\nu}_{q\to q}(p_1,p_2)\rangle\;.
  \end{split}
\end{equation}
Its scalar and purely fermionic components are given by
\begin{equation}\label{eq:pqq_munu_sc_components}
  \begin{split}
    P_{\tilde{q}\to\tilde{q}}^{\mu\nu}(p_1,p_2)
    =&\;\frac{C_F}{2}\,p_{12}^2\,S^\mu(p_1,p_2)S^\nu(p_1,p_2)\;,\\
    \langle P^{{\rm(f)}\mu\nu}_{q\to q}(p_1,p_2)\rangle
    =&\;-\frac{C_F}{2}\bigg[\,g^{\mu\nu}\bigg(\frac{z_2}{z_{12}}
    -\frac{z_2}{z_1}\frac{p_1^2}{p_{12}^2}
    -\frac{p_2^2}{p_{12}^2}\bigg)
    +\frac{p_2^\mu p_2^\nu}{p_{12}^2}\,\bigg]+\ldots\;,
  \end{split}
\end{equation}
where the dots stand for contributions proportional to the
gauge vector, $\bar{n}^\mu$ or $\bar{n}^\nu$. Such terms vanish
after multiplication by a gluon polarization tensor of the form
$d^{\mu\nu}(p_2,\bar{n})$ or $d^{\mu\nu}(\bar{p}_2,\bar{n})$,
and can therefore be dropped.

\subsubsection{Gluon initial state}
\label{sec:two-parton_tree-level_gluon}
The Feynman diagrams leading to the tree-level $g\to q\bar{q}$ and $g\to gg$
splitting tensors are shown in Figs.~\ref{fig:one-to-two_tree-diagrams}(b) and~(c),
respectively. The algebraic expressions are obtained from
Eq.~\eqref{eq:collinear_splittings_recursive} as follows~\cite{Catani:1999ss}.
\begin{equation}\label{eq:coll_gqq_ggg_step1}
  \begin{split}
    P^{\mu\nu}_{g\to q}(p_1,p_2)
    =&\;\frac{T_R}{2p_{12}^2}\,d^\mu_{\;\rho}(p_{12},\bar{n})
    {\rm Tr}[\,\slash\!\!\!\bar{p}_1
    \gamma^\rho \slash\!\!\!\bar{p}_2\gamma^\sigma\,]
    d^\nu_{\;\sigma}(p_{12},\bar{n})\;,\\
    P^{\mu\nu,\alpha\beta}_{g\to g}(p_1,p_2)
    =&\;\frac{C_A}{2p_{12}^2}\,d^\mu_{\;\lambda}(p_{12},\bar{n})\,
    \Gamma^{\kappa\alpha\lambda}(p_1,p_2)\Gamma^{\rho\beta\tau}(p_1,p_2)
      d_{\kappa\rho}(\bar{p}_1,\bar{n})d^\nu_{\;\tau}(p_{12},\bar{n})\;,
  \end{split}
\end{equation}
where $\Gamma^{\mu\nu\rho}(p,q)$ implements the Lorentz structure of the
three-gluon vertex, and where the Lorentz indices $\alpha$ and $\beta$
refer to the final-state gluon with momentum $p_2$, while the indices
$\mu$ and $\nu$ refer to the initial-state gluon. Computing the
gluon-to-quark splitting tensor is straightforward, and we obtain
\begin{equation}\label{eq:coll_gqq}
  \begin{split}
    P^{\mu\nu}_{g\to q}(p_1,p_2)
    =&\;T_R\,\bigg[\,d^{\mu\nu}(\bar{p}_{12},\bar{n})
    \bigg(1-\frac{z_{12}}{p_{12}^2}
    \bigg(\frac{p_1^2}{z_1}+\frac{p_2^2}{z_2}\bigg)\bigg)
    +p_{12}^2\frac{\bar{n}^\mu\bar{n}^\nu}{(p_{12}\bar{n})^2}
    -p_{12}^2D^\mu(p_1,p_2,\bar{n})D^\nu(p_1,p_2,\bar{n})\,\bigg]\;,
  \end{split}
\end{equation}
where we have summed over the spins of the final-state quarks.
In the final-state on-shell case Eq.~\eqref{eq:coll_gqq} reduces to
\begin{equation}\label{eq:coll_gqq_os}
  \begin{split}
    P^{\mu\nu\,\rm(os)}_{g\to q}(p_1,p_2)
    =&\;T_R\,\bigg[\,d^{\mu\nu}(p_{12},\bar{n})
    -\frac{4\,\tilde{p}_{1,2}^\mu\tilde{p}_{1,2}^\nu}{p_{12}^2}\,\bigg]\;,
    \quad\text{where}\quad
    \tilde{p}_{i,j}^\mu=\frac{z_i\,p_j^\mu-z_j\,p_i^\mu}{z_i+z_j}\;.
 \end{split}
\end{equation}
The momenta $\tilde{p}$ are generalizations of the transverse
momenta $\tilde{k}$ in Eq.~\eqref{eq:sudakov_decomposition_cg}. In particular, they fulfill
momentum conservation in the form $\tilde{p}_{i,j}+\tilde{p}_{j,i}=0$ for any $i$ and $j$.
Similarly, in the 3-parton case we have $\tilde{p}_{1,23}+\tilde{p}_{2,13}+\tilde{p}_{3,12}=0$. 
Using the standard Sudakov parametrization in
Eq.~\eqref{eq:sudakov_decomposition_cg}~\cite{Sudakov:1954sw},
taking the collinear limit, and summing over quark spins, we can write
Eq.~\eqref{eq:coll_gqq} in the familiar form of the spin-dependent
DGLAP splitting kernel
\begin{equation}\label{eq:gqq_dglap_limit}
  \begin{split}
    P^{\mu\nu\,{\rm(os)}}_{g\to q}(p_1,p_2)\to&\;T_R\,\bigg[
    -g^{\mu\nu}+4z_1z_2\frac{\tilde{k}_1^\mu \tilde{k}_1^\nu}{\tilde{k}_1^2}\,\bigg]\;.
  \end{split}
\end{equation}
Averaging over the polarizations of the initial-state gluon using conventional
dimensional regularization (CDR), $\langle P_{g\to X}(1,\ldots,m)\rangle=
 d_{\mu\nu}(p_{1\ldots m},\bar{n})P^{\mu\nu}_{g\to X}(1,\ldots,m)/(2-2\eps)$,
 yields the CDR DGLAP splitting kernel
\begin{equation}\label{eq:gqq_dglap_limit_sum}
  \begin{split}
    \langle P_{g\to q}(p_1,p_2)\rangle
    =&\;T_R\,\bigg[\,1-\frac{2}{1-\eps}\frac{z_1z_2}{z_{12}^2}\,\bigg]\;.
  \end{split}
\end{equation}
The computation of the gluon-to-gluon splitting tensor is aided by the observation
that any function multiplying this object must be symmetric in the Lorentz indices
$\mu$ and $\nu$. It is known that, in the on-shell case, this causes all interferences
between the three components of $\Gamma^{\mu\nu\rho}$ to vanish~\cite{Somogyi:2005xz}.
In the following, we derive the corresponding expression, including some of the off-shell
effects needed in Sec.~\ref{sec:tree-level_splitting_decomposition}. We assume that
$p_1^2=0$, which is sufficient to compute all factorizable components of the
three-parton splitting functions. The relation $d^{\mu\rho}(p_1,\bar{n})
d^{\nu}_{\;\rho}(p_1,\bar{n})=d^{\mu\nu}(p_1,\bar{n})$ can then be exploited 
to factorize the triple-gluon vertex functions in Eq.~\eqref{eq:coll_gqq_ggg_step1}. 
We separate the resulting splitting tensor into a symmetric and an interference part
\begin{equation}\label{eq:coll_ggg}
  P_{g\to g}^{\mu\nu,\alpha\beta}(p_1,p_2)=
  P_{g\to g,\rm(s)}^{\mu\nu,\alpha\beta}(p_1,p_2)
  +P_{g\to g,\rm(i)}^{\mu\nu,\alpha\beta}(p_1,p_2)
  +P_{g\to g,\rm(i)}^{\nu\mu,\beta\alpha}(p_1,p_2)\;.
\end{equation}
In App.~\ref{sec:gluon_splitting_asymmetric} we show that the asymmetric components vanish
up to $\mathcal{O}(\alpha_s^2)$. They can therefore be neglected for the remainder of this work.
The symmetric component is given by the sum of squared scalar emission and decay vertices
\begin{equation}\label{eq:coll_ggg_sym}
  \begin{split}
  P_{g\to g,\rm(s)}^{\mu\nu,\alpha\beta}&(p_1,p_2)
  =\frac{C_A}{2}\bigg[\,p_{12}^2\,S^\alpha(p_1,p_2)S^\beta(p_1,p_2)\,
  d^{\mu\nu}(\bar{p}_{12},\bar{n})\\
  &\;+2d^{\mu\alpha}(p_{12},\bar{n})d^{\nu\beta}(p_{12},\bar{n})
  \frac{2z_2}{z_1}\bigg(1-\frac{p_2^2}{p_{12}^2}\frac{z_{12}}{z_2}\bigg)
  +p_{12}^2\,d^{\alpha\beta}(p_{1},\bar{n})
  D^\mu(p_1,p_2,\bar{n})D^\nu(p_1,p_2,\bar{n})\,\bigg]\;.
  \end{split}
\end{equation}
In the final-state on-shell case it reduces to
\begin{equation}\label{eq:coll_ggg_sym_os}
  \begin{split}
  P_{g\to g}^{\mu\nu,\alpha\beta\,{\rm(os)}}&(p_1,p_2)
  =\frac{C_A}{2}\bigg[\,p_{12}^2\,S^\alpha(p_1,p_2)S^\beta(p_1,p_2)\,
  d^{\mu\nu}(\bar{p}_{12},\bar{n})\\
  &\;+2d^{\mu\alpha}(p_{12},\bar{n})d^{\nu\beta}(p_{12},\bar{n})
  \frac{2z_2}{z_1}+p_{12}^2\,d^{\alpha\beta}(p_{1},\bar{n})
  D^\mu(p_1,p_2,\bar{n})D^\nu(p_1,p_2,\bar{n})\,\bigg]\;.
  \end{split}
\end{equation}
Contracting Eq.~\eqref{eq:coll_ggg_sym} with the external polarization sum
for gluon $2$, we find
\begin{equation}\label{eq:coll_ggg_sym_avg}
  P_{g\to g}^{\mu\nu}(p_1,p_2)
  =2C_A\bigg[\,d^{\mu\nu}(\bar{p}_{12},\bar{n})
  \bigg(\frac{z_1}{z_2}+\frac{z_2}{z_1}-\frac{z_{12}}{z_1}\frac{p_2^2}{p_{12}^2}\bigg)
  +2(1-\eps)\,\frac{\tilde{p}_{1,2}^\mu\tilde{p}_{1,2}^\nu}{p_{12}^2}\,\bigg]\;.
\end{equation}
Using the standard Sudakov parametrization in
Eq.~\eqref{eq:sudakov_decomposition_cg}~\cite{Sudakov:1954sw},
Eq.~\eqref{eq:coll_ggg_sym_os} yields the spin-dependent
DGLAP kernel~\cite{Somogyi:2005xz}
\begin{equation}\label{eq:gqq_ggg_dglap_limit}
  \begin{split}
    P^{\mu\nu,\alpha\beta\,{\rm(os)}}_{g\to g}(p_1,p_2)
    \to&\;2C_A\,\bigg[\,\frac{z_1}{z_2}\,
    \frac{\tilde{k}_1^\alpha \tilde{k}_1^\beta}{\tilde{k}_1^2}\,g^{\mu\nu}
    +\frac{z_2}{z_1}\,g^{\mu\alpha}g^{\nu\beta}
    -d^{\alpha\beta}(p_{12},\bar{n})\,\frac{z_1z_2}{z_{12}^2}\,
    \frac{\tilde{k}_1^\mu \tilde{k}_1^\nu}{\tilde{k}_1^2}\,\bigg]\;.
  \end{split}
\end{equation}
Contracting this with the polarization tensor for gluon $2$, 
we find the familiar expression
\begin{equation}\label{eq:gqq_ggg_dglap_limit_summed}
  \begin{split}
    d_{\alpha\beta}(p_2,\bar{n})P^{\mu\nu,\alpha\beta\,{\rm(os)}}_{g\to g}(p_1,p_2)
    \to&\;2C_A\,\bigg[\,
    -g^{\mu\nu}\bigg(\frac{z_1}{z_2}+\frac{z_2}{z_1}\bigg)
    -2(1-\eps)\,\frac{z_1z_2}{z_{12}^2}\,
    \frac{\tilde{k}_1^\mu \tilde{k}_1^\nu}{\tilde{k}_1^2}\,\bigg]\;.
  \end{split}
\end{equation}
Averaging over the polarizations of the initial-state gluon yields
the DGLAP splitting kernel
\begin{equation}\label{eq:ggg_dglap_limit_sum}
  \begin{split}
    \langle P_{g\to g}(p_1,p_2)\rangle=&\;2C_A\,\bigg[\,
    \frac{z_1}{z_2}+\frac{z_2}{z_1}+\frac{z_1z_2}{z_{12}^2}\,\bigg]\;.
  \end{split}
\end{equation}
Based on the above derivation, this function can be decomposed as
\begin{equation}
    \langle P_{g\to g}(p_1,p_2)\rangle =
    P_{g\to g}^{(\mathrm{sc})}(p_1,p_2) +
    P_{g\to g}^{(\mathrm{sc})}(p_2,p_1) +
    \langle P_{g\to g}^{(\mathrm{v})}(p_1,p_2) \rangle\;,
\end{equation}
where the scalar and purely vectorial parts are defined as
\begin{equation}\label{eq:ggg_scalar_vector}
    \begin{split}
    P_{g\to g}^{(\mathrm{sc})}(p_1,p_2) &= C_A\,\frac{2z_1}{z_2}
    = \frac{C_A}{C_F}P_{\tilde{q}\to \tilde{q}}(p_1,p_2)\Big\vert_{p_1^2\to 0,p_2^2\to 0} \;,\\
    \langle P_{g\to g}^{(\mathrm{v})}(p_1,p_2)\rangle &= 2C_A\, \frac{z_1z_2}{z_{12}^2} \;.
    \end{split}
\end{equation}
Note that Eqs.~\eqref{eq:coll_gqq} and~\eqref{eq:coll_ggg_sym} are
independent of the kinematics parametrization, as the invariants have not yet been
expressed in terms of transverse momenta, and the light-cone momentum fractions
are defined unambiguously.

\subsection{One-to-three splittings}
\label{sec:three-parton_tree-level}
The leading terms of the three-parton splitting functions have been computed
in~\cite{Campbell:1997hg,Catani:1999ss}. We will re-derive them here, together with their
scalar QCD counterparts. These scalar splitting functions are needed in order to understand
the interplay between the QCD splitting functions and the QCD multipole radiators in the
dipole approximation, which will be derived in Sec.~\ref{sec:tree_level_scalar_multipoles}.
In Sec.~\ref{sec:tree-level_splitting_decomposition} we will discuss how the splitting functions
can be decomposed into terms that are captured by the scalar dipole radiators or factorized
expressions involving the functions from Sec.~\ref{sec:two-parton_tree-level},
and novel remainder functions that are only singular in the triple collinear region.
The fact that such remainder functions can be defined, and that their poles are of
lowest degree, is one of the main results of this manuscript.
We begin the discussion with the simpler quark initial states, which do not have
a spin dependence, before discussing the gluon initial states that feature
a non-trivial Lorentz structure.

\subsubsection{Quark initial state}
\label{sec:one_to_three_splittings_quark}
\begin{figure}
    \centerline{\includegraphics[scale=0.36]{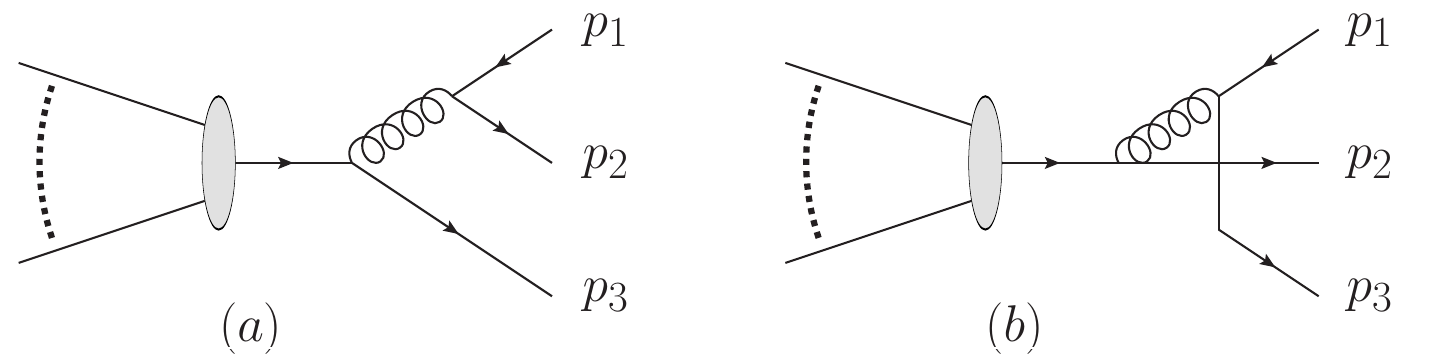}}
    \caption{Feynman diagrams leading to the $1\to 3$ quark only splitting functions
    discussed in Sec.~\ref{sec:one_to_three_splittings_quark}.  The shaded blob and lines
    to the left represent the hard process with its associated external partons.
    See the main text for details.
    \label{fig:one-to-three_splittings_quark}}
\end{figure}
The sole Feynman diagram leading to the flavor-changing quark-to-quark three-parton
splitting function is shown in Fig.~\ref{fig:one-to-three_splittings_quark}(a).
It can be obtained from the product of the off-shell quark splitting function in
Eq.~\eqref{eq:pqq_munu_sc} and the spin-dependent gluon-to-quark splitting function,
Eq.~\eqref{eq:coll_gqq}. The result is~\cite{Catani:1999ss}
\begin{equation}\label{eq:tc_qqpqpb}
  \langle P_{q\to \bar{q}'q'q}(p_1,p_2,p_3)\rangle=
  \frac{C_FT_R}{2}\frac{s_{123}}{s_{12}}\left[-\frac{t_{12,3}^2}{s_{12} s_{123}}
    +\frac{4z_3+(z_1-z_2)^2}{z_1+z_2}
    +(1-2\eps)\left(1-z_3-\frac{s_{12}}{s_{123}}\right)\right]\;,
\end{equation}
where $s_{ij}=(p_i+p_j)^2$, and where we have defined
\begin{equation}\label{eq:def_cg_t123}
    t_{12,3}=s_{123}\,S^\mu(p_3,p_{12})\, s_{12}D_\mu(p_1,p_2,\bar{n})=
    2\,\frac{z_1 2p_2p_3-z_2 2p_1p_3}{z_1+z_2}+\frac{z_1-z_2}{z_1+z_2}\,2p_1p_2\;.
\end{equation}
The corresponding scalar-to-quark splitting function is given by the product of the
scalar part of Eq.~\eqref{eq:pqq_munu_sc_components} and Eq.~\eqref{eq:coll_gqq}.
It is the component of Eq.~\eqref{eq:tc_qqpqpb} which is given in terms of the
two-quark scalar dipole radiator that will be computed in
Eq.~\eqref{eq:scalar_emission_quark_pair_individual}
\begin{equation}\label{eq:tc_qqpqpb_scalar}
  P_{\tilde{q}\to \bar{q}'q'\tilde{q}}(p_1,p_2,p_3)
  =\frac{s_{123}}{s_{12}}
  P_{\tilde{q}\to\tilde{q}}^{\mu\nu}(p_3,p_{12})P_{g\to q,\mu\nu}(p_1,p_2)
  =\frac{C_FT_R}{2}\frac{s_{123}}{s_{12}}\left[\frac{4z_3}{z_1+z_2}
  +\frac{s_{12}}{s_{123}}\bigg(1-\frac{t_{12,3}^2}{s_{12}^2}\bigg)\right]\;.
\end{equation}
As a consequence of the fact that only a single diagram contributes to
Eq.~\eqref{eq:tc_qqpqpb}, the splitting function factorizes.
A simple, yet non-trivial observation is that, in the double-soft limit, both
Eq.~\eqref{eq:tc_qqpqpb} and Eq.~\eqref{eq:tc_qqpqpb_scalar} yield, at leading
power in the soft scaling parameter
\begin{equation}\label{eq:tc_qqpqpb_ds}
  \langle P^{\rm(ds)}_{q\to \bar{q}'q'q}(p_1,p_2,p_3)\rangle
  =C_FT_R\,\frac{2(s_{13}+s_{23})}{s_{12}(z_1+z_2)}\left[1
  -\frac{(z_1 s_{23}-z_2 s_{13})^2}{(z_1+z_2)(s_{13}+s_{23})s_{12}}\,\right]\;.
\end{equation}
As discussed in Sec.~\ref{sec:motivation}, the azimuthal angle dependence
differs between Eqs.~\eqref{eq:tc_qqpqpb},~\eqref{eq:tc_qqpqpb_scalar}
and~\eqref{eq:tc_qqpqpb_ds}, because Eq.~\eqref{eq:def_cg_t123} is reduced to the
first term alone. This exemplifies how higher-order splitting functions based on
kinematic limits can obscure the origin of soft-collinear overlaps.
In the remainder of this section, we will focus on the computation of the remaining
splitting functions needed to perform a similar analysis for more complicated
final states. Their properties and decomposition in terms of lower-order results
will be presented in Sec.~\ref{sec:tree-level_splitting_decomposition}.

The three-parton quark-to-quark splitting function is obtained as the coherent sum
of the two Feynman diagrams in Figs.~\ref{fig:one-to-three_splittings_quark}(a) and~(b).
It can be written as the sum of splitting functions for identical quarks, and an
interference part~\cite{Catani:1999ss}.
\begin{equation}\label{eq:tc_qqqb}
  \langle P_{q\to \bar{q}qq}(p_1,p_2,p_3)\rangle=
  \Big(\langle P_{q\to \bar{q}'q'q}(p_1,p_2,p_3)\rangle+
  \langle P^{\rm(id)}_{q\to \bar{q}qq}(p_1,p_2,p_3)\rangle\Big)+
  \Big(2\leftrightarrow 3\Big)\;,
\end{equation}
where
\begin{equation}\label{eq:tc_qqqb_id}
  \begin{split}
  \langle P^{\rm(id)}_{q\to \bar{q}qq}(p_1,p_2,p_3)\rangle=&\;
  C_F\bigg(C_F-\frac{C_A}{2}\bigg)\bigg\{(1-\eps)\bigg(\frac{2s_{23}}{s_{12}}-\eps\bigg)\\
  &\qquad+\frac{s_{123}}{s_{12}}\bigg[\frac{1+z_1^2}{1-z_2}-\frac{2z_2}{1-z_3}
  -\eps\bigg(\frac{(1-z_3)^2}{1-z_2}+1+z_1-\frac{2z_2}{1-z_3}\bigg)-\eps^2(1-z_3)\bigg]\\
  &\qquad-\frac{s_{123}^2}{s_{12}s_{13}}\frac{z_1}{2}\bigg[\frac{1+z_1^2}{(1-z_2)(1-z_3)}
  -\eps\bigg(1+2\frac{1-z_2}{1-z_3}\bigg)-\eps^2\bigg]\bigg\}\;.
  \end{split}
\end{equation}
The scalar-to-quark splitting function corresponding to this interference term does not exist.

\begin{figure}
    \centerline{\includegraphics[scale=0.36]{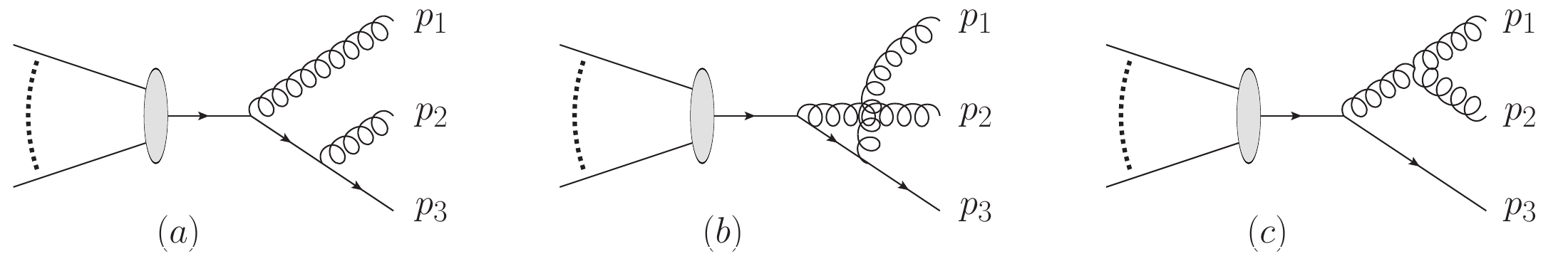}}
    \caption{Feynman diagrams leading to the $1\to 3$ quark-to-quark-gluon splitting functions
    discussed in Sec.~\ref{sec:one_to_three_splittings_quark}. The shaded blob and lines
    to the left represent the hard process with its associated external partons.
    See the main text for details.
    \label{fig:one-to-three_splittings_quark_gluon}}
\end{figure}
Next we investigate double gluon radiation. The splitting function can be separated into
abelian and purely non-abelian components~\cite{Catani:1999ss}, which are derived
from the diagrams in Figs.~\ref{fig:one-to-three_splittings_quark_gluon}(a) and~(b), and
from Figs.~\ref{fig:one-to-three_splittings_quark_gluon}(a)-(c), respectively.
The result for the abelian component is
\begin{equation}\label{eq:tc_qgg}
  \begin{split}
  \langle P^{\rm(ab)}_{q\to ggq}&\;(p_1,p_2,p_3)\rangle=C_F^2\,\bigg\{
  \frac{s_{123}^2}{2s_{13}s_{23}}\,z_3\left[\frac{1+z_3^2}{z_1z_2}
  -\eps\frac{z_1^2+z_2^2}{z_1z_2}-\eps(1+\eps)\right]
  +\eps(1-\eps)-\frac{s_{23}}{s_{13}}\,(1-\eps)^2\\
  &\;\qquad\quad+\frac{s_{123}}{s_{13}}\left[\frac{z_3(1-z_1)+(1-z_2)^3}{z_1z_2}
  +\eps^2(1+z_3)-\eps(z_1^2+z_1z_2+z_2^2)\frac{1-z_2}{z_1z_2}\right]\bigg\}
  +(1\leftrightarrow 2)\;.
  \end{split}
\end{equation} 
In the scalar case, we obtain
\begin{equation}\label{eq:tc_sqgg}
  P^{\rm(ab)}_{\tilde{q}\to gg\tilde{q}}(p_1,p_2,p_3)=C_F^2\,\bigg\{
  \frac{s_{123}^2}{s_{13}s_{23}}\frac{z_3^2}{z_1z_2}
  +\frac{s_{123}}{s_{13}}\frac{2z_3(1-z_2)}{z_1 z_2}
  +(1-\eps)\bigg\}
  +(1\leftrightarrow 2)\;.
\end{equation}
This is the first splitting function involving the seagull vertices in
Eq.~\eqref{eq:quark_current_new}.

The non-abelian part of the splitting function can be written as (see~\cite{Catani:1999ss})
\begin{equation}\label{eq:tc_qgg_nab}
  \begin{split}
  \langle P^{\rm(nab)}_{q\to ggq}(p_1,p_2,p_3)\rangle=&\;
  -\frac{C_A}{2C_F}P^{\rm(ab)}_{q\to ggq}(p_1,p_2,p_3)+C_FC_A\,\bigg\{
  \frac{1-\eps}{4}\bigg(\frac{t_{12,3}^2}{s_{12}^2}+1\bigg)-(1-\eps)^2\frac{s_{23}}{2s_{13}}\\
  &\;\qquad+\frac{s_{123}}{2s_{12}}\bigg[(1-\eps)\frac{z_1(2-2z_1+z_1^2)-z_2(6-6z_2+z_2^2)}{z_2(1-z_3)}
  +2\eps\,\frac{z_3(z_1-2z_2)-z_2}{z_2(1-z_3)}\bigg]\\
  &\;\qquad+\frac{s_{123}}{2s_{13}}\bigg[(1-\eps)\frac{(1-z_2)^3+z_3^2-z_2}{z_2(1-z_3)}
  -\eps\,\frac{2(1-z_2)(z_2-z_3)}{z_2(1-z_3)}-\eps(1-\eps)(1-z_1)\bigg]\\
  &\;\qquad+\frac{s_{123}^2}{2s_{12}s_{13}}\bigg[\frac{(1-z_3)^2(1-\eps)+2z_3}{z_2}
  +\frac{z_2^2(1-\eps)+2(1-z_2)}{1-z_3}\bigg]+(1\leftrightarrow 2)\bigg\}\;.
  \end{split}
\end{equation}
In the scalar case, we obtain instead
\begin{equation}\label{eq:tc_sqgg_nab}
  \begin{split}
  &P^{\rm(nab)}_{\tilde{q}\to gg\tilde{q}}(p_1,p_2,p_3)=
  -\frac{C_A}{2C_F}P^{\rm(ab)}_{\tilde{q}\to gg\tilde{q}}(p_1,p_2,p_3)
  +C_FC_A\,\bigg\{\frac{1-\eps}{4}\bigg(\frac{t_{12,3}^2}{s_{123}s_{12}}+1\bigg)\\
  &\;\qquad+\frac{s_{123}}{s_{12}}\bigg[\frac{z_3}{z_2}-\frac{1+3z_3}{1-z_3}\bigg]
  +\frac{s_{123}}{s_{13}}\bigg[\frac{z_3}{z_2}-\frac{1-z_2}{1-z_3}\bigg]\frac{1-z_2}{z_1}
  +\frac{s_{123}^2}{s_{12}s_{13}}\bigg[\frac{z_3}{z_2}+\frac{1-z_2}{1-z_3}\bigg]
  +(1\leftrightarrow 2)\bigg\}\;.
  \end{split}
\end{equation}
In the strongly ordered soft-collinear limit, $p_1\parallel p_2$,
one obtains the product of a soft-emission term, times a spin-correlated
decay to two gluons. This factorized form is obtained from
Eqs.~\eqref{eq:pqq_munu_sc_components} and~\eqref{eq:coll_ggg_sym_avg}.
The result is
\begin{equation}\label{eq:tc_qgg_scalar}
  \begin{split}
  P_{\tilde{q}\to gg\tilde{q}}^{\rm(nab)}(p_1,p_2,p_3)\to&\;
  \frac{s_{123}}{s_{12}}\,P_{\tilde{q}\to\tilde{q}}^{\mu\nu}(p_3,p_{12})
  P_{g\to g,\mu\nu}(p_1,p_2)\\
  =&\;C_FC_A\frac{s_{123}}{s_{12}}\bigg[
  \frac{4z_3}{z_1+z_2}\bigg(1-\frac{s_{12}}{s_{123}}\frac{1}{z_1+z_2}\bigg)
  \bigg(\frac{z_1}{z_2}+\frac{z_2}{z_1}\bigg)
  +\frac{1-\eps}{2}\frac{t_{12,3}^2}{s_{123}s_{12}}\bigg]\;.
  \end{split}
\end{equation}

\subsubsection{Gluon initial state}
\label{sec:one_to_three_splittings_gluon}
\begin{figure}
    \centerline{\includegraphics[scale=0.36]{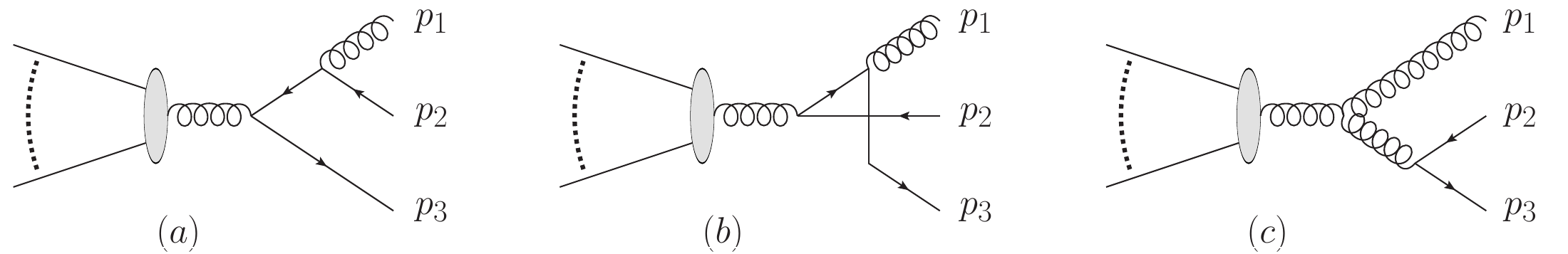}}
    \caption{Feynman diagrams leading to the $1\to 3$ gluon-to-quark-gluon splitting functions
    discussed in Sec.~\ref{sec:one_to_three_splittings_gluon}. The shaded blob and lines
    to the left represent the hard process with its associated external partons.
    See the main text for details.
    \label{fig:one-to-three_splittings_gluon_quark}}
\end{figure}
The Feynman diagrams needed to compute the three-parton gluon-to-quark splitting function
are shown in Fig.~\ref{fig:one-to-three_splittings_gluon_quark}. The abelian component
is determined from Figs.~\ref{fig:one-to-three_splittings_gluon_quark}(a) and~(b). We find
\begin{equation}\label{eq:tc_gqqb_ab}
  \begin{split}
  P^{\mu\nu\,\rm(ab)}_{g\to gq\bar{q}}(p_1,p_2,p_3)=&\;C_FT_R\bigg\{
  d^{\mu\nu}(p_{123},\bar{n})\bigg[\,\frac{2s_{123}s_{23}}{s_{12}s_{13}}
    +(1-\eps)\bigg(\frac{s_{12}}{s_{13}}+\frac{s_{13}}{s_{12}}\bigg)-2\eps\,\bigg]\\
  &\;\qquad\qquad+\frac{4s_{123}}{s_{12}s_{13}}\bigg(\tilde{p}_{2,13}^\mu\tilde{p}_{3,12}^\nu
    +\tilde{p}_{3,12}^\mu\tilde{p}_{2,13}^\nu-(1-\eps)\,\tilde{p}_{1,23}^\mu\,\tilde{p}_{1,23}^\nu
    \bigg)\bigg\}\;.
  \end{split}
\end{equation}
With the help of the Sudakov decomposition in Eq.~\eqref{eq:sudakov_decomposition_cg},
the expression can be written in terms of transverse momenta and reduced to the
leading-power result given in~\cite{Catani:1999ss}.
Similarly, the on-shell non-abelian part of the splitting tensor of a gluon
into a gluon and a $q\bar{q}$-pair is determined from the diagrams in
Figs.~\ref{fig:one-to-three_splittings_gluon_quark}(a)-(c).
\begin{equation}\label{eq:tc_gqqb_nab}
  \begin{split}
    P^{\mu\nu\,\rm(nab)}_{g\to gq\bar{q}}&(p_1,p_2,p_3)=
    -\frac{C_A}{2C_F}\,P^{\mu\nu\,\rm(ab)}_{g\to gq\bar{q}}(p_1,p_2,p_3)+\frac{C_AT_R}{4}\bigg\{
    \frac{s_{123}}{s_{23}^2}\bigg[-d^{\mu\nu}(p_{123},\bar{n})\frac{t_{23,1}^2}{s_{123}}
    -16\,\frac{1-z_1}{z_1}\,\tilde{p}_{2,3}^\mu\tilde{p}_{2,3}^\nu\,\bigg]\\
    &\;+d^{\mu\nu}(p_{123},\bar{n}) \bigg[\,\frac{2s_{13}}{s_{12}}(1-\eps)
    + \frac{2s_{123}}{s_{12}}\bigg(\frac{1-z_3}{z_1(1-z_1)}-2\bigg)
    + \frac{2s_{123}}{s_{23}}\frac{1-z_1+2z_1^2}{z_1(1-z_1)}-1\,\bigg] \\
    &\;+ \frac{s_{123}}{s_{12}s_{23}} \bigg[\,
    2s_{123}\,d^{\mu\nu}(p_{123},\bar{n})\frac{z_2(1-2z_1)}{z_1(1-z_1)}
    -16\tilde{p}_{3,12}^\mu\tilde{p}_{3,12}^\nu\frac{z_2^2}{z_1(1-z_1)}
    +8(1-\eps)\tilde{p}_{2,13}^\mu\tilde{p}_{2,13}^\nu\\
    &\qquad\;+4(\tilde{p}_{2,13}^\mu \tilde{p}_{3,12}^\nu+\tilde{p}_{3,12}^\mu \tilde{p}_{2,13}^\nu)
    \bigg( \frac{2z_2 (z_3-z_1)}{z_1 (1-z_1)}+1-\eps\bigg)\bigg]+(2\leftrightarrow 3)\bigg\}
    +P^{\mu\nu\,\rm(nab,\bar{n})}_{g\to gq\bar{q}}(p_1,p_2,p_3)\;.
  \end{split}
\end{equation}
This splitting tensor contains an explicitly $\bar{n}$-dependent contribution,
listed in App.~\ref{sec:explicit_gauge_dependence}, which does not contribute
in the triple-collinear limit, cf.\ the discussion of Eq.~\eqref{eq:tc_factorization}.
The spin-averaged splitting functions can be obtained by contracting Eqs.~\eqref{eq:tc_gqqb_ab}
and~\eqref{eq:tc_gqqb_nab} with $d_{\mu\nu}(p_{123},\bar{n})$. They are given in Eqs.~(68) and~(69)
of~\cite{Catani:1999ss}.

\begin{figure}
    \centerline{\includegraphics[scale=0.36]{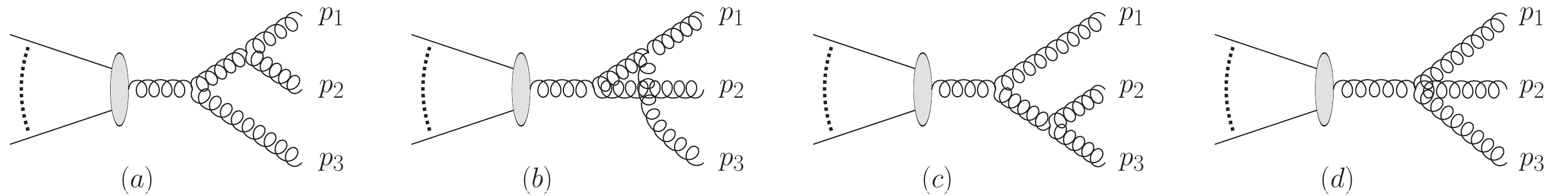}}
    \caption{Feynman diagrams leading to the $1\to 3$ gluon-to-gluon splitting function
    discussed in Sec.~\ref{sec:one_to_three_splittings_gluon}. The shaded blob and lines
    to the left represent the hard process with its associated external partons.
    See the main text for details.
    \label{fig:one-to-three_splittings_gluon_gluon}}
\end{figure}
The one-to-three gluon splitting tensor is obtained from the diagrams
in Fig.~\ref{fig:one-to-three_splittings_gluon_gluon}. It was computed
in~\cite{Catani:1999ss} and reads
\begin{equation}\label{eq:tc_ggg}
  \begin{split}
  P_{g\to ggg}^{\mu\nu}&(p_1,p_2,p_3)=C_A^2\bigg\{
  \frac{1-\eps}{4s_{12}^2}\bigg[d^{\mu\nu}(p_{123},\bar{n})
  \Big(t_{12,3}^2+3s_{12}^2\Big)+16s_{123}\frac{1-z_3}{z_3}\,
  \tilde{p}_{1,2}^\mu\tilde{p}_{1,2}^\nu\bigg]\\
  &\;\qquad-\frac{s_{123}}{s_{12}}\frac{d^{\mu\nu}(p_{123},\bar{n})}{z_3}
  \bigg[\frac{2(1-z_3)+4z_3^2}{1-z_3}-\frac{1-2z_3(1-z_3)}{z_1(1-z_1)}\bigg]\\
  &\;\qquad+\frac{s_{123}(1-\eps)}{s_{12}s_{13}}\bigg[
  2 z_1\bigg(\tilde{p}_{2,13}^\mu\tilde{p}_{2,13}^\nu\frac{1-2z_3}{z_3(1-z_3)}
  +\tilde{p}_{3,12}^\mu\tilde{p}_{3,12}^\nu\frac{1-2z_2}{z_2(1-z_2)}\bigg)\\
  &\;\qquad\quad-\frac{s_{123}}{2(1-\eps)}d^{\mu\nu}(p_{123},\bar{n})\bigg(
  \frac{4z_2z_3+2z_1(1-z_1)-1}{(1-z_2)(1-z_3)}-\frac{1-2z_1(1-z_1)}{z_2z_3}\bigg)\\
  &\;\qquad\quad+\big(\tilde{p}_{2,13}^\mu\tilde{p}_{3,12}^\nu
  +\tilde{p}_{3,12}^\mu\tilde{p}_{2,13}^\nu\big)
  \bigg(\frac{2z_2(1-z_2)}{z_3(1-z_3)}-3\bigg)\bigg]+(\text{5 permutations})\bigg\}
  +P^{\mu\nu\,\rm(\bar{n})}_{g\to ggg}(p_1,p_2,p_3)\;.
  \end{split}
\end{equation}
This splitting tensor contains an explicitly $\bar{n}$-dependent contribution,
listed in App.~\ref{sec:explicit_gauge_dependence}, which does not contribute
in the triple-collinear limit, cf.\ the discussion of Eq.~\eqref{eq:tc_factorization}.
The spin-averaged splitting function, obtained by contracting Eq.~\eqref{eq:tc_ggg}
with $d_{\mu\nu}(p_{123},\bar{n})$, is given in Eq.~(70) of~\cite{Catani:1999ss}.

\subsection{Scalar multipoles}
\label{sec:tree_level_scalar_multipoles}
In this section we focus on the radiation pattern of the scalar emitters in 
Eq.~\eqref{eq:tree_level_building_blocks}. Their gauge invariant extension
is given by the current in Eq.~\eqref{eq:scalar_current_sum}. When computing
splitting functions, the role of the charge partners that appear explicitly
in Eq.~\eqref{eq:scalar_current_sum} is played by a light-like Wilson line
that enters the calculation through the use of the axial gauge. When constructing
infrared counterterms for NNLO subtraction schemes, this simple approximation
would lead to the wrong angular radiation pattern away from the collinear limit.
It is therefore necessary to subtract the scalar part of the splitting functions
and replace it by a dipole approximation that accounts for the fact that all
color charged final-state particles radiate gluons as a coherent multipole.
Here, we derive the expressions that are necessary for this. They will allow us
to formulate the scalar results in Secs.~\ref{sec:two-parton_tree-level}
and~\ref{sec:three-parton_tree-level} as special cases of the dipole expressions.
While the calculation and its results are technically cumbersome, this section
is of central importance for the understanding of our method. It is crucial that
the identification of the scalar part of the splitting function is achieved
without invoking kinematical limits. As explained in Sec.~\ref{sec:motivation},
any such limit would obscure the structure of the splitting function remainder
due to sub-leading power corrections. We begin the discussion with the simplest
case, that of single-gluon emission. We will find that this particular case
already has some interesting features, which have been exploited in various
resummation algorithms to solve the soft double counting
problem~\cite{Catani:1990rr,Catani:1992ua}.

\subsubsection{Single gluon emission}
\label{sec:one-gluon_scalar_radiators}
The emission of a single gluon from a pair of charged scalar particles
in the fundamental representation is described by the coherent sum of
the two diagrams in Fig.~\ref{fig:scalar_single_emission}:
\begin{figure}
    \includegraphics[scale=0.36]{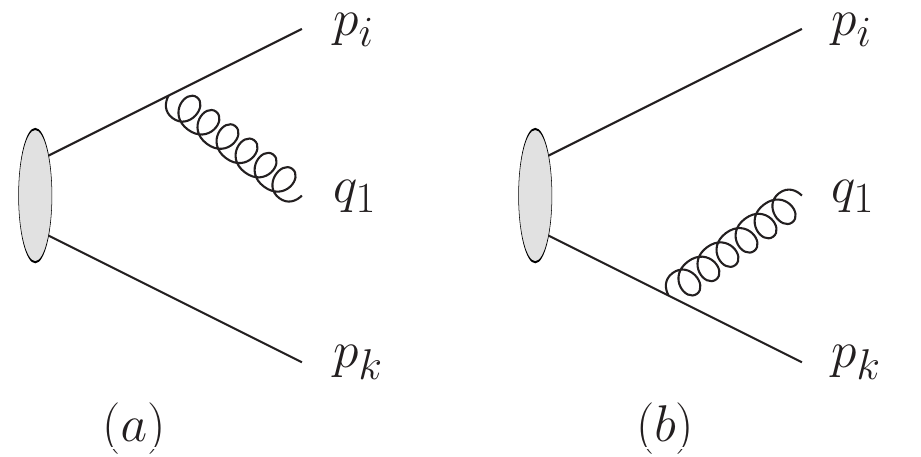}
    \caption{Diagrams contributing to the radiation of a single gluon from scalar dipoles.}
    \label{fig:scalar_single_emission}
\end{figure}
\begin{equation}\label{eq:single_gluon_emission_1}
\begin{split}
    \mathcal{S}_g^{\mp}(p_i,p_k;q_1;n) =&\; \big(T^a_{ij}S^\mu(p_i,q_1)\mp T^{a}_{lk}S^\mu(p_k,q_1)\big)
    \big(T^a_{ij}S^\nu(p_i,q_1)\mp T^{a}_{lk}S^\nu(p_k,q_1)\big)\,d_{\mu\nu}(q_1,n)\;,
\end{split}
\end{equation}
where $\mathcal{S}^-$ applies to opposite sign charges and $\mathcal{S}^+$ to same
sign charges. Note that we have dropped the coupling factors still present
in Eq.~\eqref{eq:scalar_current_sum}, to be consistent with the definition
of the splitting functions in Secs.~\ref{sec:two-parton_tree-level}
and~\ref{sec:three-parton_tree-level}. We can use the notation
of~\cite{Bassetto:1984ik,Catani:1996vz} to generalize 
Eq.~\eqref{eq:single_gluon_emission_1} to arbitrary radiators
\begin{equation}\label{eq:one-gluon_insertion_1}
  \begin{split}
  \mathcal{S}_g(\{p\};q_1;\bar{n}) = &\; \sum_{i,k}
    \hat{\bf T}_i\hat{\bf T}_k\,
    \mathcal{S}_{i;k}(q_1;\bar{n})\;.
  \end{split}
\end{equation}
The space-time dependent part of individual radiators for on-shell gluons
is given by
\begin{equation}\label{eq:one-gluon_radiator}
  \begin{split}
    \mathcal{S}_{i;k}(q_1;\bar{n})=&\;
    \frac{1}{p_{i1}^2}\frac{2z_i}{z_1}\bigg(1-\frac{p_k^2}{p_{k1}^2}\bigg)
    +\frac{1}{p_{k1}^2}\frac{2z_k}{z_1}\bigg(1-\frac{p_i^2}{p_{i1}^2}\bigg)
    -\frac{4p_ip_k}{p_{i1}^2p_{k1}^2}\;.
  \end{split}
\end{equation}
To simplify the notation in
Sec.~\ref{sec:tree-level_splitting_decomposition}, we introduce an
analogous notation for the gluon-spin dependent radiator.
\begin{equation}\label{eq:one-gluon_insertion_spin}
  \begin{split}
  \mathcal{S}^{\mu\nu}_g(\{p\};q_1)=&\;
  \sum_{i,k}
    \hat{\bf T}_i\hat{\bf T}_k\,
    \mathcal{S}^{\mu\nu}_{i;k}(q_1)\;,
    \quad&&\text{where}\quad
    &\mathcal{S}^{\mu\nu}_{i;k}(q_1)=&\;
    S^\mu(p_i,q_1)S^\nu(p_k,q_1)\;.
  \end{split}
\end{equation}
Neglecting the momentum dependence of the hard matrix element, and
assuming on-shell radiators, we can use color conservation in the form
$\hat{\bf T}_i^2=-\sum_{k\neq i}\hat{\bf T}_i\hat{\bf T}_{k}$
to eliminate the terms where $i=k$ from Eq.~\eqref{eq:one-gluon_insertion_1}.
\begin{equation}\label{eq:one-gluon_insertion_2}
  \begin{split}
  \mathcal{S}_g(\{p\};q_1;\bar{n})=-\,\sum_{i,k\neq i}
    \frac{\hat{\bf T}_i\hat{\bf T}_k}{p_iq_1}\bigg(
    \frac{p_ip_k}{p_kq_1}
    -\frac{p_i^2}{p_iq_1}
     +\frac{z_i}{z_1}
    -\frac{z_k}{z_1}\frac{p_iq_1}{p_kq_1}\bigg)\,.
  \end{split}
\end{equation}
If the gauge vector $n^\mu$ is chosen such as to define the frame
of the color multipole, this expression is simply the additively matched
radiator function in the angular ordered parton shower approach
to QCD resummation~\cite{Marchesini:1987cf,Marchesini:1989yk}.
Its properties have also been exploited for analytic
resummation~\cite{Catani:1990rr,Catani:1992ua}.
However, the more common way to write Eq.~\eqref{eq:one-gluon_insertion_2}
is to use its symmetry properties and remove the ratios of $z$ factors, making
positivity and gauge invariance manifest.
\begin{equation}\label{eq:one-gluon_insertion_3}
  \begin{split}
  \mathcal{S}_g(\{p\};q_1) = &\; -\,\sum_{i,k\neq i}
    \hat{\bf T}_i\hat{\bf T}_k\bigg(
    \frac{p_ip_k}{(p_iq_1)(p_kq_1)}
    -\frac{p_i^2/2}{(p_iq_1)^2}
    -\frac{p_k^2/2}{(p_kq_1)^2}\bigg)\;.
  \end{split}
\end{equation}
In any collinear limit, $p_i\parallel q_1$, the $i=k$ term in
Eq.~\eqref{eq:one-gluon_insertion_1} approaches $p_{i1}^{-2}4z_i/z_j$, which,
up to a color prefactor and the collinear propagator agrees with the $2z_i/z_j$
contribution of the DGLAP splitting functions in Eqs.~\eqref{eq:coll_q_to_qg_components}
and~\eqref{eq:ggg_dglap_limit_sum}. In this case the color correlations
become trivial, and the sole remaining term is proportional to the
quadratic Casimir operator. In general, we can identify the $i=k$ term
in Eq.~\eqref{eq:one-gluon_insertion_1} with the contributions proportional
to the scalar radiators squared in Eqs.~\eqref{eq:pqq_munu_sc_components}
and~\eqref{eq:coll_ggg_sym}.

\subsubsection{Quark-antiquark emission}
\label{sec:two-quark_scalar_radiators}
\begin{figure}
    \includegraphics[scale=0.36]{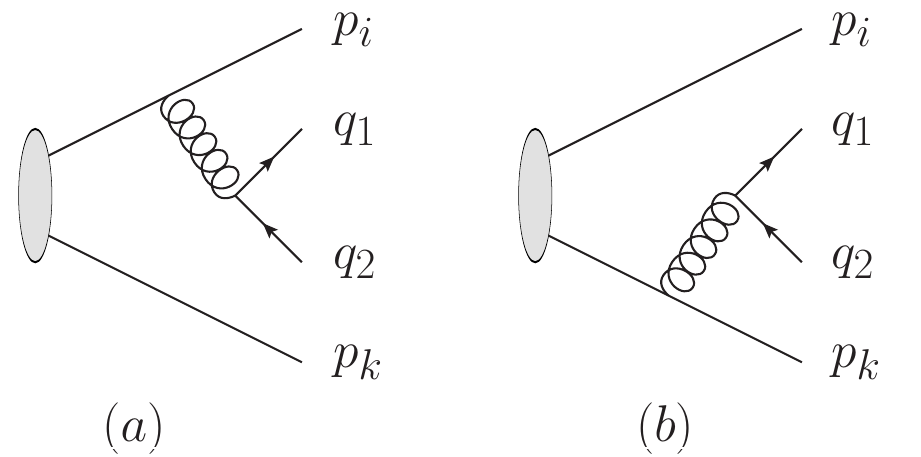}
    \caption{Double-branching diagrams contributing to the emission
      of a quark-antiquark pair from scalar dipoles.}
    \label{fig:scalar_quark_antiquark_emission}
\end{figure}
The emission of a quark-antiquark pair from a pair of scalar radiators
in the fundamental representation is given by the coherent sum of
the two diagrams in Fig.~\ref{fig:scalar_quark_antiquark_emission}.
Making use of color charge operators, this can be generalized to the emission
off a scalar current in the form of Eq.~\eqref{eq:scalar_current_sum}.
Charge conservation and the transversality of the decay vertex $F^\mu$
in Eq.~\eqref{eq:tree_level_building_blocks} can be used to simplify
the computation~\cite{Catani:1999ss}, leading to the following result
for massless final-state partons:
\begin{equation}\label{eq:scalar_emission_quark_pair}
  \begin{split}
  \mathcal{S}_{q\bar{q}}(\{p\};q_1,q_2) = &\; \sum_{i,k}
    \hat{\bf T}_i\hat{\bf T}_k\,4T_R\,\frac{
    s_{i1}s_{k2}+s_{i2}s_{k1}-s_{ik}s_{12}}{
    s_{12}^2\,s_{i12}\, s_{k12}}\;.
  \end{split}
\end{equation}
Note that in contrast to Eq.~(96) in Ref.~\cite{Catani:1999ss},
Eq.~\eqref{eq:scalar_emission_quark_pair} contains the physical propagators
rather than their eikonal counterparts. The matching to the squared amplitudes
in Sec.~\ref{sec:three-parton_tree-level} is better understood when retaining
the gauge dependence term by term. This leads to the following result
\begin{equation}\label{eq:scalar_emission_quark_pair_2}
  \begin{split}
  \mathcal{S}_{q\bar{q}}(\{p\};q_1,q_2;\bar{n}) = &\; \sum_{i,k}
    \hat{\bf T}_i\hat{\bf T}_k\,T_R\,\mathcal{S}^{(q\bar{q})}_{i;k}(q_1,q_2;\bar{n})\;,
  \end{split}
\end{equation}
where the space-time dependent part of the individual radiators is given by
\begin{equation}\label{eq:scalar_emission_quark_pair_individual}
  \begin{split}
  \mathcal{S}^{(q\bar{q})}_{i;k}(q_1,q_2;\bar{n}) = &\;
  \frac{2}{s_{i12}\,s_{k12}}\bigg[\,
    \frac{2}{s_{12}}\bigg(
    \frac{z_is_{k12}+z_ks_{i12}}{z_1+z_2}
    -s_{ik}\bigg)+1-\frac{t_{12,i}t_{12,k}}{s_{12}^2}\,\bigg]\;.
  \end{split}
\end{equation}
For $i=k$, this expression corresponds to twice the squared scalar
splitting amplitude in Eq.~\eqref{eq:tc_qqpqpb_scalar}. 

\subsubsection{Two-gluon emission}
\label{sec:two-gluon_scalar_radiators}
\begin{figure}
    \includegraphics[scale=0.36]{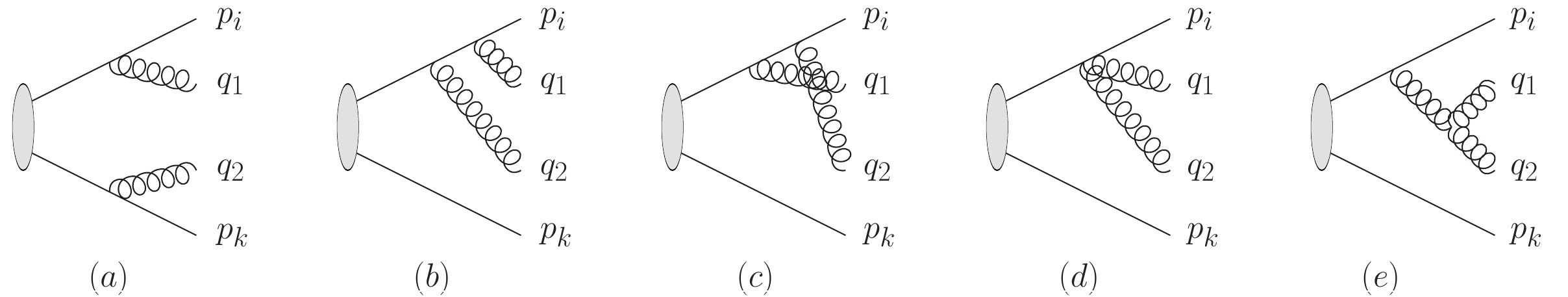}
    \caption{Types of diagrams contributing to the two-gluon current for
    emission of the charged parton pair $\{i,k\}$. The same diagrams exist
    with $i$ and $k$ interchanged.}
    \label{fig:scalar_double_gluon_radiation}
\end{figure}

The emission of a gluon pair can be described in a fashion similar to the
double-gluon soft current in~\cite{Catani:1999ss}. The diagrams contributing
to this current are shown in Fig.~\ref{fig:scalar_double_gluon_radiation}.
The main difference to the computation of the soft current lies in the form
of the one-gluon current, which is given by Eq.~\eqref{eq:scalar_current_sum}
rather than its eikonal counterpart. However, this current is not conserved
if the scalar particle is off mass-shell, a problem that is solved in the 
abelian case by the seagull diagrams in Fig.~\ref{fig:scalar_double_gluon_radiation}(d),
and in the non-abelian case by the use of a physical gauge or external ghosts.
For massless partons, we obtain the following simple result
\begin{equation}\label{eq:two-gluon_current}
    \begin{split}
        &J^{\mu\nu}_{ab}(q_1,q_2)=
        \sum_{i,k}\big\{\hat{\bf T}^a_i,\hat{\bf T}^b_k\big\}\,
          \mathcal{J}^{{\rm(ab)},\mu\nu}_{ik}(q_1,q_2)
        +\sum_i\,if^{abc}\hat{\bf T}_i^c\,
          \mathcal{J}^{{\rm(nab)},\mu\nu}_{i}(q_1,q_2)\;.
    \end{split}
\end{equation}
The color-stripped abelian and non-abelian two-gluon
currents $\mathcal{J}^{\mu\nu}_{ik}$ are defined as
\begin{equation}\label{eq:two-gluon_currents}
  \begin{split}
    &\mathcal{J}^{{\rm(ab)},\mu\nu}_{ik}(q_1,q_2)=
    \frac{1}{2}\,S^\mu(p_i,q_1)S^\nu(p_k,q_2)\\
    &\;\qquad+\frac{\delta_{ik}}{(p_i+q_{12})^2}\,\bigg[\, 
    q_1^\nu S^\mu(p_i,q_1)+q_2^\mu S^\nu(p_i,q_2)
    -q_1q_2\,S^\mu(p_i,q_1)S^\nu(p_i,q_2)-g^{\mu\nu}\bigg]\;,\\
    &\mathcal{J}^{{\rm(nab)},\mu\nu}_{i}(q_1,q_2)=
    S_\rho(p_i,q_{12})
    \Big(d^{\rho\nu}(q_{12})S^\mu(q_2,q_1)-d^{\rho\mu}(q_{12})S^\nu(q_1,q_2)\Big)\\
    &\;\qquad+\frac{1}{(p_i+q_{12})^2}\,
    \bigg[\,q_1^\nu S^\mu(p_i,q_1)-q_2^\mu S^\nu(p_i,q_2)
    +p_i(q_2-q_1)\,S^\mu(p_i,q_1)S^\nu(p_i,q_2)
    +\frac{t_{12,i}}{q_{12}^2}\,g^{\mu\nu}\,\bigg]\;.
  \end{split}
\end{equation}
In the double-soft limit, $q_{1/2}\to\lambda\, q_{1/2}$, these functions
reduce to the double-soft current in Eq.~(101) of Ref.~\cite{Catani:1999ss}. 
Making use of the color-space identity~\footnote{See also App.~B of
Ref.~\cite{Czakon:2011ve}, which contains the symmetric part of this equation.}
\begin{equation}\label{eq:two-gluon_current_squared_color_decomposition}
  \begin{split}
    &\big\{\hat{\bf T}^a_i,\hat{\bf T}^b_k\big\}
    \big\{\hat{\bf T}^a_l,\hat{\bf T}^b_m\big\}=
    \Big\{\hat{\bf T}^a_i\hat{\bf T}^a_l,
    \hat{\bf T}^b_k\hat{\bf T}^b_m\Big\}
    +i\big(2\delta_{kl}-\delta_{lm}\big)
    \big(1-\delta_{ik}\big)\big(1-\delta_{km}\big)\big(1-\delta_{im}\big)
    f^{abc}\,\hat{\bf T}^a_i\hat{\bf T}^b_k\hat{\bf T}^c_m\\
    &\;\qquad-C_A\,\bigg(\!\delta_{kl}-\frac{\delta_{lm}}{2}\!\bigg)
    \bigg(\delta_{ik}\hat{\bf T}^a_i\hat{\bf T}^a_m
    +\big(1-\delta_{ik}\big)\big(\delta_{km}
    -\delta_{im}\big)\hat{\bf T}^a_i\hat{\bf T}^a_k\bigg)
    +\bigg(\!\begin{array}{r@{\;}l}i&\leftrightarrow k\\
    l&\leftrightarrow m\end{array}\!\!\bigg)\;,
  \end{split}
\end{equation}
we obtain a relatively compact form of the two-gluon squared current.
In particular, due to its symmetries, the result is free of terms 
of the form $if^{abc}\,\hat{\bf T}^a_i\hat{\bf T}^b_k\hat{\bf T}^c_m$.
\begin{equation}\label{eq:squared_two-gluon_current}
  \begin{split}
    &\big[J^{ab}_{\mu\nu}(q_1,q_2)\big]^\dagger
    d^{\mu\rho}(q_1,\bar{n})d^{\nu\sigma}(q_2,\bar{n})
    J^{ab}_{\rho\sigma}(q_1,q_2)=2\sum_{i,k}\sum_{l,m}
    \Big\{\hat{\bf T}^a_i\hat{\bf T}^a_l,
    \hat{\bf T}^b_k\hat{\bf T}^b_m\Big\}\,
    \mathcal{S}^{\rm(ab)}_{i,k;l,m}(q_1,q_2;\bar{n})\\
    &\;\quad+2\sum_{i,k}\sum_{l}
    \Big(\Big\{\hat{\bf T}^a_i\hat{\bf T}^a_l,
    \hat{\bf T}^b_k\hat{\bf T}^b_l\Big\}+
    \Big\{\hat{\bf T}^a_l\hat{\bf T}^a_i,
    \hat{\bf T}^b_l\hat{\bf T}^b_k\Big\}\Big)\,
    \mathcal{S}^{\rm(ab)}_{i,k;l}(q_1,q_2;\bar{n})
    +2\sum_{i,l}\Big\{\hat{\bf T}^a_i\hat{\bf T}^a_l,
    \hat{\bf T}^b_i\hat{\bf T}^b_l\Big\}\,
    \mathcal{S}^{\rm(ab)}_{i;l}(q_1,q_2)\\
    &\;\quad-\sum_{i,l}\,C_A\,\hat{\bf T}^c_i\hat{\bf T}^c_l\,
    \Big[\;\mathcal{S}^{\rm(nab)}_{i;l}(q_1,q_2;\bar{n})
    -(1-2\delta_{il})\,\mathcal{S}^{\rm(ab)}_{i;l}(q_1,q_2)
    -\mathcal{S}^{\rm(ab)}_{i,i;l}(q_1,q_2;\bar{n})
    -\mathcal{S}^{\rm(ab)}_{l,l;i}(q_1,q_2;\bar{n})\\
    &\;\qquad\qquad\qquad\qquad\qquad
    +\mathcal{S}^{\rm(ab)}_{i,l;i}(q_1,q_2;\bar{n})
    +\mathcal{S}^{\rm(ab)}_{l,i;i}(q_1,q_2;\bar{n})
    +\mathcal{S}^{\rm(ab)}_{i,l;l}(q_1,q_2;\bar{n})
    +\mathcal{S}^{\rm(ab)}_{l,i;l}(q_1,q_2;\bar{n})\,\Big]\;.
  \end{split}
\end{equation}
We have defined the abelian and non-abelian radiator functions,
$\mathcal{S}(q_1,q_2;\bar{n})$, which are natural extensions of the
soft functions in Ref.~\cite{Catani:1999ss}. For on-shell radiators
they are given by
\begin{equation}\label{eq:squared_two-gluon_current_ab}
  \begin{split}
    \mathcal{S}^{\rm(ab)}_{i,k;l,m}(q_1,q_2;\bar{n})=&\;
    \frac{1}{4}\,\mathcal{S}_{i;l}(q_1;\bar{n})\mathcal{S}_{k;m}(q_2;\bar{n})
    \;,\\
    \mathcal{S}^{\rm(ab)}_{i,k;l}(q_1,q_2;\bar{n})=&\;
    \frac{1}{s_{l12}}\frac{s_{il}s_{kl}}{s_{i1}s_{k2}}
    \bigg(\frac{s_{k1}}{s_{kl}s_{l1}}+\frac{s_{i2}}{s_{il}s_{l2}}
    -\frac{s_{12}}{s_{l1}s_{l2}}\bigg)\\
    &\hspace*{-15mm}+\frac{1}{s_{l12}}\bigg[\frac{z_l}{z_2}\bigg(
    \frac{s_{il}s_{12}-s_{i2}s_{l1}}{s_{i1}s_{l1}s_{l2}}
    -\frac{z_ls_{12}}{2z_1\,s_{l1}s_{l2}}\bigg)
    +\frac{z_is_{l1}+z_ls_{i1}-z_1s_{il}}{z_2\,s_{i1}s_{l1}}
    -\frac{s_{ik}}{2s_{i1}s_{k2}}
    +\bigg(\!\!\begin{array}{r@{\,}l}i&\leftrightarrow k\\
    1&\leftrightarrow 2\end{array}\!\!\bigg)\bigg]\;,\\
    \mathcal{S}^{\rm(ab)}_{i;k}(q_1,q_2)=&\;
    \frac{(s_{i1}s_{k2}-s_{i2}s_{k1})^2
    -2s_{12}s_{ik}(s_{i1}s_{k2}+s_{i2}s_{k1})
    +s_{12}^2s_{ik}^2}{s_{i12}s_{k12}\,
    s_{i1}s_{k1}s_{i2}s_{k2}}
    +\frac{2(1-\eps)}{s_{i12}s_{k12}}\;,
  \end{split}
\end{equation}
and
\begin{equation}\label{eq:squared_two-gluon_current_nab}
  \begin{split}
    &\mathcal{S}^{\rm(nab)}_{i;k}(q_1,q_2;\bar{n})=
    \mathcal{S}^{({\rm nab},p)}_{i;k}(q_1,q_2;\bar{n})
    -2(1-\eps)\,\mathcal{S}^{\mu\nu}_{i;k}(q_{12};\bar{n})
    D_\mu(q_1,q_2,\bar{n})D_\nu(q_1,q_2,\bar{n})\\
    &\;\quad+\frac{1}{4}\,\Big(\bar{\mathcal{S}}_{i;i}(q_{12};\bar{n})
    +\bar{\mathcal{S}}_{k;k}(q_{12};\bar{n})-2\bar{\mathcal{S}}_{i;k}(q_{12};\bar{n})\Big)
    \bigg[\,\mathcal{S}_{i;i}(q_2;\bar{n})+\mathcal{S}_{1;1}(q_2;\bar{n})-2\mathcal{S}_{i;1}(q_2;\bar{n})\\
    &\;\qquad\qquad\qquad-\frac{1}{2}\Big(\mathcal{S}_{i;i}(q_2;\bar{n})+\mathcal{S}_{k;k}(q_2;\bar{n})
    -2\mathcal{S}_{i;k}(q_2;\bar{n})\Big)
    +\big(i\leftrightarrow k\big)+\big(1\leftrightarrow 2\big)
    +\bigg(\!\!\begin{array}{r@{\,}l}i&\leftrightarrow k\\
    1&\leftrightarrow 2\end{array}\!\!\bigg)\bigg]\;,
  \end{split}
\end{equation}
where
\begin{equation}\label{eq:squared_two-gluon_current_pnab}
  \begin{split}
    &\mathcal{S}^{({\rm nab},p)}_{i;k}(q_1,q_2;\bar{n})=
    \bigg[\frac{3s_{ik}}{4s_{i1}s_{k1}}+\frac{2s_{12}}{z_{12}^2}
    \bigg(\frac{z_i}{s_{i12}}-\frac{z_k}{s_{k12}}\bigg)
    \bigg(\frac{z_{i12}}{s_{i12}}-\frac{z_{k12}}{s_{k12}}\bigg)\bigg]
    \bigg(\frac{2s_{i1}}{s_{12}s_{i2}}-\frac{s_{ik}}{s_{i2}s_{k2}}\bigg)\\
    &\;\quad+\frac{1}{s_{i12}s_{k12}}\bigg\{
    \frac{3}{2}+\frac{3}{2}\frac{s_{ik}^2}{s_{i1}s_{k1}}+\frac{s_{ik}}{2s_{12}}-\frac{z_2}{z_1}
    +\frac{s_{k1}s_{i2}-s_{ik}s_{12}}{2s_{i1}s_{k2}}\bigg(1-\frac{s_{ik}}{s_{12}}\bigg)
    +\frac{s_{ik}}{s_{i1}}\bigg(1-\frac{3s_{i12}}{s_{12}}\bigg)\\
    &\;\quad\qquad
    -\frac{s_{i2}}{s_{i1}}\frac{z_i}{z_2}\bigg(2+\frac{z_1}{z_i}+\frac{s_{i12}}{s_{i2}}+\frac{s_{k12}}{s_{i2}}\bigg)
    +\frac{2s_{k1}}{s_{12}}\bigg(\frac{z_1-z_2}{z_1}\frac{z_2-z_i}{z_2}
    -\frac{z_is_{i12}}{z_2s_{i1}}+\frac{s_{i2}}{s_{i1}}-\frac{s_{i1}}{s_{i2}}\frac{2z_i+z_2}{z_1}\bigg)\\
    &\;\quad\qquad-\frac{z_{12}^2}{2z_1z_2}\frac{t_{12,i}t_{12,k}}{s_{12}^2}
    -\frac{t_{12,k}}{2s_{12}}\bigg[\frac{s_{i1}}{s_{i2}}\frac{z_{i2}}{z_1}
    \bigg(1-\frac{z_is_{12}}{z_2s_{i1}}\bigg)+\frac{2z_{12}}{z_1}\bigg(1+\frac{s_{i1}}{s_{12}}\bigg)+\frac{3z_i}{z_2}
    -\big(1\!\leftrightarrow\! 2\big)\bigg]\bigg\}\\
    &\;\quad+\frac{z_i^2}{2s_{i1}s_{i2}z_1z_2}\bigg(1+\frac{s_{i1}-s_{i2}}{s_{i12}}\frac{s_{k1}-s_{k2}}{s_{k12}}\bigg)
    +\big(i\leftrightarrow k\big)+\big(1\leftrightarrow 2\big)+\bigg(\!\!\begin{array}{r@{\,}l}i&\leftrightarrow k\\
    1&\leftrightarrow 2\end{array}\!\! \bigg)\;.
  \end{split}
\end{equation}
These radiators are related to the splitting functions in
Eqs.~\eqref{eq:tc_sqgg} and~\eqref{eq:tc_sqgg_nab} as follows\footnote{
  Note that the factor 1/4 relating the abelian and non-abelian scalar splitting function
  in Eq.~\eqref{eq:scalar_tc_correspondence} is a consequence of the color algebra
  leading to the definition of the non-abelian radiator function in
  Eq.~\eqref{eq:squared_two-gluon_current}. It therefore differs from the factor 1/2
  obtained by direct diagrammatic calculation and resulting in the relation between
  the abelian and non-abelian splitting functions in Eqs.~\eqref{eq:tc_qgg_nab}
  and \eqref{eq:tc_sqgg_nab}.}:
\begin{equation}\label{eq:scalar_tc_correspondence}
  \begin{split}
    P^{\rm(ab)}_{\tilde{q}\to gg\tilde{q}}(p_1,p_2,p_i)=&\;
    s_{i12}^2\,C_F^2\Big(\mathcal{S}^{\rm(ab)}_{i,i;i,i}(p_1,p_2;\bar{n})
    +2\mathcal{S}^{\rm(ab)}_{i,i;i}(p_1,p_2;\bar{n})+
    \mathcal{S}^{\rm(ab)}_{i;i}(p_1,p_2)\Big)\;,\\ 
    P^{\rm(nab)}_{\tilde{q}\to gg\tilde{q}}(p_1,p_2,p_i)=&\;
    -s_{i12}^2\,\frac{C_FC_A}{4}\,\mathcal{S}^{\rm(nab)}_{i;i}(p_1,p_2;\bar{n})
    -\frac{C_A}{4C_F}P^{\rm(ab)}_{\tilde{q}\to gg\tilde{q}}(p_1,p_2,p_i)\;.
  \end{split}
\end{equation}
The shifted scalar radiator function, $\bar{\mathcal{S}}$, appearing
in Eq.~\eqref{eq:squared_two-gluon_current_pnab} is defined as an
extension of Eq.~\eqref{eq:one-gluon_radiator} to off-shell gluons
\begin{equation}\label{eq:shifted_scalar_radiator}
  \bar{\mathcal{S}}_{i;k}(q_1;\bar{n})=
  \mathcal{S}_{i;k}^{\mu\nu}(q_1;\bar{n})\,d_{\mu\nu}(\bar{q}_1,\bar{n})
  =\frac{1}{p_{i1}^2}\frac{2z_i}{z_1}
  \bigg(1-\frac{p_k^2}{p_{k1}^2}-\frac{q_1^2}{p_{k1}^2}\frac{z_{k1}}{z_1}\bigg)
  -\frac{2p_ip_k}{p_{i1}^2p_{k1}^2}+(i\leftrightarrow k)\;.
\end{equation}
The shifted momentum, $\bar{q}_1$, is given by Eq.~\eqref{eq:sudakov_decomposition_cg}.
It is not related to an actual alteration of the kinematics, but simply accounts
for the fact that gluon propagators are evaluated in axial gauge. By means of
Eq.~\eqref{eq:tc_factorization} we can relate this to the fact that only 
transverse gluon polarizations appear in the factorizable components of
Eq.~\eqref{eq:squared_two-gluon_current_nab}.
  
Next we derive the radiator functions in covariant gauge. Their abelian components
are given by
\begin{equation}\label{eq:squared_two-gluon_current_ab_cov}
  \begin{split}
    \mathcal{S}^{\rm(ab)}_{i,k;l,m}(q_1,q_2)=&\;
    \frac{s_{il}+s_{il1}}{2s_{i1}s_{l1}}
    \frac{s_{km}+s_{km2}}{2s_{k2}s_{m2}}\;,\\
    \mathcal{S}^{\rm(ab)}_{i,k;l}(q_1,q_2)=&\;
    \frac{1}{s_{l12}}\bigg[\frac{s_{il}}{s_{i1}s_{l1}}\frac{s_{k1}}{s_{k2}}
    -\frac{s_{ik}}{2s_{i1}s_{k2}}+\frac{s_{i2}}{2s_{i1}s_{l2}}
    -\frac{s_{12}}{8s_{l1}s_{l2}}\bigg(1+\frac{2s_{il}}{s_{i1}}\bigg)
    \bigg(1+\frac{2s_{kl}}{s_{k2}}\bigg)\bigg]
    +\bigg(\!\!\begin{array}{r@{\,}l}i&\leftrightarrow k\\
    1&\leftrightarrow 2\end{array}\!\! \bigg)\;.
  \end{split}
\end{equation}
The computation of the non-abelian components requires the introduction
of external ghosts~\cite{Kugo:1979gm,Gottschalk:1979wq,Kunszt:1979ci,Ellis:1985er},
because the currents corresponding to gluon 1 and 2 are not conserved independently.
We find
\begin{equation}\label{eq:squared_two-gluon_current_nab_cov}
  \begin{split}
    &\mathcal{S}^{\rm(nab)}_{i;k}(q_1,q_2)=
    \bigg[\frac{3}{4}+\frac{(s_{i1}-s_{i2})(s_{k1}-s_{k2})}{4s_{i12}s_{k12}}\bigg]
    \bigg[\frac{s_{ik}}{s_{i1}s_{12}s_{k2}}+\frac{s_{ik}}{s_{i2}s_{12}s_{k1}}
    -\frac{s_{ik}^2}{s_{i1}s_{k1}s_{i2}s_{k2}}\bigg]\hspace*{-5mm}\\
    &\;\quad+\frac{1}{s_{i1}}\bigg(\frac{s_{ik}}{s_{i2}s_{k1}}
    +\frac{1}{4s_{i2}}+\frac{1}{2s_{k2}}\bigg)
    +\frac{1}{s_{i12}}\bigg\{
    \frac{2s_{i1}}{s_{12}^2}-\frac{5}{4s_{12}}
    -\frac{s_{ik}+s_{12}}{s_{i1}s_{k2}}-\frac{s_{12}}{8s_{i1}s_{i2}}
    -\frac{s_{ik}}{s_{i1}s_{k1}}\bigg(1+\frac{s_{12}}{s_{i2}}\bigg)\bigg\}\\
    &\;\quad+\frac{1}{s_{i12}s_{k12}}\bigg\{
    \frac{s_{i1}}{s_{12}}\bigg(\frac{3s_{k2}+s_{k1}}{s_{i2}}-\frac{1}{2}\bigg)
    -\frac{2s_{ik}}{s_{i1}}\bigg(1+\frac{s_{i1}}{s_{12}}\bigg)
    -\frac{s_{12}(s_{ik}-s_{12})}{2s_{i1}s_{k2}}-\frac{1}{4}\\
    &\;\qquad\qquad-\frac{1-\eps}{2}
    \frac{(s_{i1}-s_{i2})(s_{k1}-s_{k2})}{s_{12}^2}\bigg\}
    +(i\leftrightarrow k)+(1\leftrightarrow 2)
    +\bigg(\!\!\begin{array}{r@{\,}l}i&\leftrightarrow k\\
    1&\leftrightarrow 2\end{array}\!\!\bigg)\;.
  \end{split}
\end{equation}
In the double-soft limit, Eq.~\eqref{eq:squared_two-gluon_current_nab_cov}
reduces to the double-soft function in Eq.~(110) of Ref.~\cite{Catani:1999ss}.

\subsection{Composition of one-to-three splitting functions}
\label{sec:tree-level_splitting_decomposition}
In this section, we present a decomposition of the splitting functions
in Eqs.~\eqref{eq:tc_qqpqpb}, \eqref{eq:tc_qqqb}, \eqref{eq:tc_qgg} and 
\eqref{eq:tc_qgg_nab}, and of the splitting tensors in
Eqs.~\eqref{eq:tc_gqqb_ab}, \eqref{eq:tc_gqqb_nab} and \eqref{eq:tc_ggg}.
In most existing approaches to NNLO subtraction or NNLL resummation,
their singular components are identified by analyzing the strongly ordered
soft and collinear limits, as well as the double soft limit. Knowing that
the splitting functions are in fact just off-shell squared matrix elements
in a particular gauge, we find it more useful to investigate their
diagrammatic structure. This allows us to achieve a clean separation
into the off-shell splitting tensors derived in
Eqs.~\eqref{eq:pqq_munu_sc},~\eqref{eq:coll_gqq} and~\eqref{eq:coll_ggg_sym},
and the two-gluon scalar radiators in Eqs.~\eqref{eq:tc_sqgg}
and~\eqref{eq:tc_sqgg_nab}, which are related via
Eq.~\eqref{eq:scalar_tc_correspondence} to the general scalar
radiators derived in Sec.~\ref{sec:two-gluon_scalar_radiators}.
This technique follows the general principle laid out in
Sec.~\ref{sec:motivation}. This section constitutes one of two main results
of this work, the other being the corresponding decomposition
of the one-loop one-to-two splitting functions in
Sec.~\ref{sec:one-loop_splitting_decomposition}.

\subsubsection{Quark initial state}
\label{sec:decomposition_quark}
Equation~\eqref{eq:tc_qqpqpb} describes the splitting of a quark to a
quark and a distinct flavor quark pair. The result is due to the sole
Feynman diagram in Fig.~\ref{fig:one-to-three_splittings_quark}~(a),
hence the splitting function trivially factorizes and can be written as
the product of Eqs.~\eqref{eq:pqq_munu_sc} and~\eqref{eq:coll_gqq}
\begin{equation}\label{eq:split_qbpqpq_decomposition}
  \begin{split}
    \langle P_{q\to \bar{q}'q'q}(p_1,p_2,p_3)\rangle=&\;
    \frac{s_{123}}{s_{12}}
    \langle P_{q\to q}^{\mu\nu}(p_3,p_{12})\rangle\,
    P_{g\to q,\mu\nu}(p_1,p_2)\;.
  \end{split}
\end{equation}
By means of Eq.~\eqref{eq:pqq_munu_sc}, the result can be assembled
from the scalar radiator in Eq.~\eqref{eq:one-gluon_insertion_spin},
the magnetic remainder in Eq.~\eqref{eq:pqq_munu_sc_components},
and the splitting function in Eq.~\eqref{eq:coll_gqq}
\begin{equation}\label{eq:split_qbpqpq_assembly}
  \begin{split}
    \langle P_{q\to \bar{q}'q'q}(p_1,p_2,p_3)\rangle
    =&\frac{s_{123}}{s_{12}}\bigg[\,
    \frac{C_F}{2}\,s_{123}\,\mathcal{S}^{\mu\nu}_{3;3}(p_{12})
    +\langle P_{q\to q}^{{\rm(f)}\mu\nu}(p_3,p_{12})\rangle\bigg]
    P_{g\to q,\mu\nu}(p_1,p_2)\;.
  \end{split}
\end{equation}
As discussed in Secs.~\ref{sec:motivation} and~\ref{sec:three-parton_tree-level},
the azimuthal angle dependence is correctly captured if and only if no kinematical
limits are taken in the factorized expression, since otherwise Eq.~\eqref{eq:def_cg_t123}
is reduced to the first term alone. This is an important new insight from our method,
which could not have been obtained with any existing technique for higher-order
computations.

Equation~\eqref{eq:tc_qqqb} describes the splitting of a quark to a
quark and a same flavor quark pair. The result involves both Feynman
diagrams in Fig.~\ref{fig:one-to-three_splittings_quark}, and hence
it is expected that it cannot be fully factorized. The leading 
singularities do, however, originate in the two possible double-collinear
limits. One can therefore express the splitting function in terms
of its leading components and the interference term~\cite{Catani:1999ss}
\begin{equation}\label{eq:split_qbqq_decomposition}
  \begin{split}
    \langle P_{q\to \bar{q}qq}(p_1,p_2,p_3)\rangle=&\;
    \frac{s_{123}}{s_{12}}
    \langle P_{q\to q}^{\mu\nu}(p_3,p_{12})\rangle\,
    P_{g\to q,\mu\nu}(p_1,p_2)+
    \frac{s_{123}}{s_{13}}
    \langle P_{q\to q}^{\mu\nu}(p_2,p_{13})\rangle\,
    P_{g\to q,\mu\nu}(p_1,p_3)\\
    &\;+\langle P^{(p)}_{q\to \bar{q}qq}(p_1,p_2,p_3)\rangle\;,
  \end{split}
\end{equation}
where
\begin{equation}\label{eq:pure_split_qbqq}
  \begin{split}
    \langle P^{(p)}_{q\to \bar{q}qq}(p_1,p_2,p_3)\rangle=&\;
    \langle P^{\rm(id)}_{q\to \bar{q}qq}(p_1,p_2,p_3)\rangle+
    \big(2\leftrightarrow 3\big)\;.
  \end{split}
\end{equation}
We will call $\langle P^{(p)}_{q\to \bar{q}qq}\rangle$ the
(spin-averaged) pure $q\to qq\bar{q}$ splitting function,
as it cannot be reconstructed from lower-order results.
By means of Eq.~\eqref{eq:pqq_munu_sc}, the splitting function
can be assembled systematically from the scalar radiator
in Eq.~\eqref{eq:one-gluon_insertion_spin}, the magnetic
remainder in Eq.~\eqref{eq:pqq_munu_sc_components},
the splitting function in Eq.~\eqref{eq:coll_gqq},
and the pure interference contribution
\begin{equation}\label{eq:split_qbqq_assembly}
  \begin{split}
    \langle P_{q\to \bar{q}qq}(p_1,p_2,p_3)\rangle
    =&\;\bigg\{\frac{s_{123}}{s_{12}} \bigg[ \,\frac{C_F}{2}\,s_{123}\,
    \mathcal{S}^{\mu\nu}_{3;3}(p_{12})
    +\langle P_{q\to q}^{{\rm(f)}\mu\nu}(p_3,p_{12})\rangle\bigg]
    P_{g\to q,\mu\nu}(p_1,p_2)
    +\Big(2\leftrightarrow 3\Big)\,\bigg\}\\
    &\qquad\quad+\langle P^{(p)}_{q\to \bar{q}qq}(p_1,p_2,p_3)\rangle\;.
  \end{split}
\end{equation}

Equation~\eqref{eq:tc_qgg} describes the abelian component of a
splitting of a quark to a quark and a pair of gluons. The result
involves factorized contributions corresponding to the Feynman diagrams
in Fig.~\ref{fig:one-to-three_splittings_quark_gluon}(a) and~(b),
plus a non-factorizable component. It takes the simple form
\begin{equation}\label{eq:split_qgg_ab_decomposition}
  \begin{split}
    \langle P_{q\to ggq}^{\rm(ab)}(p_1,p_2,p_3)\rangle=&\;
    \frac{s_{123}}{s_{13}}\,\langle P_{q\to q}(p_{13},p_2)\rangle\,
    \langle P_{q\to q}(p_3,p_1)\rangle+
    \frac{s_{123}}{s_{23}}\,\langle P_{q\to q}(p_{23},p_1)\rangle\,
    \langle P_{q\to q}(p_3,p_2)\rangle\\
    &\;+P^{({\rm ab},p)}_{\tilde{q}\to gg\tilde{q}}(p_1,p_2,p_3)
    +\langle P^{({\rm ab},p,{\rm f})}_{q\to ggq}(p_1,p_2,p_3)\rangle\;,
  \end{split}
\end{equation}
which can be written in terms of the scalar radiator,
Eq.~\eqref{eq:one-gluon_insertion_spin}, and the magnetic remainder,
Eq.~\eqref{eq:pqq_munu_sc_components}, as follows:
\begin{equation}\label{eq:split_qgg_ab_assembly}
  \begin{split}
    \langle P_{q\to ggq}^{\rm(ab)}&(p_1,p_2,p_3)\rangle
    =\bigg[\,\frac{C_F^2}{4}\,s_{123}^2\,
    \mathcal{S}_{13;13}(p_2;\bar{n})\mathcal{S}_{3;3}(p_1;\bar{n})
    +\frac{C_F}{2}\frac{s_{123}^2}{s_{13}}\,
    \mathcal{S}_{13;13}(p_2;\bar{n})\langle P^{\rm(f)}_{q\to q}(p_3,p_1)\rangle\\
    &\quad\;+\frac{C_F}{2}\,s_{123}\,
    \langle P^{\rm(f)}_{q\to q}(p_{13},p_2)\rangle\,
    \mathcal{S}_{3;3}(p_1;\bar{n})
    +\frac{s_{123}}{ s_{13}}\langle P^{\rm(f)}_{q\to q}(p_{13},p_2)\rangle
    \langle P^{\rm(f)}_{q\to q}(p_3,p_1)\rangle
    +\Big(1\leftrightarrow 2\Big)\,\bigg]\\
    &\;+P^{({\rm ab},p)}_{\tilde{q}\to gg\tilde{q}}(p_1,p_2,p_3)
    +\langle P^{({\rm ab},p,{\rm f})}_{q\to ggq}(p_1,p_2,p_3)\rangle\;.
  \end{split}
\end{equation}
The scalar remainder function, $P^{({\rm ab},p)}_{\tilde{q}\to gg\tilde{q}}$,
is defined as the difference between Eq.~\eqref{eq:scalar_tc_correspondence}
and the factorized scalar contributions in the two different permutations induced
by the diagrams in Fig.~\ref{fig:one-to-three_splittings_quark_gluon}(a) and~(b),
with the fermion replaced by a scalar:
\begin{equation}\label{eq:pure_split_ab_sqgg}
  \begin{split}
    P^{({\rm ab},p)}_{\tilde{q}\to gg\tilde{q}}(p_1,p_2,p_3)
    =&\;C_F^2 s_{123}^2\Big[\,\mathcal{S}^{\rm(ab)}_{3,3;3,3}(p_1,p_2;\bar{n})
    +2\mathcal{S}^{\rm(ab)}_{3,3;3}(p_1,p_2;\bar{n})
    +\mathcal{S}^{\rm(ab)}_{3;3}(p_1,p_2)\\
    &\qquad\qquad
    -\frac{1}{4}\,\mathcal{S}_{13;13}(p_2;\bar{n})\mathcal{S}_{3;3}(p_1;\bar{n})
    -\frac{1}{4}\,\mathcal{S}_{23;23}(p_1;\bar{n})\mathcal{S}_{3;3}(p_2;\bar{n})\,\Big]\\ 
    =&\;C_F^2\bigg\{\frac{s_{123}^2}{s_{13}s_{23}}\frac{z_3^2}{z_1 z_2}
    -\frac{s_{123}}{s_{13}}\frac{2z_3(1-z_2)}{z_1z_2}
    +\frac{4z_3}{z_1z_2}
    +1-\eps\bigg\}
    +\Big(1\leftrightarrow 2\Big)\;.\\
  \end{split}
\end{equation}
The pure fermionic splitting contribution is defined as the overall remainder
and is given by
\begin{equation}\label{eq:pure_split_ab_qgg}
  \begin{split}
    &\langle P^{({\rm ab},p,{\rm f})}_{q\to ggq}(p_1,p_2,p_3)\rangle=
    C_F^2(1-\eps)\bigg\{\frac{s_{123}^2}{2s_{13}s_{23}}\,z_3\bigg(
    \frac{(z_1+z_2)^2}{z_1 z_2}+\eps\bigg)+\bigg(\frac{s_{12}}{s_{13}}
    +\frac{z_1z_2}{(1-z_2)^2}\bigg)(1-\eps)\\
    &\;\qquad-\frac{s_{123}}{s_{13}}\bigg[\,
    (1-z_2)\bigg(\frac{(z_1+z_2)^2}{z_1z_2}+1\bigg)
    +(1-\eps)\bigg(\frac{z_1 z_2}{1-z_2}
    -z_3\bigg)\bigg]
    +\frac{2z_1}{z_2}\bigg(\frac{1}{1-z_2}+\frac{z_3}{1-z_1}\bigg)\bigg\}
    +\Big(1\leftrightarrow 2\Big)\;.
  \end{split}
\end{equation}

To reconstruct the non-abelian part of the $q\to ggq$ splitting function,
we first note that there is an additional factorization channel, corresponding
to the diagram in Fig.~\ref{fig:one-to-three_splittings_quark_gluon}(c), which
is singular in the $1\parallel 2$ collinear limit. In this configuration,
Eq.~\eqref{eq:tc_qgg_nab} can be written as a gluon emission term, times
a spin-correlated decay to two gluons. Due to color coherence, the non-abelian 
splitting function has no leading or sub-leading singularities associated with
the $1\parallel 3$ and $2\parallel 3$ collinear limits. However, it does have
single-soft singularities associated with the interferences between the
diagrams in Fig.~\ref{fig:one-to-three_splittings_quark_gluon}(c) and~(a),
and between the diagrams in Fig.~\ref{fig:one-to-three_splittings_quark_gluon}(c) and~(b).
We therefore obtain the following result for its decomposition into lower-order components
\begin{equation}\label{eq:split_qgg_nab_assembly}
  \begin{split}
    \langle P_{q\to ggq}^{\rm(nab)}(p_1,p_2,p_3)\rangle
    =&\;\frac{C_FC_A}{4}\,s_{123}^2\Big[\,
    \bar{\mathcal{S}}_{3;3}(p_{12};\bar{n})\Big(\mathcal{S}_{1;1}(p_2;\bar{n})
    +\mathcal{S}_{2;2}(p_1;\bar{n})
    -\mathcal{S}_{1;3}(p_2;\bar{n})-\mathcal{S}_{2;3}(p_1;\bar{n})
    \Big)\\
    &\qquad\qquad\quad\;
    +2(1-\eps)\,\mathcal{S}^{\mu\nu}_{3;3}(p_{12};\bar{n})
    D_\mu(p_1,p_2,\bar{n})D_\nu(p_1,p_2,\bar{n})\,\Big]\\
    &\;+\frac{C_A}{2}\,s_{123}\Big[\,
    \langle P_{q\to q}^{\rm(f)}(p_3,p_{12})\rangle
    \Big(\mathcal{S}_{1;1}(p_2;\bar{n})+\mathcal{S}_{2;2}(p_1;\bar{n})
    -\mathcal{S}_{1;3}(p_2;\bar{n})-\mathcal{S}_{2;3}(p_1;\bar{n})\Big)\\
    &\qquad\qquad\quad\;
    +2(1-\eps)\,\langle P_{q\to q}^{\rm(f)\,\mu\nu}(p_3,p_{12})\rangle
    D_\mu(p_1,p_2,\bar{n})D_\nu(p_1,p_2,\bar{n})\,\Big]\\
    &\;-\frac{C_A}{2C_F}\Big[
    P^{({\rm ab},p)}_{\tilde{q}\to gg\tilde{q}}(p_1,p_2,p_3)
    +\langle P^{({\rm ab},p,{\rm f})}_{q\to ggq}(p_1,p_2,p_3)\rangle\Big]\\
    &\;+P^{({\rm pnab},p)}_{\tilde{q}\to gg\tilde{q}}(p_1,p_2,p_3)
    +\langle P^{({\rm pnab},p,{\rm f})}_{q\to ggq}(p_1,p_2,p_3)\rangle\;.
  \end{split}
\end{equation}

The purely non-abelian component of the scalar splitting function is defined as
the difference between the double-gluon emission result for scalar radiators,
Eq.~\eqref{eq:tc_sqgg_nab}, and the corresponding factorized contributions.
Using Eq.~\eqref{eq:scalar_tc_correspondence} to make the origin of this
splitting function in a dipole radiator manifest, we can write
\begin{equation}\label{eq:tc_sqgg_pnab_pure_definition}
  \begin{split}
  P^{({\rm pnab},p)}_{\tilde{q}\to gg\tilde{q}}&(p_1,p_2,p_3)=
  -\frac{C_FC_A}{4}s_{123}^2\Big[\,
    \mathcal{S}^{\rm(nab)}_{3;3}(p_1,p_2;\bar{n})
    +\mathcal{S}^{\rm(ab)}_{3;3}(p_1,p_2)
    +2\mathcal{S}^{\rm(ab)}_{3,3;3}(p_1,p_2;\bar{n})\\
  &\quad\;+\mathcal{S}^{\rm(ab)}_{3,3;3,3}(p_1,p_2;\bar{n})
    +\bar{\mathcal{S}}_{3;3}(p_{12};\bar{n})\Big(
    \mathcal{S}_{2;2}(p_1;\bar{n})+\mathcal{S}_{1;1}(p_2;\bar{n})
    -\mathcal{S}_{2;3}(p_1;\bar{n})-\mathcal{S}_{1;3}(p_2;\bar{n})\Big)\\
  &\quad\;+2(1-\eps)\,
    \mathcal{S}^{\mu\nu}_{3;3}(p_{12};\bar{n})\,
    D_\mu(p_1,p_2,\bar{n})D_\nu(p_1,p_2,\bar{n})\,\Big]
  +\frac{C_A}{2C_F}P^{({\rm ab},p)}_{\tilde{q}\to gg\tilde{q}}(p_1,p_2,p_3)\;.
  \end{split}
\end{equation}
The explicit form of this function is given by
\begin{equation}\label{eq:tc_sqgg_pnab_pure}
  \begin{split}
  P^{({\rm pnab},p)}_{\tilde{q}\to gg\tilde{q}}(p_1,p_2,p_3)=&\;
  C_FC_A\bigg\{\frac{1-\eps}{4}+\frac{2z_3}{z_1z_2}
  +\bigg(\frac{s_{123}}{s_{13}}\frac{z_1}{1-z_3}-1\bigg)
  \bigg(\frac{s_{123}}{s_{12}}\frac{1-z_1}{z_2}-1\bigg)-1\\
  &\;\qquad\qquad+\bigg[\frac{s_{13}-s_{23}}{s_{13}}\frac{z_1-z_2}{z_2}
  -\frac{s_{12}}{s_{13}}\bigg(\frac{1-z_3}{z_2}+\frac{2z_3}{z_1}\bigg)\bigg]\frac{z_3}{(1-z_3)^2}
  \bigg\}+(1\leftrightarrow 2)\;.
  \end{split}
\end{equation}
In analogy to the purely non-abelian component of the scalar splitting function,
we have defined a purely non-abelian component of the fermionic part of the splitting
function. It is given by
\begin{equation}\label{eq:pure_split_pnab_qgg}
  \begin{split}
   &\langle P^{({\rm pnab},p,{\rm f})}_{q\to ggq}(p_1,p_2,p_3)\rangle
    =C_FC_A\,(1-\eps)\,\bigg\{\frac{z_2}{z_1}\bigg(\frac{1}{1-z_1}+\frac{z_3}{1-z_2}\bigg)+
  \bigg(\frac{2z_2}{z_1}+1\bigg)\bigg(1-\frac{3}{4}\frac{s_{123}}{s_{12}}(1-z_3)\bigg)\\
  &\;\qquad+\frac{1-\eps}{2}\bigg[\,1+\frac{s_{12}}{s_{13}}
    +\frac{z_1}{1-z_2}\bigg(\frac{z_2}{1-z_2}-\frac{s_{123}}{s_{13}}\bigg)\bigg]
  +\frac{s_{123}}{2s_{13}}
  \bigg(\frac{s_{123}}{s_{12}}-\frac{1-z_2}{z_1}\bigg)
  \bigg(\frac{(1-z_3)^2}{z_2}+\frac{z_2^2}{1-z_3}\bigg)\\
  &\;\qquad+\bigg(\frac{s_{123}}{s_{12}}\frac{1-z_3}{z_1}
  +\frac{s_{123}}{s_{13}}\frac{1-z_2}{z_1}
  -\frac{s_{123}^2}{s_{12}s_{13}}\bigg)
  \bigg(1-z_3-\frac{s_{12}}{s_{123}}\bigg)
  -\frac{3}{4}
  \bigg\}+(1\leftrightarrow 2)\;.
  \end{split}
\end{equation}

\subsubsection{Gluon initial state}
\label{sec:decomposition_gluon}
Equations~\eqref{eq:tc_gqqb_ab} and~\eqref{eq:tc_gqqb_nab} describe
the splitting of a gluon to a gluon and a quark-antiquark pair.
The results involve factorized contributions corresponding to each
individual Feynman diagram in Fig.~\ref{fig:one-to-three_splittings_gluon_quark},
plus a non-factorizable component. For the abelian part we find
\begin{equation}\label{eq:split_gqq_ab_decomposition}
  \begin{split}
    P_{g\to gq\bar{q}}^{\mu\nu\,\rm(ab)}(p_1,p_2,p_3)=&\;
    \bigg[\frac{C_F}{2}\,s_{123}\,P_{g\to q}^{\mu\nu}(p_{12},p_3)\,
    \Big(\mathcal{S}_{2;2}(p_1;\bar{n})-\mathcal{S}_{2;3}(p_1;\bar{n})
    \Big)\\
    &\quad\;+\frac{s_{123}}{s_{12}}\,P_{g\to q}^{\mu\nu}(p_{12},p_3)
    \langle P^{\rm(f)}_{q\to q}(p_2,p_1)\rangle+(2\leftrightarrow 3)\,\bigg]
    +P^{\mu\nu\,({\rm ab},p)}_{g\to gq\bar{q}}(p_1,p_2,p_3)
    +\ldots\;.
  \end{split}
\end{equation}
Here and in the remainder of this section, the dots stand for terms proportional
to $\bar{n}^\mu$ or $\bar{n}^\nu$. These term vanish after the splitting function is
combined with a gluon current in the triple-collinear limit and can therefore be dropped,
see the discussion following Eq.~\eqref{eq:tc_factorization}.
Because a new scalar color dipole is created by the $g\to q\bar{q}$ splitting,
Eq.~\eqref{eq:pure_split_ab_ggq} contains a soft singular contribution proportional
to $P_{g\to q}^{\mu\nu}(p_2,p_3)\mathcal{S}_{2;3}(p_1;\bar{n})$.
The abelian pure $g\to gq\bar{q}$ splitting function must in fact be
interpreted as a subtracted single-gluon radiator function, similar
to $P_{\tilde{q}\to gg\tilde{q}}^{({\rm ab},p)}$, but with a non-trivial
helicity dependence. We find
\begin{equation}\label{eq:pure_split_ab_ggq}
  \begin{split}
    P^{\mu\nu\,({\rm ab},p)}_{g\to gq\bar{q}}&(p_1,p_2,p_3)=
    C_FT_R\bigg\{\frac{2\eps\,s_{123}}{s_{12}s_{13}}\,\tilde{p}_{1,23}^\mu\,\tilde{p}_{1,23}^\nu
    -d^{\mu\nu}(p_{123},\bar{n})\bigg[\frac{(s_{123}-s_{23})^2}{2s_{12}s_{13}}(1+\eps)
    -\frac{z_1(1-\eps)}{(1-z_3)^2}\bigg]\\
    &\quad-d^{\mu\nu}(p_{123},\bar{n})\bigg[
    \frac{s_{123}}{s_{12}}\frac{z_2}{z_1(1-z_2)}+\frac{s_{23}}{s_{12}}
    \bigg(1-\frac{z_2}{z_1}\frac{1+z_1}{1-z_2}\bigg)
    -\frac{1-z_1}{z_1(1-z_3)}+1\,\bigg]\\
    &\quad+\bigg(d^{\mu\nu}(p_{123},\bar{n})
    -\frac{4\tilde{p}_{12,3}^\mu\tilde{p}_{12,3}^\nu}{s_{123}}\bigg)\bigg[\,
    \frac{s_{123}}{s_{12}}
    \bigg(\frac{z_1-z_2}{z_1}-(1-\eps)\frac{z_1}{1-z_3}\bigg)
    +\frac{s_{123}}{s_{13}}\frac{1-z_2}{z_1}\bigg]
    \bigg\}+(2\leftrightarrow 3)\;.
  \end{split}
\end{equation}

To reconstruct the components of the non-abelian splitting function,
we first note that the non-abelian splitting function itself has no
leading or sub-leading singularities associated with the $1\parallel 2$
or $1\parallel 3$ collinear limits. This is a consequence of color coherence.
The $2\parallel 3$ collinear limit projects Eq.~\eqref{eq:tc_gqqb_nab}
onto the product of a gluon emission term, times a spin-correlated decay
to a quark-antiquark pair. Schematically,
\begin{equation}\label{eq:split_gqqb_nab_decomposition}
  \begin{split}
    P_{g\to gq\bar{q}}^{\mu\nu\,\rm(nab)}(p_1,p_2,p_3)=&\;
    \frac{s_{123}}{s_{23}}\,P_{g\to g,{\rm(s)}}^{\mu\nu,\alpha\beta}(p_1,p_{23})
    P_{g\to q,\alpha\beta}(p_2,p_3)
    +\text{remainder}\;,
  \end{split}
\end{equation}
where $P_{g\to g,{\rm(s)}}^{\mu\nu,\alpha\beta}(p_1,p_{23})$ is 
the symmetric part of the off-shell gluon splitting function, defined
in Eq.~\eqref{eq:coll_ggg_sym}.
This gives the final expression for the assembly of the $g\to gq\bar{q}$
splitting function
\begin{equation}\label{eq:split_gqq_nab_assembly}
  \begin{split}
    P_{g\to gq\bar{q}}^{\mu\nu\,\rm(nab)}(p_1,p_2,p_3)
    =&\;\frac{C_A}{2}\,\frac{s_{123}^2}{s_{23}}\,\Big[\,
    \mathcal{S}^{\alpha\beta}_{1;1}(p_{23})
    P_{g\to q,\alpha\beta}(p_2,p_3)d^{\mu\nu}(p_{123},\bar{n})
    +\mathcal{S}_{23;23}(p_1;\bar{n})
    P^{\mu\nu}_{g\to q}(p_2,p_3)\,\Big]\\
    &\;+\frac{C_A}{2}\,\frac{s_{123}^2}{s_{23}}\,
    2(1-\eps)D^\mu(p_1,p_{23},\bar{n})D^\nu(p_1,p_{23},\bar{n})\,
    \langle P_{g\to q}(p_2,p_3)\rangle\\
    &\;+\frac{C_A}{4}\,s_{123}\,\Big[\,
    P_{g\to q}^{\mu\nu}(p_{12},p_3)
    +P_{g\to q}^{\mu\nu}(p_2,p_{13})\,\Big]\,
    \mathcal{S}_{2;3}(p_1;\bar{n})\\
    &\;-\frac{C_A}{2C_F}\,
    P^{\mu\nu\,({\rm ab},p)}_{g\to gq\bar{q}}(p_1,p_2,p_3)
  +P^{\mu\nu\,({\rm pnab},p)}_{g\to gq\bar{q}}(p_1,p_2,p_3)+\ldots\;.
  \end{split}
\end{equation}
The purely non-abelian component is given by
\begin{equation}\label{eq:pure_split_gqqb_pnab_2}
  \begin{split}
    &P^{\mu\nu\,({\rm pnab},p)}_{g\to gq\bar{q}}(p_1,p_2,p_3)=
    \frac{C_AT_R}{2}\bigg\{-d^{\mu\nu}(p_{123},\bar{n})\,
    (1-\eps)\bigg(\frac{s_{123}}{s_{12}}\frac{z_1}{1-z_3}-\frac{s_{13}}{s_{12}}
    -\frac{z_1}{(1-z_3)^2}\bigg)\\
    &\quad\;-d^{\mu\nu}(p_{123},\bar{n})\bigg[
    \frac{1-2z_1}{z_1}\bigg(\frac{s_{123}}{s_{12}}\frac{1-z_3}{1-z_1}
    +\frac{s_{123}}{s_{23}}-\frac{s_{123}^2}{s_{12}s_{23}}\frac{z_2}{1-z_1}\bigg)
    -\frac{2}{z_1}-\frac{2z_2}{z_1(1-z_3)}+1\,\bigg]\\
    &\quad\;+\frac{2(\tilde{p}_{2,13}^\mu\tilde{p}_{3,12}^\nu
    +\tilde{p}_{3,12}^\mu \tilde{p}_{2,13}^\nu)}{s_{23}}
    \frac{s_{123}}{s_{12}}\bigg[\frac{2z_2}{z_1}\frac{z_3-z_1}{1-z_1}+1-\eps\bigg]
    +\frac{4\tilde{p}_{1,23}^\mu\tilde{p}_{1,23}^\nu}{s_{23}}
    \bigg[\,\frac{2z_2z_3}{(1-z_1)^2}-(1-\eps)\bigg]\\
    &\quad\;-\frac{4\tilde{p}_{3,12}^\mu\tilde{p}_{3,12}^\nu}{s_{23}}
    \bigg[\,\frac{2z_2}{z_1}\bigg(\frac{s_{123}}{s_{12}}\frac{z_2}{1-z_1}-\frac{s_{23}}{s_{12}}\bigg)
    -(1-\eps)\bigg(\frac{s_{123}}{s_{13}}+\frac{s_{23}}{s_{12}}\frac{z_1}{1-z_3}\bigg)\bigg]
    -\frac{2}{z_1}\frac{4\tilde{p}_{2,3}^\mu\tilde{p}_{2,3}^\nu}{s_{23}}\bigg\}+(2\leftrightarrow 3)\;.
  \end{split}
\end{equation}
This expression is regular in the single-soft gluon limit, as well as
in all double-collinear limits, see Sec.~\ref{sec:tree-level_singularity_structure}.

Equation~\eqref{eq:tc_ggg} describes the splitting of a gluon
to three gluons. The result involves factorized contributions
corresponding to each Feynman diagram containing a propagator,
plus a non-factorizable component. To reconstruct the splitting function,
we first note that only one transverse projector is needed, while all
remaining terms can be inferred from the symmetry. We obtain
\begin{equation}\label{eq:split_ggg_decomposition}
  \begin{split}
    P_{g\to ggg}^{\mu\nu}&(p_1,p_2,p_3)=
    \bigg[\frac{1}{2}\frac{s_{123}}{s_{12}}\,P_{g\to g}^{\mu\nu,\alpha\beta}(p_3,p_{12})
    P_{g\to g,\alpha\beta}(p_1,p_2)
    +\frac{C_A^2}{4C_F^2}\,
    P_{\tilde{q}\to gg\tilde{q}}^{({\rm ab},p)}(p_1,p_2,p_3)d^{\mu\nu}(p_{123},\bar{n})\\
    &\quad+\frac{C_A}{2C_F}\,P_{\tilde{q}\to gg\tilde{q}}^{({\rm pnab},p)}(p_1,p_2,p_3)
    d^{\mu\nu}(p_{123},\bar{n})+(5\;\text{permutations})\bigg]
    +\text{remainder}\;.\\
  \end{split}
\end{equation}
As in the case of the abelian and non-abelian $q\to ggq$ splitting functions,
we need to account for the fact that the full single-gluon dipole radiation pattern
includes interference terms which appear explicitly in the three-parton splitting
functions. Color coherence dictates that this overlap must factorize into the gluon
splitting tensor and a single-gluon dipole radiator, while Bose symmetry requires
that it appears in all permutations. The $g\to ggg$ splitting function can therefore
be assembled as
\begin{equation}\label{eq:split_ggg_assembly}
  \begin{split}
    P_{g\to ggg}^{\mu\nu}&(p_1,p_2,p_3)
    =\bigg\{ \frac{C_A}{4}\,s_{123}\,\bigg[\,
    \frac{s_{123}}{s_{12}}\mathcal{S}_{12;12}(p_3;\bar{n})
    P^{\mu\nu}_{g\to g}(p_1,p_2)\\
    &\quad\;+\bigg(\frac{s_{123}}{s_{12}}\,S_{3;3}^{\alpha\beta}(p_{12})\,
    P_{g\to g,\alpha\beta}(p_1,p_2)
    -C_A\,s_{123}\,\bar{\mathcal{S}}_{3;3}(p_{12};\bar{n})
    \mathcal{S}_{2;3}(p_1;\bar{n})\bigg)
    d^{\mu\nu}(p_{123},\bar{n})\\
    &\quad\;+2(1-\eps)D^\mu(p_{12},p_3,\bar{n})D^\nu(p_{12},p_3,\bar{n})\,
    \bigg(\frac{s_{123}}{s_{12}}\,\langle P_{g\to g}(p_1,p_2)\rangle
    -\frac{C_A}{2}\,s_{123}\,\mathcal{S}_{2;3}(p_1;\bar{n})\bigg)\,\bigg]\\
    &\quad\;+\frac{C_A^2}{4C_F^2}\,
    P_{\tilde{q}\to gg\tilde{q}}^{({\rm ab},p)}(p_1,p_2,p_3)d^{\mu\nu}(p_{123},\bar{n})
    +\frac{C_A}{2C_F}\,P_{\tilde{q}\to gg\tilde{q}}^{({\rm pnab},p)}(p_1,p_2,p_3)
    d^{\mu\nu}(p_{123},\bar{n})\\
    &\quad\;+(5\;\text{permutations})\,\bigg\}
    +P^{\mu\nu\,(p)}_{g\to ggg}(p_1,p_2,p_3)+\ldots\;.
  \end{split}
\end{equation}
The relative prefactors of the abelian and pure non-abelian
pure splitting functions,
$P_{\tilde{q}\to gg\tilde{q}}^{({\rm ab},p)}$ and
$P_{\tilde{q}\to gg\tilde{q}}^{({\rm pnab},p)}$,
are due to the color structure of the scalar multipole radiator
in the octet case, cf.\ Sec.~\ref{sec:tree_level_scalar_multipoles},
and the remainder function is defined as
\begin{equation}\label{eq:pure_split_ggg}
  \begin{split}
    P_{g\to ggg}^{\mu\nu\,(p)}(p_1,p_2,p_3)=&\;C_A^2\bigg\{
    -d^{\mu\nu}(p_{123},\bar{n})\bigg[\frac{s_{123}}{s_{12}}\frac{z_2}{1-z_1}
    +\frac{s_{123}^2}{s_{12}s_{13}}\bigg(\frac{1-z_2^2-z_3^2}{2(1-z_2)(1-z_3)}-1\bigg)
    +\frac{4z_3}{z_1z_2}\bigg]\\
    &\qquad\;+(1-\eps)\frac{4\tilde{p}_{12,3}^\mu\tilde{p}_{12,3}^\nu}{s_{123}}
    \bigg[\frac{s_{123}^2}{s_{12}s_{13}}\bigg(\frac{z_1}{z_2}-\frac{z_1}{1-z_2}-\frac{1}{2}\bigg)
    +\frac{s_{123}}{2s_{13}}\frac{1-z_2}{z_1}\\
    &\qquad\qquad\qquad\qquad\qquad\qquad-\frac{s_{123}}{s_{12}}\bigg(\frac{3(1-z_3)^2}{4z_1z_2}
    +\frac{z_1z_2}{(1-z_3)^2}-2\bigg)\bigg]\\
    &\qquad\;+(1-\eps)\frac{\tilde{p}_{12,3}^\mu\tilde{p}_{2,13}^\nu
    +\tilde{p}_{2,13}^\mu\tilde{p}_{12,3}^\nu}{s_{123}} \frac{s_{123}^2}{s_{12}s_{13}}
    \bigg(3-\frac{2z_2(1-z_2)}{z_3(1-z_3)}\bigg)\bigg\}\\
    &\qquad\;+\frac{C_A}{z_3}P^{\mu\nu}_{g\to g}(p_1,p_2)
    +(5\;\text{permutations})\;.
  \end{split}
\end{equation}

\subsubsection{Singularity structure of the remainder functions}
\label{sec:tree-level_singularity_structure}
\begin{table}[t]
  \renewcommand{\arraystretch}{1.5}
  \begin{tabular}{c|c|c|c|c|c}\hline
  \parbox{15ex}{\centering Function} & \parbox{15ex}{\centering Definition} &
  \multicolumn{4}{c}{Scaling behavior for $\lambda\to 0$}\\\cline{3-6}
  $\times s_{123}^{-2}$ & & \parbox{23ex}{\centering $p_1\to\lambda p_1$, $p_2\to\lambda p_2$} &
  \parbox{23ex}{\centering $p_1\to\lambda p_1$} &
  \parbox{23ex}{\centering $\tilde{p}_{1,2}\to\lambda\tilde{p}_{1,2}$} &
  \parbox{23ex}{\centering $\tilde{p}_{2,3}\to\lambda\tilde{p}_{2,3}$}\\\hline
  $P_{\tilde{q}\to q'\bar{q}'\tilde{q}}$ & Eq.~\eqref{eq:tc_qqpqpb_scalar} &
  $\varpropto\lambda^{-4}$ & -- & $\varpropto\lambda^{-2}$ & --\\
  $\langle P_{q\to q\bar{q}q}^{(p)}\rangle$ & Eq.~\eqref{eq:pure_split_qbqq} &
  $\varpropto\lambda^{-3}$ & -- & $\varpropto\lambda^{0}$ & $\varpropto\lambda^{0}$\\\hline
  $P_{\tilde{q}\to gg\tilde{q}}^{\rm(ab)}$ & Eq.~\eqref{eq:tc_sqgg} &
  $\varpropto\lambda^{-4}$ & $\varpropto\lambda^{-2}$ & -- & $\varpropto\lambda^{-2}$\\
  $P_{\tilde{q}\to gg\tilde{q}}^{({\rm ab},p)}$ & Eq.~\eqref{eq:pure_split_ab_sqgg} &
  $\varpropto\lambda^{-4}$ & $\varpropto\lambda^{-1}$ & -- & $\varpropto\lambda^0$\\\hline
  $\langle P_{q\to ggq}^{\rm(ab)}\rangle$ & Eq.~\eqref{eq:tc_qgg} &
  $\varpropto\lambda^{-4}$ & $\varpropto\lambda^{-2}$ & -- & $\varpropto\lambda^{-2}$\\
  $\langle P_{q\to ggq}^{({\rm ab},p,f)}\rangle$ & Eq.~\eqref{eq:pure_split_ab_qgg} &
  $\varpropto\lambda^{-2}$ & $\varpropto\lambda^{-1}$ & -- & $\varpropto\lambda^0$\\\hline
  $P_{\tilde{q}\to gg\tilde{q}}^{\rm(nab)}$ & Eq.~\eqref{eq:tc_sqgg_nab} &
  $\varpropto\lambda^{-4}$ & $\varpropto\lambda^{-2}$ & $\varpropto\lambda^{-2}$ & --\\
  $P_{\tilde{q}\to gg\tilde{q}}^{({\rm pnab},p)}$ & Eq.~\eqref{eq:tc_sqgg_pnab_pure} &
  $\varpropto\lambda^{-4}$ & $\varpropto\lambda^{-1}$ & $\varpropto\lambda^0$ & -- \\\hline
  $\langle P_{q\to ggq}^{\rm(nab)}\rangle$ & Eq.~\eqref{eq:tc_qgg_nab} &
  $\varpropto\lambda^{-4}$ & $\varpropto\lambda^{-2}$ & $\varpropto\lambda^{-2}$ & -- \\
  $\langle P_{q\to ggq}^{({\rm pnab},p,f)}\rangle$ & Eq.~\eqref{eq:pure_split_pnab_qgg} &
  $\varpropto\lambda^{-2}$ & $\varpropto\lambda^{-1}$ & $\varpropto\lambda^0$ & -- \\\hline
  $P_{g\to gq\bar{q}}^{\mu\nu\,\rm(ab)}$ & Eq.~\eqref{eq:tc_gqqb_ab} &
  -- & $\varpropto\lambda^{-2}$ & -- & $\varpropto\lambda^{-2}$ \\
  $P_{g\to gq\bar{q}}^{\mu\nu\,({\rm ab},p)}$ & Eq.~\eqref{eq:pure_split_ab_ggq} &
  -- & $\varpropto\lambda^{-1}$ & -- & $\varpropto\lambda^0$ \\\hline
  $P_{g\to gq\bar{q}}^{\mu\nu\,\rm(nab)}$ & Eq.~\eqref{eq:tc_gqqb_nab} &
  $\varpropto\lambda^{-4}$ & $\varpropto\lambda^{-2}$ & -- & $\varpropto\lambda^{-2}$ \\
  $P_{g\to gq\bar{q}}^{\mu\nu\,({\rm pnab},p)}$ & Eq.~\eqref{eq:pure_split_gqqb_pnab_2} &
  $\varpropto\lambda^{-3}$ & $\varpropto\lambda^{-1}$ & -- & $\varpropto\lambda^0$ \\\hline
  $P_{g\to ggg}^{\mu\nu}$ & Eq.~\eqref{eq:tc_ggg} &
  $\varpropto\lambda^{-4}$ & $\varpropto\lambda^{-2}$ & \multicolumn{2}{c}{$\varpropto\lambda^{-2}$} \\
  $P_{g\to ggg}^{\mu\nu\,(p)}$ & Eq.~\eqref{eq:pure_split_ggg} &
  $\varpropto\lambda^{-3}$ & $\varpropto\lambda^{-1}$ & \multicolumn{2}{c}{$\varpropto\lambda^0$} \\\hline
  \end{tabular}
  \caption{Scaling behavior of the tree-level splitting functions and their pure components.
    See the main text for details.
  \label{tab:limits}}
\end{table}
Here we summarize the singularity structure of the various three-parton splitting functions
introduced in Secs.~\ref{sec:three-parton_tree-level} and~\ref{sec:tree-level_splitting_decomposition}.
The kinematical limits are taken according to Ref.~\cite{Catani:1999ss} and are parametrized
in terms of a scaling parameter $\lambda$. In the limit where particle $i$ becomes soft,
$\lambda$ enters the computation as $p_i\to \lambda p_i$. In the limit where particles $i$ and $j$
both become soft, $\lambda$ enters the computation as $p_i\to \lambda p_i$ and $p_j\to \lambda p_j$.
In the limit where particles $i$ and $j$ become collinear, $\lambda$ enters the computation
as $\tilde{p}_{i,j}\to\lambda\tilde{p}_{i,j}$. The results are given in Tab.~\ref{tab:limits}.
A few comments are in order
\begin{itemize}
\item The function $\langle P_{q\to q\bar{q}q}^{\rm(id)}\rangle$ has a leading pole
in the double-collinear limits because it constitutes a new type of interference
term for which no unique Born splitting function exists at the next lower order.
This singularity is canceled in the symmetrized splitting function
$\langle P_{q\to q\bar{q}q}^{(p)}\rangle$, and the remainder term is sub-leading.
\item All other pure splitting functions are sub-leading in the single soft limits,
and sub-sub-leading in the double-collinear limits. This is achieved by consistently
subtracting the single-soft and double-collinear enhanced components, including
spin correlations. We emphasize that the subtraction must be performed in the same
axial gauge that has been used to determine the splitting functions, in order
to avoid remainders which would create spurious singularities
that cannot be removed systematically. 
\item The functions $P_{\tilde{q}\to gg\tilde{q}}^{({\rm ab},p)}$
and $P_{\tilde{q}\to gg\tilde{q}}^{({\rm pnab},p)}$ have leading poles
in the double-soft limit. By means of Eqs.~\eqref{eq:pure_split_ab_sqgg}
and~\eqref{eq:tc_sqgg_pnab_pure_definition}, these two splitting functions are 
identified as scalar radiators rather than genuine splitting functions.
The double-soft singularities arise from interferences in the two-gluon emission pattern,
which occur for the first time in the computation of the three-parton final state.
\item All other pure splitting functions are sub-leading in the double-soft limit.
As in the single-gluon radiator case, it is important that the scalar components
are subtracted using the same gauge that was used to compute the full splitting
function. In particular, using Eqs.~\eqref{eq:scalar_tc_correspondence} with
radiator functions in covariant gauge would lead to an inconsistent result
that contains spurious sub-leading poles.
\end{itemize}

\section{One-loop expressions}
\label{sec:loop_mes}
The one-to-two parton splitting functions at one loop have been discussed and computed
in multiple ways~\cite{Curci:1980uw,Furmanski:1980cm,Heinrich:1997kv,Bassetto:1998uv,
  Bern:1998sc,Bern:1999ry,Kosower:1999rx}, and their soft limits have been analyzed
in great detail~\cite{Bern:1995ix,Bern:1999ry,Kosower:1999rx}. In this section we
derive a decomposition of the corresponding splitting amplitudes in terms of scalar
and spin-dependent components, as needed for a correspondence with the treatment
of the tree-level splitting functions in Sec.~\ref{sec:three-parton_tree-level}.

Several results in this section are novel and potentially interesting beyond the fact that
they can be used to decompose the splitting functions. In particular, we derive a generalization
of the one-loop soft current in~\cite{Catani:2000pi} and discuss its relation to the one-loop
splitting amplitudes. In Ref.~\cite{Catani:2000pi} the authors found that a certain class of
diagrams does not contribute to the soft current, which could be proven with the help of
partial fractioning identities that are effectively related to the Ward identities which the
amplitudes satisfy in the soft-gluon limit. Here we provide the generalization of this
important result. By using the Background Field Method, we are able to prove that the cancelation
is not a coincidence specific to the soft-gluon approximation, but a consequence of the Ward
identities that are satisfied by the $n$-point Greens functions in a physical scheme.
As a consequence, we are able to also compute the abelian component of the one-loop
scalar current, which is sub-leading in the soft limit, but necessary in order to fully
decompose the one-loop splitting amplitudes. Details of this calculation
are discussed in Sec.~\ref{sec:one-loop_scalar_multipoles}.

\subsection{One-to-two splittings}
\label{sec:one-loop_splittings}
This subsection presents a recomputation of the known one-loop splitting functions in order
to provide a baseline for their subsequent decomposition in Sec.~\ref{sec:one-loop_splitting_decomposition}. 
All calculations are performed using the techniques of Ref.~\cite{Kosower:1999rx,Bern:2004cz}.
We emphasize the structure of the results in terms of scalar components and spin-dependent remainders.
As in the tree-level case, we make use of the conventional dimensional regularization scheme
in $D=4-2\eps$ dimensions. Taking gauge invariance of on-shell amplitudes as a guiding principle,
we evaluate $t$-channel gluon propagators in Feynman gauge, while $s$-channel gluon propagators
are in light-like axial gauge. The calculation of one-loop splitting amplitudes was performed in 
\textsc{Form}~\cite{Tentyukov:2007mu,Kuipers:2012rf,Ruijl:2017dtg},
with all standard and light-cone integrals cross-checked against the literature with
the help of \textsc{Mathematica}~\cite{Cole:1981uy}.
The reduction of tensor integrals was carried out using a dedicated implementation of the 
Passarino-Veltman scheme~\cite{Passarino:1978jh} using the 
\textsc{FeynCalc}~\cite{Hahn:1998yk,Hahn:2000kx,Hahn:2000jm} package.
All scalar and tensor integrals needed for the computation are listed
in App.~\ref{sec:one-loop_integrals}. Following the notation of \cite{Kosower:1999rx},
we express the one-loop splitting amplitudes in terms of the basis functions
\begin{align}\label{eq:def_f1_f2}
  f_1(z) &= \frac{2c_\Gamma}{\eps^2}\bigg[
  -\Gamma(1-\eps)\Gamma(1+\eps)\frac{1}{z}\left(\frac{1-z}{z}\right)^\eps
  - \frac{1}{z} + \frac{(1-z)^\eps}{z}\,_2F_1(\eps,\eps,1+\eps,z)\bigg]\;,\\
  f_2 &= -\frac{c_\Gamma}{\eps^2}\;,
\end{align}
where
\begin{equation}\label{eq:def_cgamma}
    c_\Gamma=(4\pi)^\eps\,\frac{\Gamma(1+\eps)\Gamma^2(1-\eps)}{\Gamma(1-2\eps)}\;.
\end{equation}
Note that, in contrast to \cite{Kosower:1999rx}, we have chosen to pull out a factor of $1/(4\pi)^2$ from the definition of $c_\Gamma$.

Unlike the tree-level $1\to2$ splittings considered earlier, where it was important to track the momentum fractions $z_1$ and $z_2$ in order to use them as building blocks for $1 \to 3$ splitting functions, in this section we simply identify $z_1 \equiv z$ and $z_2 \equiv 1-z$.

\subsubsection{Quark initial state}
\label{sec:one_to_two_splittings_quark_oneloop}
\begin{figure}[t]
    \includegraphics[scale=0.36]{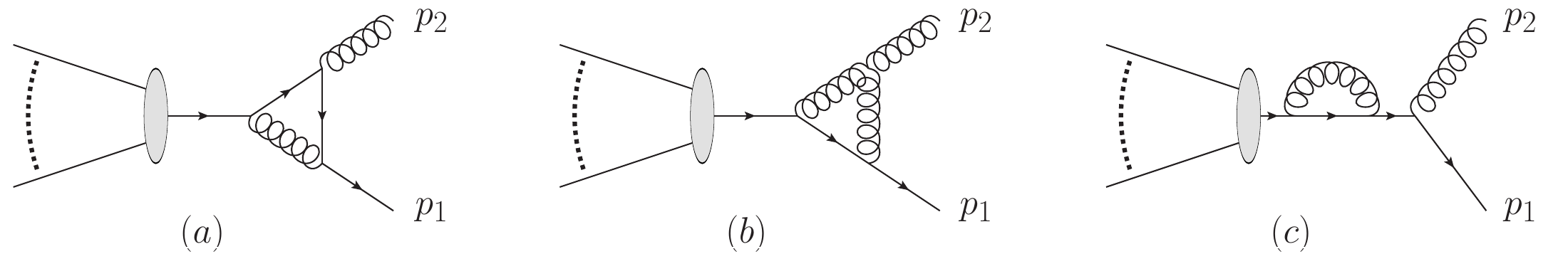}
    \caption{Feynman diagrams leading to the $1\to2$ quark splitting function
    at one loop discussed in Sec.~\ref{sec:one_to_two_splittings_quark_oneloop}.
      \label{fig:one-to-two_splittings_quark_one-loop}}
\end{figure}
The $q\to qg$ splitting amplitude at one-loop order is determined from the diagrams
shown in Fig.~\ref{fig:one-to-two_splittings_quark_one-loop}. 
It has been shown that due to the gauge invariance of the sum of cut diagrams,
any $t$-channel gluon propagators can be evaluated in Feynman 
gauge~\cite{Kosower:1999rx,Bern:2004cz}, which substantially simplifies the calculation.
For future reference, we introduce the tree-level quark-splitting amplitude,
\begin{equation}\label{eq:tree-level_amplitude_q2qg}
    \mathcal{P}_{q\to qg}^{(0)}(p_1,p_2) = g_sT^a_{ij}\,\bar{u}(p_1)\slash\!\!\!\varepsilon^*(p_2)\slash\!\!\!p_{12}u(n) \;.
\end{equation}

Using the results of App.~\ref{sec:one-loop_integrals},
the integration of the abelian contribution in
Fig.~\ref{fig:one-to-two_splittings_quark_one-loop}(a) yields
\begin{equation}
\begin{split}
    \mathcal{P}_{q\to qg}^{(1),a}(p_1,p_2) &= \frac{g_s^2}{16\pi^2}\, \bigg(C_F-\frac{C_A}{2}\bigg)
    \left(\!-\frac{\mu^2}{s_{12}}\!\right)^\eps\,\left[\,zf_1(z)
    + \left(\frac{2}{1-2\eps}-\eps\right)f_2\,\right]\,
    \mathcal{P}_{q\to qg}^{(0)}(p_1,p_2) \\
    &\qquad -\frac{g_s^3}{16\pi^2}T^a_{ij} \bigg(C_F-\frac{C_A}{2}\bigg)\,
    \left(\!-\frac{\mu^2}{s_{12}}\!\right)^\eps\,
    \frac{2}{s_{12}}\frac{\eps^2}{1-2\eps}f_2\,
    \bar{u}(p_1)\slash\!\!\!p_2\slash\!\!\!p_{12}u(n) (p_1\varepsilon^*(p_2)) \,.
\end{split}
\end{equation}
The non-abelian contribution in Fig.~\ref{fig:one-to-two_splittings_quark_one-loop}(b)
is given by
\begin{equation}
\begin{split}
    \mathcal{P}_{q\to qg}^{(1),b}(p_1,p_2) &=
    \frac{g_s^2}{16\pi^2}\, \frac{C_A}{2}\left(\!-\frac{\mu^2}{s_{12}}\!\right)^\eps
    \left[\,(1-z)f_1(1-z) + \frac{4-5\eps}{1-2\eps}f_2\,\right]\,
    \mathcal{P}_{q\to qg}^{(0)}(p_1,p_2) \\
    &\qquad + \frac{g_s^3}{16\pi^2}T^a_{ij}\, \frac{C_A}{2}\left(\!-\frac{\mu^2}{s_{12}}\!\right)^\eps\,
    \frac{2}{s_{12}}\frac{\eps^2}{1-2\eps}f_2\,
    \bar{u}(p_1)\slash\!\!\!p_2\slash\!\!\!p_{12}u(n) (p_1\varepsilon^*(p_2)) \,.
\end{split}
\end{equation}
The self-energy in Fig.~\ref{fig:one-to-two_splittings_quark_one-loop}(c) yields
\begin{equation}
    \mathcal{P}_{q\to qg}^{(1),c}(p_1,p_2) = -\frac{g_s^2}{16\pi^2}\, C_F
    \left(\!-\frac{\mu^2}{s_{12}}\!\right)^\eps\,
    \frac{(4-\eps)(1-\eps)}{1-2\eps}f_2\,
    \mathcal{P}_{q\to qg}^{(0)}(p_1,p_2) \,.
\end{equation}
The complete splitting amplitude is given by the sum of the above terms and reads~\cite{Kosower:1999rx,Bern:1999ry}
\begin{equation}\label{eq:one-loop_qqg_splitting_amplitude}
  \begin{split}
    &\mathcal{P}_{q\to qg}^{(1)}(p_1,p_2) = \frac{g_s^2}{16\pi^2}\,\frac{C_A}{2}
    \left(\!-\frac{\mu^2}{s_{12}}\!\right)^\eps\,
    \bigg[\,(1-z)f_1(1-z)-\frac{1}{N_C^2}\Big(zf_1(z)-2f_2\Big)\bigg]
    \mathcal{P}_{q\to qg}^{(0)}(p_1,p_2) \\
    &\qquad-\frac{g_s^3}{16\pi^2}T^a_{ij}\,
    \bigg(\frac{N_C}{2}+\frac{1}{2N_C}\bigg)\left(\!-\frac{\mu^2}{s_{12}}\!\right)^\eps\, \bigg[
    \bar{u}(p_1)\slash\!\!\!\varepsilon^*(p_2)\slash\!\!\!p_{12}u(n)
    -\frac{2}{s_{12}}\bar{u}(p_1)\slash\!\!\!p_2
    \slash\!\!\!p_{12}u(n) (p_1\varepsilon^*(p_2))\bigg]\frac{\eps^2}{1-2\eps}f_2\;.
  \end{split}
\end{equation}

The scalar one-loop splitting amplitude corresponding to the $q\to qg$ case is determined
from the three diagrams in Fig.~\ref{fig:one-to-two_splittings_quark_one-loop}
and the bubble-type diagrams involving seagull vertices. However, the latter vanish in light-like axial gauge.
Upon integration of the loop momentum, the abelian contribution corresponding to
the scalar analogue of Fig.~\ref{fig:one-to-two_splittings_quark_one-loop}(a) is given by
\begin{equation}\label{eq:one-loop_amplitude_scalar_f1}
    \mathcal{P}_{\tilde{q}\to \tilde{q}g}^{(1),a}(p_1,p_2) = \frac{g_s^2}{16\pi^2}\,
    \bigg(C_F-\frac{C_A}{2}\bigg)\left(\!-\frac{\mu^2}{s_{12}}\!\right)^\eps\,
    \left[\,zf_1(z) + \frac{2}{1-2\eps}f_2\,\right]\,\mathcal{P}_{\tilde{q}\to\tilde{q}g}^{(0)}(p_1,p_2)\;,
\end{equation}
where we have introduced the tree-level (anti-)triplet scalar amplitude
\begin{equation}\label{eq:tree-level_amplitude_scalar}
    \mathcal{P}_{\tilde{q}\to\tilde{q}g}^{(0)}(p_1,p_2) = g_sT^a_{ij}\,\big(2p_1^\mu\varepsilon^*_\mu(p_2)\big)\;.
\end{equation}
The non-abelian contribution, corresponding to the scalar version of
Fig.~\ref{fig:one-to-two_splittings_quark_one-loop}(b), reads
\begin{equation}\label{eq:one-loop_amplitude_scalar_f2}
    \mathcal{P}_{\tilde{q}\to \tilde{q}g}^{(1),b}(p_1,p_2) = \frac{g_s^2}{16\pi^2}\,
    \frac{C_A}{2}\left(\!-\frac{\mu^2}{s_{12}}\!\right)^\eps
    \left[\,(1-z)f_1(1-z) + \frac{4(1-\eps)}{1-2\eps}f_2\,\right]\,
    \mathcal{P}_{\tilde{q}\to\tilde{q}g}^{(0)}(p_1,p_2) \;,
\end{equation}
while the scalar counterpart of the self-energy in
Fig.~\ref{fig:one-to-two_splittings_quark_one-loop}(c) yields
\begin{equation}\label{eq:one-loop_amplitude_scalar_f3}
    \mathcal{P}_{\tilde{q}\to \tilde{q}g}^{(1),c}(p_1,p_2) = -\frac{g_s^2}{16\pi^2}\,
    C_F \left(\!-\frac{\mu^2}{s_{12}}\!\right)^\eps\,
    \frac{4(1-\eps)}{1-2\eps}f_2\, \mathcal{P}_{\tilde{q}\to\tilde{q}g}^{(0)}(p_1,p_2) \;.
\end{equation}
Combining the three contributions gives the (anti-)triplet scalar one-loop
splitting amplitude
\begin{equation}\label{eq:one-loop_scalar_split}
    \mathcal{P}_{\tilde{q}\to\tilde{q}g}^{(1)}(p_1,p_2)
    =\frac{g_s^2}{16\pi^2}\, \frac{C_A}{2}\left(\!-\frac{\mu^2}{s_{12}}\!\right)^\eps
    \bigg[\,(1-z)f_1(1-z) - \frac{1}{N_C^2}\Big(zf_1(z)-2f_2\Big)\bigg]\,
    \mathcal{P}_{\tilde{q}\to\tilde{q}g}^{(0)}(p_1,p_2)\;.
\end{equation}
We note that this result has the same singularity structure as
$\mathcal{P}_{q\to qg}^{(1)}$, given by the first line of
Eq.~\eqref{eq:one-loop_qqg_splitting_amplitude}.

\subsubsection{Gluon initial state}
\label{sec:one_to_two_splittings_gluon_oneloop}
\begin{figure}[t]
    \includegraphics[scale=0.36]{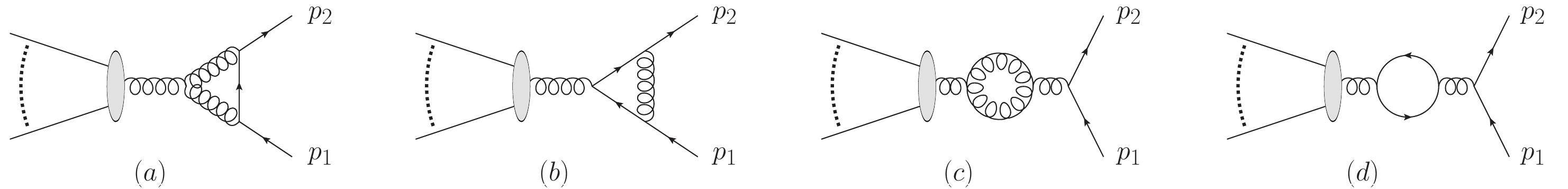}
    \caption{Feynman diagrams leading to the $1\to2$ gluon-to-quark splitting function at one loop discussed in Sec.~\ref{sec:one_to_two_splittings_gluon_oneloop}.
      \label{fig:one-to-two_splittings_gluon_quark_one-loop}}
\end{figure}

There are two one-to-two splittings initiated by a gluon, $g\to q\bar{q}$ and $g \to gg$.
The calculation of the corresponding splitting functions is again performed in light-like
axial gauge, as this gauge choice reflects the physical degrees of freedom in the
gluon propagators.
As before, due to the gauge invariance of the on-shell amplitude arising from the sum of cut diagrams,
any $t$-channel gluon propagator can be evaluated in Feynman gauge~\cite{Kosower:1999rx,Bern:2004cz}.

The one-loop $g\to q\bar{q}$ splitting function is determined by the four diagrams
in Fig.~\ref{fig:one-to-two_splittings_gluon_quark_one-loop}. 
In diagrams (a) and (c), the $s$-channel gluon loop momenta are labeled by $k$ and $p_1+p_2-k$,
in diagrams (b) and (d) the same denote the quark and anti-quark loop momenta.
Integrands involving two light-cone denominators are rewritten using the partial fractioning,
\begin{equation}
    \frac{1}{(kn)((p_1+p_2-k)n)} = \frac{1}{n(p_1+p_2)}\left(\frac{1}{kn}+\frac{1}{(p_1+p_2-k)n}\right)\,,
\end{equation} 
and a subsequent shift of the loop momentum $k\to -k+p_1+p_2$ in the second term. 
This allows us to express all amplitudes in terms of the loop integrals in App.~\ref{sec:one-loop_integrals}.

For the triangle contributions in Figs.~\ref{fig:one-to-two_splittings_gluon_quark_one-loop}(a) and \ref{fig:one-to-two_splittings_gluon_quark_one-loop}(b), integration of the loop momentum leads to
\begin{align}
    \mathcal{P}_{g\to q\bar{q}}^{(1),a}(p_1,p_2) &= \frac{g_s^2}{16\pi^2}\,C_A
    \left(\!-\frac{\mu^2}{s_{12}}\!\right)^\eps\,
    \left[\,zf_1(z)+(1-z)f_1(1-z)+\frac{4-5\eps}{1-2\eps}\,f_2\,\right]\,
    \mathcal{P}_{g\to q\bar{q}}^{(0)}(p_1,p_2) \;,\\
    \mathcal{P}_{g\to q\bar{q}}^{(1),b}(p_1,p_2) &= -\frac{g_s^2}{16\pi^2}\,(C_A-2C_F)
    \left(\!-\frac{\mu^2}{s_{12}}\!\right)^\eps\,\left(\frac{2}{1-2\eps}-\eps\right)\,
    f_2\, \mathcal{P}_{g\to q\bar{q}}^{(0)}(p_1,p_2) \;,
\end{align}
with the tree-level gluon-to-quark splitting amplitude
\begin{equation}\label{eq:tree-level_amplitude_g2qq}
    \mathcal{P}_{g\to q\bar{q}}^{(0)}(p_1,p_2) = g_sT^a_{ij}\, \bar{u}(p_2)\slash\!\!\!\varepsilon(p_{12})v(p_1)\;.
\end{equation}
The vacuum polarization diagrams in Fig.~\ref{fig:one-to-two_splittings_gluon_quark_one-loop}(c) and \ref{fig:one-to-two_splittings_gluon_quark_one-loop}(d) are given by
\begin{align}
    \mathcal{P}_{g\to q\bar{q}}^{(1),c}(p_1,p_2) &= -\frac{g_s^2}{16\pi^2}\,C_A\left(\!-\frac{\mu^2}{s_{12}}\!\right)^\eps\,\frac{6(1-\eps)(4-3\eps)}{(1-2\eps)(3-2\eps)}f_2\, \mathcal{P}_{g\to q\bar{q}}(p_1,p_2) \,,\\
    \mathcal{P}_{g\to q\bar{q}}^{(1),d}(p_1,p_2) &= \frac{g_s^2}{16\pi^2}\,T_Rn_f\left(\!-\frac{\mu^2}{s_{12}}\!\right)^\eps\,\frac{4(1-\eps)\eps}{(1-2\eps)(3-2\eps)}f_2\, \mathcal{P}_{g\to q\bar{q}}^{(0)}(p_1,p_2) \,.
\end{align}
The full splitting function is given by the sum of the terms above~\cite{Kosower:1999rx,Bern:1999ry},
\begin{equation}\label{eq:one-loop_gqq_splitting_amplitude}
    \begin{split}
    \mathcal{P}_{g\to q\bar{q}}^{(1)}(p_1,p_2;n) &= \frac{g_s^2}{16\pi^2}\left(\!-\frac{\mu^2}{s_{12}}\!\right)^\eps\,\Bigg[C_A\left(zf_1(z)+(1-z)f_1(1-z)-\left(2+\frac{3(2-\eps)}{(1-2\eps)(3-2\eps)}\right)f_2\right)\\
    &\qquad - \frac{1}{N_C}\left(\frac{2}{1-2\eps}-\eps\right)f_2 + T_Rn_f\frac{4(1-\eps)\eps}{(1-2\eps)(3-2\eps)}f_2\Bigg]\, \mathcal{P}_{g\to q\bar{q}}^{(0)}(p_1,p_2) \,.
    \end{split}
\end{equation}

\begin{figure}[t]
    \includegraphics[scale=0.36]{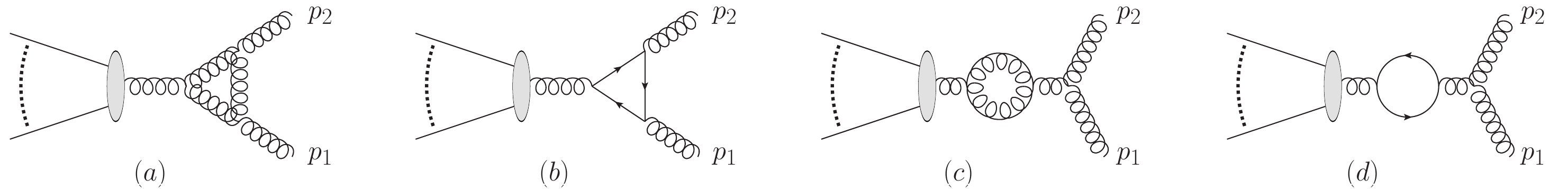}
    \caption{Feynman diagrams leading to the $1\to2$ gluon-to-gluon splitting function at one loop discussed in Sec.~\ref{sec:one_to_two_splittings_gluon_oneloop}.
      \label{fig:one-to-two_splittings_gluon_gluon_one-loop}}
\end{figure}
The diagrams leading to the one-loop $g\to gg$ splitting amplitude
are shown in Fig.~\ref{fig:one-to-two_splittings_gluon_gluon_one-loop}.
The additional bubble-like amplitude involving the four-gluon vertex vanishes in light-like axial gauge and is not shown.
In diagrams (a) and (c), the $s$-channel gluon loop momenta are again denoted
by $k$ and $p_1+p_2-k$, whereas these label the $s$-channel quark-loop momenta in 
diagram (b) and (d). The tree-level splitting amplitude is given by
\begin{equation}\label{eq:Pggg-tree-level}
    \mathcal{P}_{g\to gg}^{(0)}(p_1,p_2) = 2 g_s f^{abc} \left[
    \varepsilon^*(p_1)\cdot\varepsilon^*(p_2)\varepsilon(p_{12})\cdot p_2 + \varepsilon(p_{12})\cdot\varepsilon^*(p_1)\varepsilon^*(p_2)\cdot p_1
    - \varepsilon(p_{12})\cdot\varepsilon^*(p_2)\varepsilon^*(p_1)\cdot p_2 \right] \,.
\end{equation}
After integrating over the loop momentum, the triangle diagrams Fig.~\ref{fig:one-to-two_splittings_gluon_gluon_one-loop}(a) and Fig.~\ref{fig:one-to-two_splittings_gluon_gluon_one-loop}(b) yield
\begin{align}
    \begin{split}
    \mathcal{P}_{g\to gg}^{(1),a}&(p_1,p_2;n) = \frac{g_s^2}{16\pi^2}\frac{C_A}{2}\, 
    \left(\!-\frac{\mu^2}{s_{12}}\!\right)^\eps\, \bigg[\,
    \bigg(zf_1(z)+(1-z)f_1(1-z) \\
    &\qquad\qquad\qquad\qquad\qquad\qquad + \left(\frac{6(1-\eps)(4-3\eps)}{(1-2\eps)(3-2\eps)} - 2\right)f_2 \bigg)\, \mathcal{P}_{g\to gg}^{(0)}(p_1,p_2)
    \\
    &\quad + g_sf^{abc} \, \frac{2\eps^2}{(1-2\eps)(3-2\eps)} f_2\,
    \bigg(\varepsilon^*(p_1)\cdot\varepsilon^*(p_2) - \frac{2\varepsilon^*(p_2)\cdot p_1 \varepsilon^*(p_1)\cdot p_2}{s_{12}}\bigg)\varepsilon(p_{12})\cdot(p_1-p_2) \bigg] \,,
    \end{split}\\
    \begin{split}
    \mathcal{P}_{g\to gg}^{(1),b}&(p_1,p_2;n) = \frac{g_s^2}{16\pi^2}T_Rn_f\, \left(-\frac{\mu^2}{s_{12}}\right)^\eps\, \Bigg[
    -\frac{2(1-\eps)\eps}{(1-2\eps)(3-2\eps)}f_2\, \mathcal{P}^{(0)}_{g\to gg}(p_1,p_2) \\
    &\quad - g_sf^{abc}\,\frac{\eps^2}{(1-\eps)(1-2\eps)(3-2\eps)} f_2\,
    \bigg(\varepsilon^*(p_1)\cdot\varepsilon^*(p_2)-\frac{2\varepsilon^*(p_2)\cdot p_1 \, \varepsilon^*(p_1)\cdot p_2}{s_{12}}\bigg)\varepsilon(p_{12})\cdot(p_1-p_2)\Bigg]\;.
    \end{split}
\end{align}
The two vacuum-polarization diagrams Fig.~\ref{fig:one-to-two_splittings_gluon_gluon_one-loop}(c)
and \ref{fig:one-to-two_splittings_gluon_gluon_one-loop}(d) read
\begin{align}
    \mathcal{P}_{g\to gg}^{(1),c}(p_1,p_2;n) &= -\frac{g_s^2}{16\pi^2}\frac{C_A}{2}\, \left(-\frac{\mu^2}{s_{12}}\right)^\eps\, \frac{6(1-\eps)(4-3\eps)}{(1-2\eps)(3-2\eps)}f_2\, \mathcal{P}_{g\to gg}^{(0)}(p_1,p_2)\,,\\
    \mathcal{P}_{g\to gg}^{(1),d}(p_1,p_2;n) &= \frac{g_s^2}{16\pi^2}T_Rn_f\, \left(-\frac{\mu^2}{s_{12}}\right)^\eps\, \frac{2(1-\eps)\eps}{(1-2\eps)(3-2\eps)}f_2\, \frac{1}{2} \mathcal{P}^{(0)}_{g\to gg}(p_1,p_2)\,.
\end{align}
The complete splitting amplitude is given by the sum of the above terms
and reads~\cite{Kosower:1999rx,Bern:1999ry}
\begin{equation}\label{eq:one-loop_ggg_splitting_amplitude}
  \begin{split}
  \mathcal{P}_{g\to gg}^{(1)}(p_1,p_2) &= \frac{g_s^2}{16\pi^2}\frac{C_A}{2}\, \left(\!-\frac{\mu^2}{s_{12}}\!\right)^\eps\, \bigg[\, \big(zf_1(z)+(1-z)f_1(1-z)-2f_2\big)\, \mathcal{P}_{g\to gg}^{(0)}(p_1,p_2) \\
  &\qquad\qquad + g_sf^{abc}\left(1-\frac{T_Rn_f}{N_C}\frac{1}{1-\eps}\right)\frac{2\eps^2}{(1-2\eps)(3-2\eps)} f_2\, \\
  &\qquad\qquad\qquad\qquad \times \bigg(\varepsilon^*(p_1)\cdot\varepsilon^*(p_2)-\frac{2\varepsilon^*(p_2)\cdot p_1 \, \varepsilon^*(p_1)\cdot p_2}{s_{12}}\bigg)\varepsilon(p_{12})\cdot(p_1-p_2)\bigg]\,.
  \end{split}
\end{equation}

In order to extract the scalar component of the $g\to gg$ splitting amplitude, we decompose the triple-gluon vertex according to Eq.~\eqref{eq:rearranged_gluon_vertex}.
For diagram \ref{fig:one-to-two_splittings_gluon_gluon_one-loop}(a), the scalar can be routed in two ways, which correspond to the fermion flow of
Fig.~\ref{fig:one-to-two_splittings_quark_one-loop}(a) and~(b). 
Their contributions agree with Eqs.~\eqref{eq:one-loop_amplitude_scalar_f1}
and~\eqref{eq:one-loop_amplitude_scalar_f2} up to color factors.
Diagram \ref{fig:one-to-two_splittings_gluon_gluon_one-loop}(c) also gives two contributions, corresponding to routing the scalar piece on each propagator of the loop.  
These are trivially identical and the
sum agrees with Eq.~\eqref{eq:one-loop_amplitude_scalar_f3} up to color factors.
Diagrams \ref{fig:one-to-two_splittings_gluon_gluon_one-loop}(b) and \ref{fig:one-to-two_splittings_gluon_gluon_one-loop}(d) do not contribute, as they do not contain triple-gluon vertices linking the initial- and final-state gluons. 
The scalar decomposition of the amplitudes corresponding to Fig.\ref{fig:one-to-two_splittings_gluon_gluon_one-loop} can therefore be identified as
\begin{align}
    \begin{split}
    \mathcal{P}_{g\to gg}^{(1,\mathrm{sc})}(p_1,p_2;n) &= \frac{g_s^3}{16\pi^2}f^{abc}C_A\, \left(\!-\frac{\mu^2}{s_{12}}\!\right)^\eps\,
    \bigg[ zf_1(z)+(1-z)f_1(1-z)-2f_2 \bigg] \\
    &\qquad \times 
    \left(\varepsilon(p_{12}) \cdot \varepsilon^*(p_{1}) \, \varepsilon^*(p_2) \cdot p_1
    -\varepsilon(p_{12}) \cdot \varepsilon^*(p_{2}) \, \varepsilon^*(p_1) \cdot p_2 \right) \,.
     \end{split}
\end{align}
As expected this reproduces the pole structure of the full splitting function,
$\mathcal{P}_{g\to gg}^{(1)}$, i.e., the first line of Equation~\eqref{eq:one-loop_ggg_splitting_amplitude}.
Accounting for the symmetry of the two-gluon final state
(see Eq.~\eqref{eq:rearranged_gluon_vertex} and Fig.~\ref{fig:examples}(b)),
we can read off the one-loop splitting function for a color adjoint scalar,
\begin{align}\label{eq:one-loop_scalar_split_adjoint}
    \begin{split}
    \mathcal{P}_{g\to g}^{(1,\mathrm{sc})}(p_1,p_2;n) &= 
    \frac{g_s^3}{16\pi^2}f^{abc}\frac{C_A}{2}\, \left(\!-\frac{\mu^2}{s_{12}}\!\right)^\eps\,
    \bigg[ zf_1(z)+(1-z)f_1(1-z)-2f_2 \bigg]\,2p_1^\mu\varepsilon_\mu^*(p_2)\,.
    \end{split}
\end{align}

\subsection{Scalar multipoles}
\label{sec:one-loop_scalar_multipoles}
The starting point for the computation of the one-loop scalar radiation pattern
is an extension of the decomposition of the soft one-loop amplitude in Eq.~(38)
of Ref.~\cite{Catani:2000pi}. It states that, in the soft gluon limit, one can
write the matrix element as
\begin{equation}\label{eq:cg_decomposition_soft}
  |\mathcal{M}^{(1)}_{\rm soft}(q,\{p\})\rangle
  =\mu^{2\eps}\varepsilon^\mu(q){\bf J}_\mu(q)|\mathcal{M}^{(1)}\rangle
  +\Big(|\mathcal{M}^{(1)}_{\rm soft}(q,\{p\})\rangle
  -\mu^{2\eps}\varepsilon^\mu(q){\bf J}_\mu(q)|\mathcal{M}^{(1)}(\{p\})\rangle\Big)\;.
\end{equation}
The non-factorizable one-loop soft corrections are contained in
$|\mathcal{M}^{(1)}_{\rm soft}(q,\{p\})\rangle$. They can be
computed using eikonal Feynman rules.
Their explicit poles follow a dipole radiation pattern described
by Catani's infrared singularity operators~\cite{Catani:1998bh,Sterman:2002qn}.
The natural extension of Eq.~\eqref{eq:cg_decomposition_soft}
to the case of scalar radiators is therefore given by a dipole
approximation to the hard process producing the scalar particles.
As the matrix elements in this approximation may not include all
Feynman diagrams, a systematic and manifestly gauge-invariant
approach to the one-loop computation is indispensable.
This problem is solved by the background field method~\cite{
  DeWitt:1967ub,Honerkamp:1972fd,Kluberg-Stern:1974nmx,
  Abbott:1980hw,Abbott:1981ke,Abbott:1983zw}. 
The technique allows to obtain individually renormalizable
$n$-point Greens functions that obey the naive Ward identities
and are thus physically meaningful quantities.
The extension of Eq.~\eqref{eq:cg_decomposition_soft} reads
\begin{equation}\label{eq:one-loop_decomposition_scalar}
  |\mathcal{M}^{(1)}_{\rm approx}(q,\{p\})\rangle
  =\mu^{2\eps}\varepsilon^\mu(q){\bf J}_\mu(q)|\mathcal{M}^{(1)}\rangle
  +\Big(|\mathcal{M}^{(1)}_{\rm BGF,dip}(q,\{p\})\rangle
  -\mu^{2\eps}\varepsilon^\mu(q){\bf J}_\mu(q)|\mathcal{M}^{(1)}(\{p\})\rangle\Big)\;,
\end{equation}
where, $|\mathcal{M}^{(1)}_{\rm BGF,dip}(q,\{p\})\rangle$ consists of
a sum over all dipole contributions and one-particle reducible 
contributions in the background field technique.
We will compute them in the following subsections, using a number of 
standard methods~\cite{Passarino:1978jh,Denner:2005nn} and tools~\cite{
  Hahn:1998yk,Hahn:2000kx,Hahn:2000jm,Christensen:2008py,Alloul:2013bka,
  Patel:2015tea,Patel:2016fam,Tentyukov:2007mu,Kuipers:2012rf,Ruijl:2017dtg}.
Due to the spin-independence of the soft-gluon limit of the one-loop
matrix element, Eq.~\eqref{eq:one-loop_decomposition_scalar} reproduces
Eq.~\eqref{eq:cg_decomposition_soft}, but it includes additional
terms at sub-leading power in the soft scaling parameter.

We begin with a single term in the dipole approximation, where the
antenna is formed by massless scalars with momenta $p_i$ and $p_k$
and color indices $i$ and $k$. They radiate a gluon of momentum
$q_1$ and color index $a$. All momenta are considered outgoing.
To simplify the notation, we define the invariants
\begin{equation}\label{eq:def_invariants_one-loop_scalar}
  s=(p_i+p_k)^2\;,\qquad
  t=(p_i+q_1)^2\;,\qquad
  u=(p_k+q_1)^2\;,\qquad
  \text{and}\qquad
  Q^2=(p_i+p_k+q_1)^2\;.
\end{equation}
The leading-order diagrams are shown
in Fig.~\ref{fig:scalar_single_emission}, and the
leading-order dipole component of the current is defined in 
Eq.~\eqref{eq:scalar_current_sum}.

\subsubsection{Factorizable contributions}
\begin{figure}
    \includegraphics[scale=0.36]{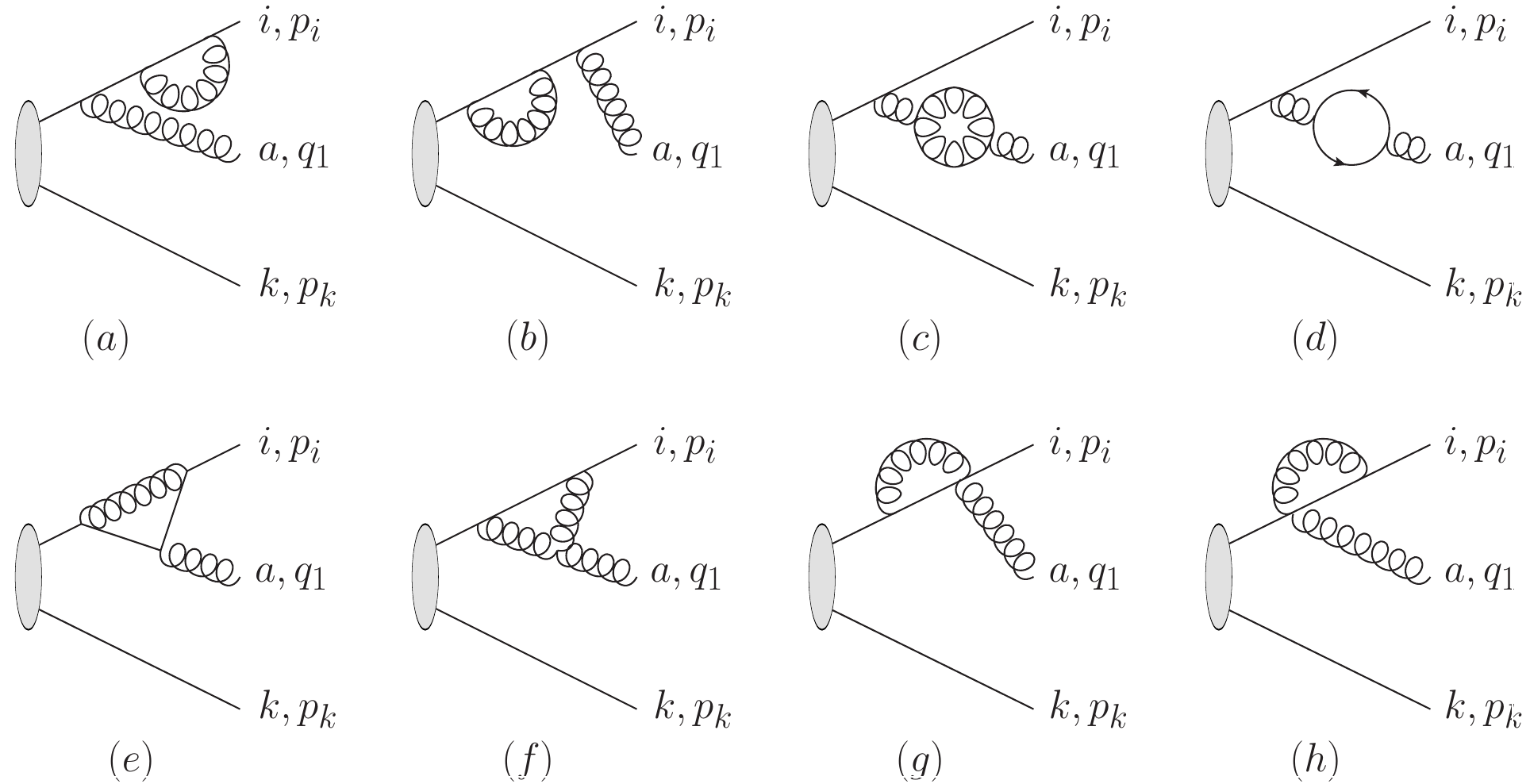}
    \caption{Factorizable one-loop corrections to the scalar-scalar interference.
      \label{fig:factorized_contributions}}
\end{figure}
We first derive the factorizable one-loop corrections to the production of the scalars.
The relevant diagrams are shown in Fig.~\ref{fig:factorized_contributions}.
The one-loop scalar integrals are listed in App.~\ref{sec:one-loop_integrals}.
Diagrams of type~(a) and~(b) are given by the leading-order current, times
the scalar self energy, which, for a scalar of virtuality $p^2$, is given by
\begin{equation}\label{eq:one-loop_scalar_self_energy}
    \hat{\Sigma}_{ij}(p)=-\frac{g_s^2}{16\pi^2}\,C_F\delta_{ij}\,
    2p^2\,\hat{I}_2\big(p^2\big)\;.
\end{equation}
Diagrams of type~(c) are given by the leading-order current times the $C_A$ part
of the vacuum polarization, which, for a gluon of virtuality $q^2$ is given by
\begin{equation}\label{eq:one-loop_vacuum_polarization_ca}
  \hat{\Pi}_{\mu\nu}^{ab}(q)=-\frac{g_s^2}{16\pi^2}\,C_A\delta^{ab}\,
  P_{\mu\nu}(q)\,q^2\,\hat{I}_2(q^2)\,
  \frac{11-7\eps}{3-2\eps}\;,
  \quad\text{where}\quad
  P^{\mu\nu}(q)=-g^{\mu\nu}+\frac{q^\mu q^\nu}{q^2}\;.
\end{equation}
Diagrams of type~(d) are given by the leading-order current times the
fermionic ($n_f$ part) of the vacuum polarization, which, for a gluon
of virtuality $q^2$, is given by
\begin{equation}\label{eq:one-loop_vacuum_polarization_nf}
  \hat{\Pi}_{\mu\nu}^{ab\,\rm(f)}(q)=-\frac{g_s^2}{16\pi^2}\,T_Rn_f\delta^{ab}\,
  P_{\mu\nu}(q)\,q^2\,\hat{I}_2(q^2)\,
  \frac{4-4\eps}{3-2\eps}\;.
\end{equation}
This contribution is needed as the virtual correction corresponding to
Eq.~\eqref{eq:scalar_emission_quark_pair_individual}, which is
schematically shown in Fig.~\ref{fig:scalar_quark_antiquark_emission}.
Diagrams of type~(e),~(f),~(g) and~(h) are given in terms of the leading-order current,
times the vertex correction to the scalar-scalar-gluon vertex, which is given by
\begin{equation}\label{eq:one-loop_scalar_vertex_corr}
  \hat{\Gamma}^{\mu,a}_{S,ij}(q,p,k)=
  \frac{g_s^3}{16\pi^2}\,2C_F\,T^a_{ij}\,\frac{p^\mu}{pk}\,
  \Big(q^2\,\hat{I}_2\big(q^2\big)-p^2\,\hat{I}_2\big(p^2\big)\Big)\;,
  \qquad\text{where}\qquad
  q^\mu=p^\mu+k^\mu\;.
\end{equation}
As the vertex corrections in the background field method satisfy the
naive Ward identities, their abelian contribution cancels the self
energies from diagrams of type~(a) and~(b), while the non-abelian
component vanishes identically. This provides the generalization
of the soft-gluon result for diagrams of class A in Sec.~4.2
of Ref.~\cite{Catani:2000pi}.

In addition to the self energy, vacuum polarization and vertex correction,
we obtain the factorizable one-loop corrections in diagrams of type~(c)
and~(e) of Fig.~\ref{fig:box_integrals}. They are quantum corrections
to the production of the charged scalars, times a leading-order current.
The result for diagrams of type~(c) is given by
\begin{equation}\label{eq:factorizable_hard_correction}
   I_{h,f}(Q^2,t)=-\frac{g_s^2}{16\pi^2}\,\hat{\bf T}_i^b\hat{\bf T}_k^b\,
   \Big(\big(2Q^2-t\big)\hat{I}_3^{2m}(Q^2,t)+\hat{I}_2(Q^2)-\hat{I}_2(t)\Big)\;.
\end{equation}
In the limit $t\to 0$, the term in parentheses reduces to 
$2Q^2\,\hat{I}_3^{1m}(Q^2)+\hat{I}_2(Q^2)$, which yields
the result for diagrams of type~(e).

\subsubsection{Box-type contributions}
\begin{figure}
    \includegraphics[scale=0.36]{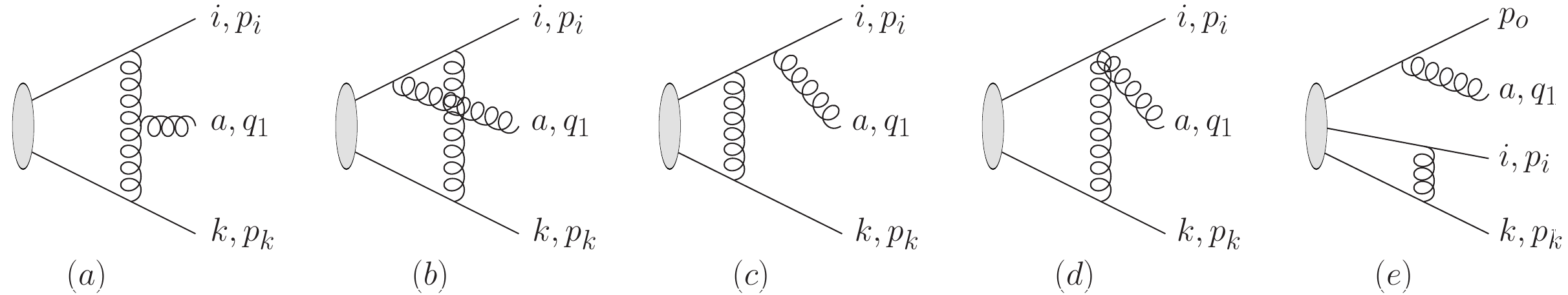}
    \caption{Box-type diagrams contributing to the scalar-scalar interference at one loop.
      \label{fig:box_integrals}}
\end{figure}
To obtain a gauge invariant set of box-type diagrams, we combine the box and
seagull diagrams in Fig.~\ref{fig:box_integrals}(a),~(b) and~(d) and add the
factorizable contributions in diagrams~(c) and~(e). In diagrams of type~(e),
the final-state gluon is emitted off a color charged particle other than
$i$ or $j$, hence the result only depends on $s$ and a leading-order current.
\begin{equation}
  \label{eq:scalar_scalar_1l_uncorrelated}
  I_{f}(s,\{\vec{p}\,\})=\frac{g_s^3}{16\pi^2}\,\hat{\bf T}_i^b\hat{\bf T}_k^b\,
  \Big[\,2s\,\hat{I}_3^{1m}(s)+\hat{I}_2(s)\,\Big]
  \sum_{l\neq i,k}\hat{\bf T}^a_l\,\frac{p_l^\mu}{2p_lq_1}\epsilon_\mu^*(q_1)\;.
\end{equation}
We combine these terms with the factorized contribution
$-\mu^{2\eps}\varepsilon^\mu(q){\bf J}_\mu(q)|\mathcal{M}^{(1)}(\{p\})\rangle$
in Eq.~\eqref{eq:one-loop_decomposition_scalar}, which can be derived from the
hard vertex correction, Eq.~\eqref{eq:factorizable_hard_correction}, and the
leading-order current in Eq.~\eqref{eq:scalar_current_sum}~\cite{Catani:2000pi}. 
Using the identity $2\hat{\bf T}_i^a\hat{\bf T}_i^b=
if^{abc}\,\hat{\bf T}_i^c+\{\hat{\bf T}^a,\hat{\bf T}^b\}_i$,
we separate the remainder into an abelian and a purely non-abelian part.
The results are
\begin{equation}\label{eq:factorization_counterterm}
  \begin{split}
  C^{\rm(nab)}(s,t,u)=&\;\frac{ig_s^3}{16\pi^2}\,f^{abc}\,\hat{\bf T}_i^b\hat{\bf T}_k^c\,
  \Big[\,2s\,\hat{I}_3^{1m}(s)+\hat{I}_2(s)\,\Big]
  \bigg(\frac{p_i^\mu}{t}-\frac{p_k^\mu}{u}\bigg)\epsilon_\mu^*(p_2)\;,\\
  C^{\rm(ab)}(s,t,u)=&\;\frac{g_s^3}{16\pi^2}\,
  \big\{\hat{\bf T}^a,\hat{\bf T}^b\big\}_i\hat{\bf T}_k^b\,
  \Big[\,2s\,\hat{I}_3^{1m}(s)+\hat{I}_2(s)\,\Big]\,\frac{p_i^\mu}{t}\epsilon_\mu^*(q_1)+
  \Bigg(\!\begin{array}{r@{\,}l}p_i&\leftrightarrow p_k\\
  i&\leftrightarrow k\\t&\leftrightarrow u\end{array}\!\!\Bigg)\;.
  \end{split}
\end{equation}
The sum of diagrams in Fig.~\ref{fig:box_integrals} can be split into
abelian and non-abelian corrections. The non-abelian part has a dipole
structure and factorizes onto the leading-order dipole matrix element.
After subtracting off the non-abelian counterterm in
Eq.~\eqref{eq:factorization_counterterm}, it is given by the expression
\begin{equation}\label{eq:scalar_scalar_1l_dipole}
  \begin{split}
  I_{4}^{{\rm(nab)}}(s,t,u)=&\;-\frac{ig_s^3}{16\pi^2}\,
  f^{abc}\,\hat{\bf T}_i^{b}\hat{\bf T}_k^{c}\bigg[\;\hat{I}_2(Q^2)-\hat{I}_2(s)
    +2t\,\hat{I}_3^{1m}(t)+2u\,\hat{I}_3^{1m}(u)\\
    &\;\qquad\qquad\qquad\quad
    -s\,t\,\hat{I}_4^{1m}(s,t,u)-u\,s\,\hat{I}_4^{1m}(u,s,t)
       +2\,t\,u\,\hat{I}_4^{1m}(t,u,s)\,\bigg]
  \bigg(\frac{p_i^\mu}{t}-\frac{p_k^\mu}{u}\bigg)\epsilon_\mu^*(q_1)\;.
  \end{split}
\end{equation}
In the soft limit, Eq.~\eqref{eq:scalar_scalar_1l_dipole} reduces to a dipole
contribution to the 1-loop soft current~\cite{Bern:1998sc,Bern:1999ry,Catani:2000pi}
\begin{equation}\label{eq:scalar_scalar_1l_dipole_sl}
  \begin{split}
  I_{4}^{{\rm(nab)}}(s,t,u)\overset{q_1\to 0}{\to}&\;
  \frac{g_s^3}{16\pi^2}\,f^{abc}\,\hat{\bf T}_i^{b}\hat{\bf T}_k^{c}\,\frac{c_\Gamma}{\eps^2}\,
  \frac{4\,\pi\eps}{\sin(\pi\eps)}\bigg(\!\!-\frac{\mu^2 s}{t\,u}\!\bigg)^\eps
  \bigg(\frac{p_i^\mu}{t}-\frac{p_k^\mu}{u}\bigg)\epsilon_\mu^*(q_1)\;.
  \end{split}
\end{equation}
For a single dipole, the color factor simplifies to $-i\hat{\bf T}^a_{ik}\,C_A/2$,
and the result remains otherwise unchanged.\\
In the collinear limit, $t\to 0$, $s\to z\,Q^2$ and $u\to(1-z)\,Q^2$, we obtain
\begin{equation}\label{eq:scalar_scalar_1l_dipole_cl}
  \begin{split}
  &I_{4}^{{\rm(nab)}}(s,t,u)\overset{t\to0}{\to}
  \frac{g_s^3}{16\pi^2}\,f^{abc}\,\hat{\bf T}_i^{b}\hat{\bf T}_k^{c}\,
  \frac{c_\Gamma}{\eps^2}\,\bigg[\,
  (1-z^{-\eps})\frac{2-3\eps}{1-2\eps}\left(\!-\frac{\mu^2}{Q^2}\!\right)^\eps\\
  &\qquad+2
  \left(\!-\frac{\mu^2}{t}\!\right)^\eps\Big(
    2(1-z)\,_2F_1(1,1;1-\eps;z)
    -z\,_2F_1(1,1;1-\eps;1-z)+1\Big)\,\bigg]
    \frac{p_i^\mu}{t}\,\epsilon_\mu^*(q_1)\;,
  \end{split}
\end{equation}
where the first term in the square bracket contributes a sub-leading pole
due to the scale difference between the 1-loop hard function in
Eq.~\eqref{eq:factorizable_hard_correction} and the counterterm
in Eq.~\eqref{eq:factorization_counterterm}.
A conversion to the functions defined in \cite{Kosower:1999rx} is
obtained by using the following relation between hypergeometric functions
\begin{equation}\label{eq:collinear_box_conversion}
    \, _2F_1(1,1;1-\eps;1-z)=\Gamma(1-\eps)\Gamma(1+\eps)\,z^{-1-\eps}(1-z)^\eps
    +\frac{1}{z}-\frac{(1-z)^\eps}{z}\, _2F_1(\eps,\eps;1+\eps;z)\;,
\end{equation}
which gives $f_1(z)=-2c_\Gamma/\eps^2 \,_2F_1(1,1;1-\eps;1-z)$.
In the soft-collinear limit, $z\to 1$, Eq.~\eqref{eq:scalar_scalar_1l_dipole_cl}
simplifies to Eq.~\eqref{eq:scalar_scalar_1l_dipole_sl}. The entire
non-factorizable contribution then comes from the term proportional to
$-2(1-z)\,_2F_1(1,1;1-\eps;z)$, which can be written as
$(1-z)f_1(1-z)\,\eps^2/c_\Gamma$. This agrees with the soft limit
of Eqs.~(4.2) and~(4.18) in~\cite{Kosower:1999rx}, up to color factors.
The collinear limit $u\to 0$ of Eq.~\eqref{eq:scalar_scalar_1l_dipole}
is obtained from Eq.~\eqref{eq:scalar_scalar_1l_dipole_cl} by
interchanging $t$ and $u$.

The abelian contribution is symmetric in the particle momenta
and color indices, and it factorizes on the leading-order current.
After subtracting off the abelian counterterm in
Eq.~\eqref{eq:factorization_counterterm}, we obtain
\begin{equation}\label{eq:scalar_scalar_1l_ab_osc_all}
  \begin{split}
  &I_{4}^{{\rm(ab)}}(s,t,u)=\frac{g_s^3}{16\pi^2}\frac{
    \big\{\hat{\bf T}^b,\hat{\bf T}^a\big\}_i\hat{\bf T}_k^b\, u
    -\hat{\bf T}_i^b\big\{\hat{\bf T}^b,\hat{\bf T}^a\big\}_k\, t}{u+t}\,
    \Big[\hat{I}_2(Q^2)-\hat{I}_2(s)\Big]
    \bigg(\frac{p_i^\mu}{t}-\frac{p_k^\mu}{u}\bigg)\,\epsilon_\mu^*(q_1)\\
    &\;\quad+\frac{g_s^3}{16\pi^2}
    \big\{\hat{\bf T}^a,\hat{\bf T}^b\big\}_i\hat{\bf T}_k^b\bigg[\;
    2Q^2\,\hat{I}_3^{1m}(Q^2)-2s\,\hat{I}_3^{1m}(s)-2t\,\hat{I}_3^{1m}(t)
    +s\,t\,\hat{I}_4^{1m}(s,t,u)\;\bigg]
    \bigg(\frac{p_i^\mu}{t}-\frac{p_k^\mu}{u}\bigg)\,\epsilon_\mu^*(q_1)\\
    &\;\quad-\frac{g_s^3}{16\pi^2}
    \hat{\bf T}_i^b\big\{\hat{\bf T}^a,\hat{\bf T}^b\big\}_k\bigg[\;
    2Q^2\,\hat{I}_3^{1m}(Q^2)-2s\,\hat{I}_3^{1m}(s)-2u\,\hat{I}_3^{1m}(u)
    +u\,s\,\hat{I}_4^{1m}(u,s,t)\;\bigg]
    \bigg(\frac{p_i^\mu}{t}-\frac{p_k^\mu}{u}\bigg)\,\epsilon_\mu^*(q_1)\;.
  \end{split}
\end{equation}
In the soft limit, Eq.~\eqref{eq:scalar_scalar_1l_ab_osc_all} vanishes,
in agreement with the non-abelian structure of the soft-gluon current
at 1-loop order~\cite{Catani:2000pi}.
In the $t\to 0$, $s\to z\,Q^2$, $u\to(1-z)\,Q^2$ limit,
Eq.~\eqref{eq:scalar_scalar_1l_ab_osc_all} gives
\begin{equation}\label{eq:collinear_limit_abelian_osc}
  \begin{split}
    I_{4}^{\rm(ab)}(s,t,u)\overset{t\to 0}{\to}&\;
    \frac{ig_s^3}{16\pi^2}\,\frac{c_\Gamma}{\eps^2}
    \big\{\hat{\bf T}^a,\hat{\bf T}^b\big\}_i\hat{\bf T}_k^b\\
    &\quad\times\bigg[\,2\left(\!-\frac{\mu ^2}{t}\!\right)^\eps
    \Big(z\,_2F_1(1,1;1-\eps;1-z)-1\Big)
    +(1-z^{-\eps})\frac{2-3\eps}{1-2\eps}\left(\!-\frac{\mu ^2}{Q^2}\!\right)^\eps\,\bigg]
    \frac{p_i^\mu}{t}\,\epsilon_\mu^*(q_1)\;.
  \end{split}
\end{equation}
Again, this can be converted to the conventions used in~\cite{Kosower:1999rx}
by means of Eq.~\eqref{eq:collinear_box_conversion}.

\subsubsection{Other contributions}
In scalar QCD, there are also corrections to the scalar-scalar interference,
which involve an $s$-channel gluon decaying into the two final-state scalars.
These corrections factorize on the two-gluon production process
and are individually gauge invariant. They do not appear in a spin
decomposition of the one-loop splitting amplitudes. The bubble-type corrections
to a color conserving hard function are individually gauge invariant as well
and vanish identically.
We therefore find that Eqs.~\eqref{eq:one-loop_vacuum_polarization_ca},%
~\eqref{eq:one-loop_vacuum_polarization_nf},~\eqref{eq:scalar_scalar_1l_dipole} 
and~\eqref{eq:scalar_scalar_1l_ab_osc_all} are the only relevant
corrections to the dipole radiators at one-loop level.

\subsection{Composition of one-loop splitting functions}
\label{sec:one-loop_splitting_decomposition}
In this subsection we provide the main result on the one-loop splitting functions.
As in the tree-level case, we perform a decomposition into scalar radiators,
computed in Sec.~\ref{sec:one-loop_scalar_multipoles}, and pure splitting remainders.
We also quote the actual splitting functions, which are obtained from the splitting
amplitudes computed in Sec.~\ref{sec:one-loop_splittings} by interference with the
tree-level expressions.

In order to match the splitting functions to the scalar radiators,
we use the expressions for a color dipole and employ color conservation.
The abelian and non-abelian one-loop scalar currents in the collinear limit
are given by Eq.~\eqref{eq:scalar_scalar_1l_dipole_cl}
and~\eqref{eq:collinear_limit_abelian_osc}.
After subtracting the contributions due to the
hard vertex correction, and the counterterms in
Eq.~\eqref{eq:factorization_counterterm}, we find
\begin{equation}\label{eq:scalar_one-loop_matching_f}
  I_{4,f}(s,t,u)\overset{t\to0}{\to}
  \frac{g_s^3}{16\pi^2}\,C_A\,T^a_{ik}\,
  \bigg[\,(1-z)\,f_1(1-z)-\frac{1}{N_C^2}\Big(z\,f_1(z)-2f_2\Big)
    \,\bigg]\left(\!-\frac{\mu^2}{t}\!\right)^\eps
  \frac{p_i^\mu}{t}\,\epsilon_\mu^*(q_1)\;,
\end{equation}
for color triplet scalars, and
\begin{equation}\label{eq:scalar_one-loop_matching_a}
  I_{4,a}(s,t,u)\overset{t\to 0}{\to}
  \frac{g_s^3}{16\pi^2}\,C_A\,F^a_{ik}\,
  \bigg[\,(1-z)\,f_1(1-z)+z\,f_1(z)-2f_2\,\bigg]
  \left(\!-\frac{\mu^2}{t}\!\right)^\eps\frac{p_i^\mu}{t}\,\epsilon_\mu^*(q_1)\;,
\end{equation}
for color octet scalars.
These correspond to the scalar components of 
Eqs.~\eqref{eq:one-loop_qqg_splitting_amplitude}
and~\eqref{eq:one-loop_ggg_splitting_amplitude}, which are given in 
Eqs.~\eqref{eq:one-loop_scalar_split}
and~\eqref{eq:one-loop_scalar_split_adjoint} respectively.
In particular, we find the following relation which
makes the correspondence between the collinear limit
of the multipole radiator and the splitting amplitude
in the axial gauge manifest~\cite{Bern:2004cz}.
\begin{equation}
  \begin{split}
  \lim_{t\to0}I_{4,f}\big(zQ^2,t,(1-z)Q^2\big)
  &=\frac{1}{t}\,
  \mathcal{P}_{\tilde{q}\to\tilde{q}g}^{(1)}(p_i,q_1)\;,\\
  \lim_{t\to0}I_{4,a}\big(zQ^2,t,(1-z)Q^2\big)
  &=\frac{1}{t}\,
  \mathcal{P}_{g\to gg}^{(1),\mathrm{sc}}(p_i,q_1)\;.\\
  \end{split}
\end{equation}

\subsubsection{Quark initial state}
The full one-loop quark splitting function is determined by the product of Eq.~\eqref{eq:one-loop_qqg_splitting_amplitude} and the tree-level quark splitting amplitude in Eq.~\eqref{eq:tree-level_amplitude_q2qg}, in analogy to Eq.~\eqref{eq:coll_q_to_qg},
\begin{equation}
    P_{q\to q}^{(1)\, ss'} = 2\mathrm{Re}\bigg\{\,\delta^{ss'}\sum\limits_{\mathrm{pol}}
    \mathcal{P}_{q\to qg}^{(1)}(p_1,p_2)\mathcal{P}_{q\to qg}^{(0)\,*}(p_1,p_2)\,\bigg\} \left(\frac{16\pi^2}{g_s^4}\right)\,.
\end{equation}
The spin-averaged unrenormalized one-loop quark splitting function therefore reads
\begin{equation}\label{eq:one-loop_q2qg_splitting_function}
    \begin{split}
        \langle P_{q\to q}^{(1)}(p_1,p_2)\rangle &= P_{\tilde{q}\to \tilde{q}}^{(1)}(p_1,p_2) 
        + \langle P_{q\to q}^{(1,p)}(p_1,p_2)\rangle\;,
    \end{split}
\end{equation}
where the scalar one-loop splitting function is given by the product of
Eq.~\eqref{eq:one-loop_scalar_split} and a tree-level splitting amplitude, c.f. Eq.~\eqref{eq:coll_q_to_qg_components} (note that $p_1^2=p_2^2=0$):
\begin{equation}\label{eq:one-loop_sq2sqg_splitting_function}
    P_{\tilde{q}\to \tilde{q}}^{(1)}(p_1,p_2) = C_A\,\left(\!-\frac{\mu^2}{s_{12}}\!\right)^\eps\,\left[(1-z)f_1(1-z) - \frac{1}{N_C^2}\left(zf_1(z)-2f_2\right)\right]\, P_{\tilde{q}\to\tilde{q}}(p_1,p_2)\;.
\end{equation}
The spin-dependent remainder is given by
\begin{equation}\label{eq:one-loop_q2qg_remainder_function}
    \begin{split}
        \langle P_{q\to q}^{(1,p)}(p_1,p_2)\rangle =&\;
        C_A\,\left(\!-\frac{\mu^2}{s_{12}}\!\right)^\eps\,\left[(1-z)f_1(1-z) - \frac{1}{N_C^2}\left(zf_1(z)-2f_2\right)\right]\, \langle P_{q\to q}^{(\mathrm{f})}(p_1,p_2)\rangle \\
        &\qquad - C_A\left(1+\frac{1}{N_C^2}\right)\,\left(\!-\frac{\mu^2}{s_{12}}\!\right)^\eps\,\frac{\eps^2}{1-2\eps}f_2\,
        \langle P_{q\to q}^{(\mathrm{f},1)}(p_1,p_2)\rangle\;,
    \end{split}
\end{equation}
which consists of two terms, one proportional to the purely fermionic tree-level contribution of the tree-level splitting function given in Eq.~\eqref{eq:coll_q_to_qg_components} and a second, new spin structure first appearing at one loop,
\begin{equation}
    \langle P_{q\to q}^{(\mathrm{f},1)}(p_1,p_2) \rangle = \langle P_{q\to q}(p_1,p_2)\rangle - C_F\frac{(1+z)z}{1-z} = C_F\big(1-\eps(1-z)\big)\;.
\end{equation}
Hence, only the first term in the one-loop quark splitting function accounts for the semi-classical radiation pattern. 
As expected, this includes the leading contribution in terms of combined explicit $1/\eps$, and implicit $1/(1-z)$ singularities. All other terms
constitute the purely fermionic contribution.

\subsubsection{Gluon initial state}
The full one-loop gluon splitting tensors are determined from the polarization sum over the product of a one-loop splitting amplitude with the respective tree-level amplitude, in analogy to the tree-level case discussed in Eq.~\eqref{eq:coll_gqq_ggg_step1},
\begin{equation}
    P_{g\to X}^{(1)\,\mu\nu}(p_1,p_2) = 2\mathrm{Re}\bigg\{\sum\limits_{\mathrm{pol}} \big(\varepsilon^{*\,\mu}(p_{12})\mathcal{P}_{g\to X}^{(1)}(p_1,p_2)\big)\big(\mathcal{P}_{g\to X}^{(0)\,*}(p_1,p_2)\varepsilon^\nu(p_{12})\big)\bigg\}  \left(\frac{16\pi^2}{g_s^4}\right) \;.
\end{equation}

Averaging over the polarizations of the initial-state gluon in the polarization sum over the product of Eq.~\eqref{eq:one-loop_ggg_splitting_amplitude} with the tree-level splitting amplitude in Eq.~\eqref{eq:Pggg-tree-level} then yields the complete unrenormalized spin-averaged gluon-to-gluon splitting function,
\begin{equation}\label{eq:one-loop_g2gg_splitting_function}
    \begin{split}
    \langle P_{g\to g}^{(1)}(p_1,p_2)\rangle &= P_{g\to g}^{(1,\mathrm{sc})}(p_1,p_2) + P_{g\to g}^{(1,\mathrm{sc})}(p_2,p_1)
    +\langle P_{g\to g}^{(1,p)}(p_1,p_2)\rangle\;,
    \end{split}
\end{equation}
where the adjoint scalar and vector tree-level splitting functions are defined in Eq.~\eqref{eq:ggg_scalar_vector}. The adjoint scalar one-loop splitting function is given by the product of Eq.~\eqref{eq:one-loop_scalar_split_adjoint} and the tree-level adjoint scalar splitting amplitude as,
\begin{equation}\label{eq:one-loop_scalar_gluon_splitting_function}
    P_{g\to g}^{(1,\mathrm{sc})}(p_1,p_2) = C_A\,\left(\!-\frac{\mu^2}{s_{12}}\!\right)^\eps\,\big[(1-z)f_1(1-z)+zf_1(z)-2f_2\big]\, P_{g\to g}^{(\mathrm{sc})}(p_1,p_2)\;.
\end{equation}
The remainder function is given by
\begin{equation}\label{eq:one-loop_g2gg_remainder_function}
    \begin{split}
    \langle P_{g\to g}^{(1,p)}(p_1,p_2)\rangle &= 
    C_A\, \left(\!-\frac{\mu^2}{s_{12}}\!\right)^\eps\,\big[(1-z)f_1(1-z)+zf_1(z)-2f_2\big]\, \langle P_{g\to g}^{(\mathrm{v})}(p_1,p_2)\rangle \\
    &\qquad + \left(C_A-\frac{T_Rn_f}{1-\eps}\right)\, \left(\!-\frac{\mu^2}{s_{12}}\!\right)^\eps\,\frac{2\eps^2}{(1-2\eps)(3-2\eps)} f_2\, \langle P_{g\to g}^{(\mathrm{v},1)}(p_1,p_2)\rangle \;,
    \end{split}
\end{equation}
where we have introduced the purely spin-dependent component first appearing at one loop,
\begin{equation}
    \langle P_{g\to g}^{(\mathrm{v},1)}(p_1,p_2)\rangle = -C_A\,\frac{1-2\eps z(1-z)}{1-\eps}\;.
\end{equation}

The complete spin-averaged one-loop gluon-to-quark splitting tensor is determined from averaging over the polarizations of the initial-state gluon in the polarization sum of the product of Eq.~\eqref{eq:one-loop_gqq_splitting_amplitude} with the tree-level splitting amplitude of Eq.~\eqref{eq:tree-level_amplitude_g2qq}. It fully factorizes onto the tree-level splitting function in Eq.~\eqref{eq:gqq_dglap_limit_sum} and does not contain a scalar contribution. It reads
\begin{equation}\label{eq:one-loop_g2q_splitting_function}
    \begin{split}
    \langle P_{g\to q}^{(1)}(p_1,p_2)\rangle &= C_A\,\left(\!-\frac{\mu^2}{s_{12}}\!\right)^\eps\,\Bigg[\left(zf_1(z)+(1-z)f_1(1-z)-\left(2+\frac{3(2-\eps)}{(1-2\eps)(3-2\eps)}\right)f_2\right)\\
    &\qquad - \frac{1}{N_C^2}\left(\frac{2}{1-2\eps}-\eps\right)f_2 + \frac{T_Rn_f}{N_C}\frac{4(1-\eps)\eps}{(1-2\eps)(3-2\eps)}f_2\Bigg]\, \langle P_{g\to q}(p_1,p_2)\rangle \;.
    \end{split}
\end{equation}

\subsubsection{Singularity structure of the remainder functions}
\label{sec:loop-level_singularity_structure}
\begin{table}[t]
  \renewcommand{\arraystretch}{1.5}
  \begin{tabular}{c|c|c|c}\hline
  \parbox{15ex}{\centering Function} & \parbox{15ex}{\centering Definition} &
  \multicolumn{2}{c}{Scaling behavior for $\lambda\to 0$}\\\cline{3-4}
  $\times s_{12}^{-1}$ & & \parbox{23ex}{\centering $p_{2}\to\lambda p_{2}$} &
  \parbox{23ex}{\centering $p_{1}\to\lambda p_{1}$}\\\hline
  $P_{\tilde{q}\to\tilde{q}}^{(1)}$ & Eq.~\eqref{eq:one-loop_sq2sqg_splitting_function} &
  $\varpropto\lambda^{-2-2\eps}/\eps^2$ & $\varpropto\lambda^{-\eps}/\eps^2$\\
  $P_{g\to g}^{(1,sc)}$ & Eq.~\eqref{eq:one-loop_scalar_gluon_splitting_function} &
  $\varpropto\lambda^{-2-2\eps}/\eps^2$ & $\varpropto\lambda^{-2\eps}/\eps^2$\\\hline
  $\langle P_{q\to q}^{(1)}\rangle$ & Eq.~\eqref{eq:one-loop_q2qg_splitting_function} &
  $\varpropto\lambda^{-2-2\eps}/\eps^2$ & $\varpropto\lambda^{-1-\eps}/\eps^2$\\
  $\langle P_{q\to q}^{(1,p)}\rangle$ & Eq.~\eqref{eq:one-loop_q2qg_remainder_function} &
  $\varpropto\lambda^{-2\eps}/\eps^2$ & $\varpropto\lambda^{-1-\eps}/\eps^2$\\\hline
  $\langle P_{g\to g}^{(1)}\rangle$ & Eq.~\eqref{eq:one-loop_g2gg_splitting_function} &
  \multicolumn{2}{c}{$\varpropto\lambda^{-2-2\eps}/\eps^2$}\\
  $\langle P_{g\to g}^{(1,p)}\rangle$ & Eq.~\eqref{eq:one-loop_g2gg_remainder_function} &
  \multicolumn{2}{c}{$\varpropto\lambda^{-2\eps}/\eps^2$}\\\hline
  $\langle P_{g\to q}^{(1)}\rangle$ & Eq.~\eqref{eq:one-loop_g2q_splitting_function} &
  \multicolumn{2}{c}{$\varpropto\lambda^{-1-\eps}/\eps^2$}\\\hline
  \end{tabular}
  \caption{Limits of the various one-loop splitting functions and their pure components.
    See the main text for details.
  \label{tab:limits_one-loop}}
\end{table}
Here we summarize the singularity structure of the various one-loop splitting functions
introduced in Secs.~\ref{sec:one-loop_splittings} and~\ref{sec:one-loop_splitting_decomposition}.
The kinematical limit where particle $i$ becomes soft is parametrized as $p_i\to \lambda p_i$. 
We can use the relation
\begin{equation}
    _2F_1(1,1;1-\eps;z)=(1-z)^{-1-\eps}\, _2F_1(-\eps,-\eps;1-\eps;z)
    =(1-z)^{-1-\eps}\Big(1+\eps^2{\rm Li}_2(z)+\mathcal{O}(\eps^3)\Big)
\end{equation}
to convert the hypergeometric functions in $f_1$ to powers of $z$ and $1-z$, times a remainder
that is finite for all values of $z$ and does not contribute sub-leading poles. This allows
the dominant scaling behavior in the soft limit to be extracted.
The results are given in Tab.~\ref{tab:limits_one-loop}. 
Here we identify only the leading singular structure in terms of combined explicit and implicit poles.

\section{Outlook}
\label{sec:outlook}
In this work we re-examined the tree-level and one-loop splitting functions used to capture
the infrared singular behavior of QCD scattering amplitudes in next-to-to-next-to-leading order
QCD calculations and next-to-next-to-leading logarithmic resummation. We made use of the fact
that variants of scalar QCD are minimal extensions of the semi-classical approximation, and avoided
any kinematical approximations. We demonstrated that all QCD splitting functions can be obtained from
dipole radiator functions computed in scalar QCD, combined with spin-dependent remainders that exist
only in splitting configurations and have (sub-)sub-leading singularities in the kinematical limits.
Using the background field method to construct a dipole approximation to scalar multipole
radiators, we have achieved the same decomposition at one loop order.

The significance of this result is twofold. On the one hand, we are able to show that neither
the soft nor the collinear kinematical limit is needed in order to derive universally applicable
subtraction terms for higher-order calculations. This is ultimately due to the fact that a semi-classical
limit of the theory must exist. On the other hand, we anticipate that the results presented here
will be used to devise a subtraction or resummation scheme which does not suffer from overlaps
between soft and collinear sectors, as well as overlaps between single- and double-unresolved limits.
Such a scheme is urgently needed for example to construct a fully differential parton shower
at next-to-next-to leading logarithmic precision. For fixed-order calculations, the reorganization
and classification of infrared singularities should aid the construction of a generic subtraction
scheme at NNLO precision that will have an improved convergence compared to existing methods based on
leading power approximations in the soft regions.

\section*{Acknowledgments}
We thank Robert Szafron for discussions on sub-leading power contributions
to the triple-collinear splitting functions and Bogdan Dobrescu for discussions
on the Gordon identity.
This manuscript has been authored by Fermi Forward Discovery Group, LLC
under Contract No. 89243024CSC000002 with the U.S.\ Department of Energy,
Office of Science, Office of High Energy Physics.
The work of J.M.C., S.H. and M.K. was supported by the U.S. Department of Energy,
Office of Science, Office of Advanced Scientific Computing Research,
Scientific Discovery through Advanced Computing (SciDAC-5) program,
grant “NeuCol”. D.R.\ is supported by the European Union under
the HORIZON program in Marie Sk{\l}odowska-Curie project No. 101153541.
D.R.\ was further supported by the STFC under grant agreement ST/P006744/1
and by Durham Universities Physics Department Developing Talents Award
during the initial stages of this work. 

\appendix

\section{The polarization tensor in axial gauge}
\label{sec:polarization_tensor}
In this appendix we briefly recall the relation between the gluon polarization tensor
in a light-like axial gauge, and an explicit sum over physical (i.e.\ transverse)
gluon polarizations in the helicity formalism~\cite{Dixon:1996wi}. While this
connection is never explicitly used in the main text, we find it useful
for guiding the decomposition of splitting functions with intermediate
off-shell gluons, in particular Eq.~\eqref{eq:shifted_scalar_radiator}.

For on-shell gluons, the light-like axial gauge allows a simple interpretation
of Eq.~\eqref{eq:axial_gauge} in terms of polarizations of massless particles
in the helicity formalism~\cite{Dixon:1996wi}\footnote{At higher orders
  in the perturbative expansion, this identity requires the four-dimensional
  helicity scheme~\cite{Bern:1991aq,Bern:2002zk}}:
\begin{equation}\label{eq:axial_gauge_onshell}
  d^{\mu\nu}(p,\bar{n})=\frac{1}{4p\bar{n}}\,\sum_{\lambda=\pm}
  {\rm Tr}\Big[\slash\!\!\!p\gamma^\mu\slash\!\!\!\bar{n}\gamma^\nu P_\lambda\Big]
  =\sum_{\lambda=\pm}
  \epsilon_\lambda^\mu(p,\bar{n})\epsilon_\lambda^{\nu\,*}(p,\bar{n})\;,
  \quad\text{where}\quad
  \epsilon_\pm^\mu(p,\bar{n})=\pm\frac{\langle \bar{n}^\mp|\gamma^\mu|p^\mp\rangle}{
  \sqrt{2}\langle \bar{n}^\mp|p^\pm\rangle}\;.
\end{equation}
Here $P_{\pm}=(1\pm\gamma^5)/2$ are the left- and right-chiral projectors.
Equation~\eqref{eq:axial_gauge_onshell} allows to relate our technique to
the helicity-formalism based methods of Refs.~\cite{Duhr:2008wc,Cohen:2024xuf}.
For off-shell vector bosons, one can use the shifted momentum defined
in Eq.~\eqref{eq:sudakov_decomposition_cg}, which leads to the following
generalization of Eq.~\eqref{eq:axial_gauge_onshell}:
\begin{equation}\label{eq:tc_factorization}
  d^{\mu\nu}(p,\bar{n})=d^{\mu\nu}(\bar{p},\bar{n})
  +p^2\,\frac{\bar{n}^\mu\bar{n}^\nu}{(p\bar{n})^2}
  =-d^{\mu\rho}(p,\bar{n})d_\rho^{\;\;\nu}(p,\bar{n})
  +p^2\,\frac{\bar{n}^\mu\bar{n}^\nu}{(p\bar{n})^2}\;.
\end{equation}
We have implicitly used this relation in Eq.~\eqref{eq:collinear_splittings_recursive}
to define the collinear splitting functions. In the $m$-particle collinear limit,
$\tilde{k}_i\to \lambda\tilde{k}_i$, $\lambda\to 0$, the second term on the 
right-hand side of Eq.~\eqref{eq:tc_factorization} scales as $\lambda^2$.
It can therefore be neglected in the computation of the leading-power
gluon splitting functions. The factorized form of the first term allows us
to absorb one of the polarization tensors into the splitting amplitude,
while the other is associated with the gluon current produced by the
hard matrix element.
Any $\bar{n}^\mu$-dependent contribution to the gluon splitting function
in Eq.~\eqref{eq:collinear_splittings_recursive} will then vanish upon
multiplication by this current and can therefore be dropped. We use
this particular feature of the collinear limit
in Sec.~\ref{sec:decomposition_gluon}.

\section{Tree-level recursion}
\label{sec:recursion}
In this appendix, we show how the Berends-Giele recursion can be reformulated
in terms of expressions for a scalar theory, and a spin-dependent remainder.
Note that this does not directly translate into a separation of scalar and
remainder components at the amplitude squared level.

\subsection{The quark current at tree level}
\label{sec:quark_current_tree_level}
We begin the general discussion by deriving the structure of the 
quark-gluon vertex in terms of scalar and magnetic interactions.
A similar approach was presented in~\cite{Bern:1993tz}. Here we focus
in particular on a suitable formulation in terms of building blocks
for higher-order scattering matrix elements.

We use the color-dressed~\cite{Duhr:2006iq} Berends-Giele 
recursion~\cite{Berends:1987me,Berends:1988yn,Berends:1990ax}
to describe an off-shell (anti-)quark of momentum $p_\alpha$ by a current
which is obtained from the propagator times vertex for the
interaction of a fermion of momentum $p_\beta$
with a gluon of momentum $p_\gamma$. We sum over all partitions
$\{\beta,\gamma\}$ of the set of indices, $\alpha$, into two
disjoint subsets.\footnote{For details on the notation,
see Ref.~\cite{Duhr:2006iq}.} The set of all possible partitions
is denoted by $P(\alpha,2)$.
\begin{equation}\label{eq:sc_decomposition_fermion_1}
  \begin{split}
    \Psi_i(p_\alpha,\pm m)=&\;
    i\,\frac{\slash\!\!\!p_\alpha\pm m}{p_\alpha^2-m^2}\,
    \sum_{\substack{\{\beta,\gamma\}\in\\ P(\alpha,2)}}(-ig_sT^a_{ij}\gamma^\mu)\;
    J_\mu^a(p_\gamma)\Psi_j(p_\beta,\pm m)\\
    =&\;\frac{g_sT^a_{ij}}{p_\alpha^2-m^2}
    \sum_{\substack{\{\beta,\gamma\}\in\\ P(\alpha,2)}}
    \bigg[\,(p_\alpha+p_\beta)^\mu J_\mu^a(p_\gamma)
    +i\sigma^{\mu\nu}J_\mu^a(p_\gamma)p_{\gamma,\nu}
    -\gamma^\mu J_\mu^a(p_\gamma)(\slash\!\!\!p_\beta\mp m)\,
    \bigg]\Psi_j(p_\beta,\pm m)\;.
  \end{split}
\end{equation}
Here, $\Psi$ is the (anti-)quark current, and $J_\mu$ is
a gluon current. These currents can be either on-shell or off-shell.
The first term in the square bracket on the right-hand side describes
the interaction of the gluon field with the scalar current while
the second term represents the magnetic interaction.
The third term plays a special role.
If $\Psi_j(p,\pm m)$ is an external wave function, we can use the equation
of motion to show that it vanishes, i.e.\ $(\slash\!\!\!p\mp m)\Psi_j(p,\pm m)=0$.
If $\Psi_j(p,\pm m)$ is itself an off-shell current, it is obtained
as a sum of terms of the form of Eq.~\eqref{eq:sc_decomposition_fermion_1}.
We can simplify the combined expression as follows
\begin{equation}\label{eq:prop_to_seagull}
  \begin{split}
    -\sum_{\substack{\{\beta,\gamma\}\in\\ P(\alpha,2)}}\,g_sT^a_{ij}
    \gamma^\mu J_\mu^a(p_\gamma)\,(\slash\!\!\!p_\beta\mp m)\Psi_j(p_\beta,\pm m)
    =&\;-g_s^2T^a_{ik}T^b_{kj}
    \sum_{\substack{\{\beta,\gamma\}\in\\ P(\alpha,2)}}
    \sum_{\substack{\{\delta,\epsilon\}\in\\ P(\gamma,2)}}\gamma^\mu\gamma^\nu\,
    J_\mu^a(p_\delta)J_\nu^b(p_\epsilon)\Psi_j(p_\beta,\pm m)\;.
  \end{split}
\end{equation}
The factor $\gamma^\mu\gamma^\nu$ on the right-hand side of this equation
can be decomposed into two types of four-point vertices, by using the relation
$\gamma^\mu\gamma^\nu=g^{\mu\nu}-i\sigma^{\mu\nu}$. The first term gives
the seagull vertex of the scalar theory for which
Eq.~\eqref{eq:scalar_dicurrent} describes the single-emission current.
It is needed in order for this theory to satisfy the Ward identities.
The second term vanishes in the complete sum over partitions due to its antisymmetry.
Using this methodology, we can separate the recursive formula into a scalar piece
and a spin-dependent remainder.
\begin{equation}\label{eq:quark_current}
  \begin{split}
    \Psi_i(p_\alpha)=&\;
    \frac{ig^{\mu\nu}}{p_\alpha^2-m^2}
    \sum_{\substack{\{\beta,\gamma\}\in\\P(\alpha,2)}}
    \bigg[-ig_sT^a_{ij}(p_\beta+p_\alpha)_\nu J_\mu^a(p_\gamma)
    +\sum_{\substack{\{\delta,\epsilon\}\in\\ OP(\gamma,2)}}ig_s^2\,\big\{T^a,T^b\big\}_{ij}\,
    J_\mu^a(p_\delta)J_\nu^b(p_\epsilon)\bigg]\Psi_j(p_\beta)\\
    &\;+\frac{\,\sigma^{\mu\nu}}{p_\alpha^2-m^2}
    \sum_{\substack{\{\beta,\gamma\}\in\\P(\alpha,2)}}
    (-ig_sT^a_{ij})\,(p_\beta-p_\alpha)_\nu J_\mu^a(p_\gamma)\Psi_j(p_\beta)\;.
  \end{split}
\end{equation}
We have made use of the Bose symmetry in the gluon fields to rewrite the sum over
unordered gluon currents in the square brackets into a sum over ordered partitions
of the set $\gamma$ into two disjoint subsets. The complete set of such partitions
is denoted by $OP(\gamma,2)$. In an abelian theory, this rearrangement would
induce a simple symmetry factor in the seagull vertex, consistent with
the Feynman rules of scalar QED. Note that there is no seagull vertex for the
magnetic interaction, which is a consequence of the antisymmetry of $\sigma^{\mu\nu}$.

\subsection{The gluon current at tree level}
Next we analyze the gluon current in the color-dressed Berends-Giele approach.
It can be written as
\begin{equation}\label{eq:gluon_current_step1}
  J_\mu^a(p_\alpha)=\bar{J}_\mu^a(p_\alpha)+\tilde{J}_\mu^a(p_\alpha)\;.
\end{equation}
where the gluon-induced contribution, $\tilde{J}_\mu$, arises from triple
and quartic gluon interactions and is given by
\begin{equation}\label{eq:gluon_gluon_current_step1}
  \begin{split}
    \tilde{J}_\mu^a(p_\alpha)=&\;
    i\,\frac{d_{\mu\nu}(p_\alpha)}{p_\alpha^2}\,
    \sum_{\substack{\{\beta,\gamma\}\in\\ OP(\alpha,2)}}
    (-g_sf^{abc})\bigg[\,g^{\rho\sigma}(p_\beta-p_\gamma)^\nu
    +g^{\sigma\nu}(2p_\gamma+p_\beta)^\rho-g^{\nu\rho}(2p_\beta+p_\gamma)^\sigma\bigg]\,
    J_\rho^b(p_\beta)J_\sigma^c(p_\gamma)\\
    &\;+i\,\frac{d_{\mu\nu}(p_\alpha)}{p_\alpha^2}\,
    \sum_{\substack{\{\beta,\gamma\}\in\\ OP(\alpha,2)}}
    \sum_{\substack{\{\delta,\epsilon\}\in\\ OP(\gamma,2)}}
    ig_s^2\bigg[\quad\; f^{abe}f^{ecd}\big(g^{\nu\tau}g^{\rho\sigma}-g^{\nu\sigma}g^{\tau\rho}\big)\\[-5mm]
    &\hspace*{5cm}+f^{ace}f^{edb}\big(g^{\nu\rho}g^{\sigma\tau}-g^{\nu\tau}g^{\rho\sigma}\big)\\
    &\hspace*{5cm}+f^{ade}f^{ebc}\big(g^{\nu\sigma}g^{\tau\rho}-g^{\nu\rho}g^{\sigma\tau}\big)\,\bigg]\,
    J_\rho^b(p_\beta)J_\sigma^c(p_\delta)J_\tau^d(p_\epsilon)\;.
  \end{split}
\end{equation}
By summing over all partitions and relabeling gluon momenta, we can reduce this expression to the simple form
\begin{equation}\label{eq:gluon_gluon_current}
  \begin{split}
    \tilde{J}_\mu^a(p_\alpha)=&\;
    i\,\frac{d_{\mu\nu}(p_\alpha)}{p_\alpha^2}\,
    \sum_{\substack{\{\beta,\gamma\}\in\\ P(\alpha,2)}}
    (-g_sf^{acb})\bigg[\,g^{\nu\rho}(2p_\beta+p_\gamma)^\sigma
    -\frac{1}{2}g^{\rho\sigma}(p_\beta-p_\gamma)^\nu\bigg]\,
    J_\rho^b(p_\beta)J_\sigma^c(p_\gamma)\\
    &\;+i\,\frac{d_{\mu\nu}(p_\alpha)}{p_\alpha^2}\,
    \sum_{\substack{\{\beta,\gamma\}\in\\ P(\alpha,2)}}
    \sum_{\substack{\{\delta,\epsilon\}\in\\ P(\gamma,2)}}
    ig_s^2f^{ace}f^{edb}g^{\nu\rho}g^{\sigma\tau}\,
    J_\rho^b(p_\beta)J_\sigma^c(p_\delta)J_\tau^d(p_\epsilon)\\
    =&\;\frac{ig^{\sigma\tau}}{p_\alpha^2}\,
    \sum_{\substack{\{\beta,\gamma\}\in\\ P(\alpha,2)}}
    \bigg[-ig_sF^c_{ab}(p_\beta+p_\alpha)_\tau J_\sigma^c(p_\gamma)
    +\sum_{\substack{\{\delta,\epsilon\}\in\\ OP(\gamma,2)}}
    ig_s^2\big\{F^c,F^d\big\}_{ab}\,J_\sigma^c(p_\delta)J_\tau^d(p_\epsilon)\bigg]\,
    (-d_{\mu}^{\;\;\rho}(p_\alpha))J_\rho^b(p_\beta)\\
    &\;-i\,\frac{d_{\mu\nu}(p_\alpha)}{p_\alpha^2}\,
    \sum_{\substack{\{\beta,\gamma\}\in\\ OP(\alpha,2)}}
    g_sf^{abc}(p_\beta-p_\gamma)^\nu\,g^{\rho\sigma}
    J_\rho^b(p_\beta)J_\sigma^c(p_\gamma)\;.
  \end{split}
\end{equation}
We have defined $F^c_{ab}=if^{acb}$ to make the radiation pattern explicit.
The quark-induced contribution to the gluon current, $\bar{J}_\mu$, is given by
\begin{equation}\label{eq:gluon_quark_current}
  \begin{split}
    \bar{J}_\mu^a(p_\alpha)=&\;
    i\,\frac{d_{\mu\nu}(p_\alpha)}{p_\alpha^2}\,
    \sum_{\substack{\{\beta,\gamma\}\in\\ P(\alpha,2)}}
    \bar{\Psi}_i(p_\gamma,\pm m)(-ig_sT^a_{ij}\gamma^\nu)\Psi_j(p_\beta,\pm m)\;.
  \end{split}
\end{equation}
We use the identity $\{\slash\!\!\!p,\slash\!\!\!n\}/(2pn)=1$,
with $n$ an auxiliary vector, to rewrite this in the following form
\begin{equation}\label{eq:gluon_quark_current_2}
  \begin{split}
    \bar{J}_\mu^a(p_\alpha)=&\;
    i\,\frac{d_{\mu\nu}(p_\alpha)}{p_\alpha^2}\,
    \sum_{\substack{\{\beta,\gamma\}\in\\ P(\alpha,2)}}
    \bar{\Psi}_i(p_\gamma,\pm m)(-ig_sT^a_{ij})
    \frac{1}{2}\bigg[\frac{\gamma^\nu\slash\!\!\!p_\alpha\slash\!\!\!n
     +\gamma^\nu\slash\!\!\!n\slash\!\!\!p_\alpha}{2p_\alpha n}+
    \frac{\slash\!\!\!p_\alpha\slash\!\!\!n\gamma^\nu
    +\slash\!\!\!n\slash\!\!\!p_\alpha\gamma^\nu}{2p_\alpha n}\bigg]
    \Psi_j(p_\beta,\pm m)\;.
  \end{split}
\end{equation}
Using $\gamma^\mu\slash\!\!\!n+\slash\!\!\!n\gamma^\mu=2n^\mu$,
and working in an axial gauge (see the discussion in Sec.~\ref{sec:spin_decomposition}),
this expression simplifies to a purely magnetic interaction between the pseudo-particle
described by the gauge vector $n$ and the actual fermion described by $p_\beta$.
The complete gluon current then reads
\begin{equation}\label{eq:gluon_current}
  \begin{split}
    &J_\mu^a(p_\alpha,n)=\;
    \frac{ig^{\sigma\tau}}{p_\alpha^2}\,
    \sum_{\substack{\{\beta,\gamma\}\in\\ P(\alpha,2)}}
    \bigg[-ig_sF^c_{ab}(p_\beta+p_\alpha)_\tau J_\sigma^c(p_\gamma,n)\\
    &\;\qquad\qquad\qquad\qquad\qquad\quad
    +\sum_{\substack{\{\delta,\epsilon\}\in\\ OP(\gamma,2)}}
    ig_s^2\{F^c,F^d\}_{ab}\,J_\sigma^c(p_\delta,n)
    J_\tau^d(p_\epsilon,n)\bigg]\,
    (-d_{\mu}^{\;\;\rho}(p_\alpha,n))J_\rho^b(p_\beta,n)\\
    &\;\quad+i\,\frac{d_{\mu\nu}(p_\alpha,n)}{p_\alpha^2}\,
    \sum_{\substack{\{\beta,\gamma\}\in\\ P(\alpha,2)}}
    \bigg[\,-ig_s\frac{F^a_{bc}}{2}\,(p_\beta-p_\gamma)^\nu g^{\rho\sigma}
    J_\rho^b(p_\beta,n)J_\sigma^c(p_\gamma,n)\\
    &\;\qquad\qquad\qquad\qquad\qquad\quad
    -ig_sT^a_{ij}\,(p_\beta+p_\gamma)_\rho\bar{\Psi}_i(p_\gamma)
    \frac{\slash\!\!\!n i\sigma^{\rho\nu}
    +i\sigma^{\nu\rho}\slash\!\!\!n}{2(p_\beta+p_\gamma)n}
    \Psi_j(p_\beta)\bigg]\;.
  \end{split}
\end{equation}
The scalar parts of Eq.~\eqref{eq:quark_current} and~\eqref{eq:gluon_current}
are similar to the eikonal interaction Hamiltonians obtained,
for example in~\cite{Kulish:1970ut,Catani:1985xt}, but they 
include effects of kinematical recoil, which is essential
when computing higher-point splitting functions.

\section{Asymmetric component of the gluon splitting tensor}
\label{sec:gluon_splitting_asymmetric}
In this appendix, we derive the interference component of the gluon splitting
tensor in Eq.~\eqref{eq:coll_ggg}. It is better understood by contracting
$P_{g\to g,\rm(i)}^{\mu\nu,\gamma\delta}(p_1,p_2)$ with the
polarization tensors for the decay of gluon $2$. The definitions
in Eqs.~\eqref{eq:tree_level_building_blocks} lead to
\begin{equation}
  \begin{split}
    S_\nu(p_1,p_2)d^{\mu\nu}(p_2,\bar{n})=
    \frac{2}{p_{12}^2}\,\frac{\tilde{p}_{1,2}^\mu}{z_2/z_{12}}+\ldots\;,
    \quad D^\mu(p_1,p_2,\bar{n})=\frac{2}{p_{12}^2}\,\tilde{p}_{1,2}^\mu+\ldots\;,
  \end{split}
\end{equation}
where the ellipses represent terms proportional to $\bar{n}^\mu$ that vanish
upon multiplication with $d^\rho_{\;\mu}(p,\bar{n})$ for any momentum, $p$.
We can use these identities to simplify the interference contribution as follows
\begin{equation}\label{eq:coll_ggg_asym}
  \begin{split}
  &P_{g\to g,\rm(i)}^{\mu\nu,\gamma\delta}(p_1,p_2)\,
  d^\alpha_{\;\gamma}(p_2,\bar{n})d^\beta_{\;\delta}(p_2,\bar{n})
  =\frac{4C_A}{z_1z_2}\,\bigg(z_{12}\tilde{p}_{1,2}^\alpha
  -\frac{p_2^2\bar{n}^\alpha}{2p_{12}\bar{n}}\bigg)
  z_{12}\tilde{p}_{1,2}^\rho\,d^{\nu\sigma}(p_{12},\bar{n})\,
  d^\mu_{\;\rho}(p_{12},\bar{n})\,d^{\beta}_{\;\sigma}(p_2,\bar{n})\\
  &\quad-\frac{4C_A}{z_1z_2}
  \bigg(z_{12}\tilde{p}_{1,2}^\nu-\frac{p_2^2\bar{n}^\nu}{2p_{12}\bar{n}}\bigg)
  \bigg[z_2z_{12}\tilde{p}_{1,2}^\sigma d^{\rho\alpha}(p_2,\bar{n})
  +z_1\bigg(z_{12}\tilde{p}_{1,2}^\alpha
  -\frac{p_2^2\bar{n}^\alpha}{2p_{12}\bar{n}}\bigg)
  d^{\rho\sigma}(p_1,\bar{n})\bigg]
  d^{\mu}_{\;\rho}(p_{12},\bar{n})d^{\beta}_{\;\sigma}(p_2,\bar{n})\;.
  \end{split}
\end{equation}
In the strongly ordered soft and collinear limits, $\tilde{p}_{i,j}^\nu
d^\mu_{\;\nu}(p,\bar{n})$ reduces to $\tilde{p}_{i,j}^\mu$, and we find
that the sum of asymmetric terms in Eq.~\eqref{eq:coll_ggg} vanishes 
when combined with a tensor that is symmetric in $\mu$ and $\nu$.
The complete gluon splitting tensor is then effectively given by
the sum of squared diagrams that led to Eq.~\eqref{eq:coll_ggg_sym}.
This implies that the first interference contributions arise at
$\mathcal{O}(\alpha_s^3)$, i.e.\ beyond the accuracy considered here.

\section{One-loop integrals}
\label{sec:one-loop_integrals}
In this section we collect the one-loop integrals required to perform the
calculations in the main text.  Note that we have extracted
an overall factor in Eq.~\eqref{eq:def_cgamma} that differs from some of the previous literature by
a factor of $1/(4\pi)^2$, leading to the appearance of the same factor in the definitions below.

\subsection{Scalar basis integrals}
The standard scalar bubble and triangle integrals used
in Sec.~\ref{sec:loop_mes} are given by~\cite{Bern:1993kr}
\begin{align}\label{eq:one-loop_integrals_scalar}
  \begin{split}
    \hat{I}_2(s_{12})=&\;16\pi^2 \mu^{2\eps}
    \int\frac{d^{4-2\eps}k}{(2\pi)^{4-2\eps}}\, \frac{1}{k^2(k-p_1-p_2)^2}
    =\frac{ic_\Gamma}{\eps(1-2\eps)}\,
    \bigg(\!\!-\frac{\mu^2}{s_{12}}\!\bigg)^\eps
    = -\frac{i\eps}{1-2\eps} \bigg(\!\!-\frac{\mu^2}{s_{12}}\!\bigg)^\eps f_2\;,
  \end{split}\\
  \begin{split}
    \hat{I}_3^{1m}(s_{12})=&\;16\pi^2 \mu^{2\eps}
    \int\frac{d^{4-2\eps}k}{(2\pi)^{4-2\eps}}\, \frac{1}{k^2(k-p_1)^2(k-p_1-p_2)^2}
    =\frac{ic_\Gamma}{\eps^2}\,
    \frac{1}{s_{12}}\bigg(\!\!-\frac{\mu^2}{s_{12}}\!\bigg)^\eps
    = -\frac{i}{s_{12}} \left(\!-\frac{\mu^2}{s_{12}}\!\right)^\eps f_2\;.
  \end{split}
\end{align}
The corresponding integrals needed for the computation in light-like
axial gauge are given by~\cite{Kosower:1999rx}
\begin{align}\label{eq:one-loop_light-cone_integrals_scalar}
  \begin{split}
    \hat{J}_2(s_{12}) =&\; 16\pi^2 \mu^{2\eps} \int\frac{d^{4-2\eps}k}{(2\pi)^{4-2\eps}}\, \frac{1}{k^2(k-p_1-p_2)^2(kn)} 
    = \frac{i}{n(p_1+p_2)}\left(\!-\frac{\mu^2}{s_{12}}\!\right)^\eps f_2 \,,
  \end{split}\\
  \begin{split}
    \hat{J}_1(s_{12},z) =&\; 16\pi^2 \mu^{2\eps} \int\frac{d^{4-2\eps}k}{(2\pi)^{4-2\eps}}\, \frac{1}{k^2(k-p_1)^2(k-p_1-p_2)^2(kn)} 
    = -\frac{i}{n(p_1+p_2) s_{12}}\left(\!-\frac{\mu^2}{s_{12}}\!\right)^\eps f_1(z)\;.
  \end{split}
\end{align}
In addition, we need the two-mass triangle and one-mass box integrals~\cite{Bern:1993kr,Haug:2022hkr}
\begin{align}
  \begin{split}
    \hat{I}_3^{2m}(s,t)=&\;\frac{ic_\Gamma}{\eps^2}\,
    \frac{1}{s-t}\bigg[\bigg(\!\!-\frac{\mu^2}{s}\!\bigg)^\eps
    -\bigg(\!\!-\frac{\mu ^2}{t}\!\bigg)^\eps\,\bigg]\;,
  \end{split}\\
  \begin{split}
    \hat{I}_4^{1m}(s,t,u)=&\;\frac{2ic_\Gamma}{\eps^2}\frac{1}{s\,t}\bigg[
    \bigg(\!\!-\frac{s+u}{s\, t/\mu^2}\!\bigg)^\eps
    \,_2F_1\bigg(\!-\eps ,-\eps ;1-\eps ;1-\frac{s}{s+u}\bigg)\\
    &\;\qquad\qquad+\bigg(\!\!-\frac{t+u}{s\, t/\mu^2}\!\bigg)^\eps
    \,_2F_1\bigg(\!-\eps ,-\eps ;1-\eps;1-\frac{t}{t+u}\bigg) \\
    &\;\qquad\qquad-\bigg(\!\!-\frac{(s+u) (t+u)}{s\, t\, (s+t+u)/\mu^2}\!\bigg)^\eps
    \,_2F_1\bigg(\!-\eps ,-\eps ;1-\eps;1-\frac{s\, t}{(s+u) (t+u)}\bigg)\bigg]\;.    
  \end{split}
\end{align}

\subsection{Tensor integrals}
The tensor one-loop integrals needed for the computation of the
splitting functions in Sec.~\ref{sec:one-loop_splittings}
are given by~\cite{Kosower:1999rx}
\begin{align}
    \begin{split}
    \hat{I}_{2a}^\mu(s_{12}) &= 16\pi^2 \mu^{2\eps}\int\frac{d^{4-2\eps}k}{(2\pi)^{4-2\eps}}\, \frac{k^\mu}{k^2(k-p_1-p_2)^2}
    = -\frac{i\eps}{2(1-2\eps)} \left(\!-\frac{\mu^2}{s_{12}}\!\right)^\eps f_2\, (p_1+p_2)^\mu \,,
    \end{split}\\
    \begin{split}
    \hat{I}_{2b}^{\mu\nu}(s_{12}) &= 16\pi^2 \mu^{2\eps}\int\frac{d^{4-2\eps}k}{(2\pi)^{4-2\eps}}\, \frac{k^\mu k^\nu}{k^2(k-p_1-p_2)^2} \\
    &= \frac{i\eps}{2(3-2\eps)(1-2\eps)}\left(\!-\frac{\mu^2}{s_{12}}\!\right)^\eps f_2\, \left(\frac{s_{12}}{2}g^{\mu\nu} - (2-\eps)(p_1+p_2)^\mu(p_1+p_2)^\nu\right) \,.
    \end{split}
\end{align}
and
\begin{align}
    \begin{split}
    \hat{I}_{3a}^\mu(s_{12}) &= 16\pi^2 \mu^{2\eps}\int\frac{d^{4-2\eps}k}{(2\pi)^{4-2\eps}}\, \frac{k^\mu}{k^2(k-p_1)^2(k-p_1-p_2)^2} 
    = -\frac{i}{s_{12}}\left(\!-\frac{\mu^2}{s_{12}}\!\right)^\eps f_2\,\Bigg[\frac{1-\eps}{1-2\eps}\, p_1^\mu - \frac{\eps}{1-2\eps}\, p_2^\mu\Bigg] \,,
    \end{split}\\
    \begin{split}
    \hat{I}_{3b}^{\mu\nu}(s_{12}) &= 16\pi^2 \mu^{2\eps}\int\frac{d^{4-2\eps}k}{(2\pi)^{4-2\eps}}\, \frac{k^\mu k^\nu}{k^2(k-p_1)^2(k-p_1-p_2)^2} \\
    &= -\frac{i}{s_{12}}\left(\!-\frac{\mu^2}{s_{12}}\!\right)^\eps f_2\, \Bigg[s_{12}\frac{\eps}{4(1-\eps)(1-2\eps)}\, g^{\mu\nu} + \frac{2-\eps}{2(1-2\eps)}\, p_1^\mu p_1^\nu \\
    &\qquad\qquad\qquad\qquad\qquad
    - \frac{\eps}{2(1-2\eps)}\, p_2^\mu p_2^\nu - \frac{\eps(2-\eps)}{2(1-\eps)(1-2\eps)}\, \left(p_1^\mu p_2^\nu + p_2^\mu p_1^\nu\right)\Bigg] \,,
    \end{split}\\
    \begin{split}
    \hat{I}_{3c}^{\kappa\mu\nu}(s_{12}) &= 16\pi^2 \mu^{2\eps}\int\frac{d^{4-2\eps}k}{(2\pi)^{4-2\eps}}\, \frac{k^\kappa k^\mu k^\nu}{k^2(k-p_1)^2(k-p_1-p_2)^2} \\
    &= -\frac{i}{s_{12}}\left(\!-\frac{\mu^2}{s_{12}}\!\right)^\eps f_2\, \Bigg[\frac{(2-\eps)(3-\eps)}{2(1-2\eps)(3-2\eps)}p_1^\kappa p_1^\mu p_1^\nu - \frac{(2-\eps)\eps}{2(1-2\eps)(3-2\eps)}p_2^\kappa p_2^\mu p_2^\nu \\
    &\qquad\qquad\qquad\qquad\qquad
    - \frac{(3-\eps)\eps}{2(1-2\eps)(3-2\eps)}\left(p_1^\kappa p_2^\mu p_2^\nu + p_1^\mu p_2^\nu p_2^\kappa + p_1^\nu p_2^\kappa p_2^\mu\right) \\
    &\qquad\qquad\qquad\qquad\qquad
    - \frac{(2-\eps)(3-\eps)\eps}{2(1-\eps)(1-2\eps)(3-2\eps)}\left(p_2^\kappa p_1^\mu p_1^\nu + p_2^\mu p_1^\nu p_1^\kappa + p_2^\nu p_1^\kappa p_1^\mu\right) \\
    &\qquad\qquad\qquad\qquad\qquad
    + s_{12}\frac{(2-\eps)\eps}{4(1-\eps)(1-2\eps)(3-2\eps)}(p_1^\kappa g^{\mu\nu} + p_1^\mu g^{\nu\kappa} + p_1^\nu g^{\kappa\mu}) \\
    &\qquad\qquad\qquad\qquad\qquad
    + s_{12}\frac{\eps}{4(1-2\eps)(3-2\eps)}(p_2^\kappa g^{\mu\nu} + p_2^\mu g^{\nu\kappa} + p_2^\nu g^{\kappa\mu})
    \Bigg] \,.
    \end{split}
\end{align}
The required light-cone bubble integrals read
\begin{align}
    \begin{split}
    \hat{J}_4^\mu(s_{12}) &= 16\pi^2 \mu^{2\eps}\int\frac{d^{4-2\eps}k}{(2\pi)^{4-2\eps}}\, \frac{k^\mu}{k^2(k-p_1-p_2)^2(kn)} \\
    &= \frac{i}{n(p_1+p_2)}\left(\!-\frac{\mu^2}{s_{12}}\!\right)^\eps f_2\, \Bigg[-\frac{\eps}{1-2\eps}\, (p_1+p_2)^\mu + \frac{s_{12}}{2n(p_1+p_2)}\frac{1}{1-2\eps}\, n^\mu\Bigg] \,,
    \end{split} \\
    \begin{split}
    \hat{J}_6^{\mu\nu}(s_{12}) &= 16\pi^2 \mu^{2\eps}\int\frac{d^{4-2\eps}k}{(2\pi)^{4-2\eps}}\, \frac{k^\mu k^\nu}{k^2(k-p_1-p_2)^2(kn)} \\
    &= \frac{i}{n(p_1+p_2)}\left(\!-\frac{\mu^2}{s_{12}}\!\right)^\eps f_2\,\Bigg[\;s_{12}\frac{\eps}{4(1-\eps)(1-2\eps)}g^{\mu\nu}  -\frac{\eps}{2(1-2\eps)}(p_1+p_2)^\mu(p_1+p_2)^\nu \\
    &\qquad\qquad\qquad\qquad\qquad\qquad\; + \left(\frac{s_{12}}{n(p_1+p_2)}\right)^2\frac{1}{4(1-\eps)(1-2\eps)} n^\mu n^\nu \\
    &\qquad\qquad\qquad\qquad\qquad\qquad\;
    - \frac{s_{12}}{n(p_1+p_2)}\frac{\eps}{4(1-\eps)(1-2\eps)}\big((p_1+p_2)^\mu n^\nu+(p_1+p_2)^\nu n^\mu\big) \Bigg]\,.
    \end{split}
\end{align}
The required light-cone triangle integrals are (note that there is a typo in Eq.~(3.15) of \cite{Kosower:1999rx})
\begin{align}
    \begin{split}
    \hat{J}_3^\mu(s_{12},z) &= 16\pi^2 \mu^{2\eps}\int\frac{d^{4-2\eps}k}{(2\pi)^{4-2\eps}}\, \frac{k^\mu}{k^2(k-p_1)^2(k-p_1-p_2)^2(kn)} \\
    &= -\frac{i}{2n(p_1+p_2) s_{12}}\left(\!-\frac{\mu^2}{s_{12}}\!\right)^\eps\, \Bigg[ f_1(z)\, p_1^\mu + \frac{2f_2 - zf_1(z)}{1-z}\, p_2^\mu - \frac{s_{12}}{2n(p_1+p_2)}\frac{2f_2 - f_1(z)}{1-z}\, n^\mu \Bigg]
    \end{split}\\
    \begin{split}
    \hat{J}_5^{\mu\nu}(s_{12},z) &= 16\pi^2 \mu^{2\eps}\int\frac{d^{4-2\eps}k}{(2\pi)^{4-2\eps}}\, \frac{k^\mu k^\nu}{k^2(k-p_1)^2(k-p_1-p_2)^2(kn)} \\
    &= -\frac{i}{n(p_1+p_2) s_{12}}\left(\!-\frac{\mu^2}{s_{12}}\!\right)^\eps\, \Bigg[\;C_{5g}\,g^{\mu\nu} + C_{511}\,p_1^\mu p_1^\nu + C_{522}\, p_2^\mu p_2^\nu + C_{512}\,\left(p_1^\mu p_2^\nu + p_1^\nu p_2^\mu\right) \\
    &\qquad\qquad\qquad\qquad\qquad\qquad\qquad 
    + C_{51n}\,\left(p_1^\mu n^\nu + p_1^\nu n^\mu\right) + C_{52n}\,\left(p_2^\mu n^\nu + p_2^\nu n^\mu\right) + C_{5n}\, n^\mu n^\nu \Bigg]
    \end{split}
\end{align}
with the coefficients
\begin{equation}
\begin{split}
    C_{5g}  &= s_{12}\left(\frac{z}{4(1-2\eps)(1-z)}f_1(z) - \frac{1}{2(1-2\eps)(1-z)}f_2\right)\,,\\
    C_{511} &= \frac{1-\eps}{2(1-2\eps)}f_1(z)\,,\\
    C_{522} &= \frac{(1-\eps)z^2}{2(1-2\eps)(1-z)^2}f_1(z) - \frac{\eps+(1-2\eps)z}{(1-2\eps)(1-z)^2}f_2\,,\\
    C_{512} &= -\frac{(1-\eps)z}{2(1-2\eps)(1-z)}f_1(z) + \frac{1-\eps}{(1-2\eps)(1-z)}f_2\,,\\
    C_{51n} &= \frac{s_{12}}{2n(p_1+p_2)}\left(-\frac{\eps}{2(1-2\eps)(1-z)}f_1(z) + \frac{\eps}{(1-2\eps)(1-z)}f_2\right)\,,\\
    C_{52n} &= \frac{s_{12}}{2n(p_1+p_2)}\left(-\frac{(1-\eps)z}{2(1-2\eps)(1-z)^2}f_1(z) + \frac{1-\eps z}{(1-2\eps)(1-z)^2}f_2\right)\,,\\
    C_{5n}  &= \left(\frac{s_{12}}{2n(p_1+p_2)}\right)^2\left(\frac{1-\eps}{2(1-2\eps)(1-    z)^2}f_1(z) - \frac{2-\eps-z}{(1-2\eps)(1-z)^2}f_2\right)\,.
\end{split}
\end{equation}

\section{Explicitly gauge dependent parts of three-particle off-shell currents}
\label{sec:explicit_gauge_dependence}
The one-to-three gluon-to-quark splitting tensor is given by Eq.~\eqref{eq:tc_gqqb_nab}.
Its explicitly $\bar{n}$-dependent component reads
\begin{equation}\label{eq:tc_gqqb_nab_offshell}
  \begin{split}
    P^{\mu\nu\,\rm(nab,\bar{n})}_{g\to gq\bar{q}}(p_1,p_2,&\,p_3)=
    \frac{C_AT_R}{2}\frac{s_{123}}{s_{23}}\bigg\{\,
    \frac{\tilde{p}_{1,23}^\mu\bar{n}^\nu+\tilde{p}_{1,23}^\nu\bar{n}^\mu}{(z_2+z_3)\,p_{123}\bar{n}}
    \bigg[\frac{s_{123}}{s_{13}}\frac{z_3}{z_1}-\frac{s_{23}}{s_{13}}\frac{z_2}{1-z_1}-1\,\bigg]\\
    &\;-\frac{\tilde{p}_{2,3}^\mu\bar{n}^\nu+\tilde{p}_{2,3}^\nu\bar{n}^\mu}{(z_2+z_3)\,p_{123}\bar{n}}
    \bigg[\frac{2s_{123}}{s_{23}}\frac{z_3}{z_1}+\frac{2s_{12}}{s_{23}}\frac{1-z_1}{z_1}+
    \frac{s_{23}}{s_{12}}\bigg]-\frac{s_{123}}{z_1(1-z_1)}\frac{\bar{n}^\mu\bar{n}^\nu}{(p_{123}\bar{n})^2}
    +(2\leftrightarrow 3)\bigg\}\;.
  \end{split}
\end{equation}

The one-to-three all-gluon splitting tensor is given by Eq.~\eqref{eq:tc_ggg}.
Its explicitly $\bar{n}$-dependent component is given by
\begin{align}\label{eq:tc_ggg_offshell}\nonumber
    &P^{\mu\nu\,\rm(\bar{n})}_{g\to ggg}(p_1,p_2,p_3)=C_A^2\bigg\{
    -\frac{\tilde{p}_{1,2}^\mu\bar{n}^\nu+\tilde{p}_{1,2}^\nu\bar{n}^\mu}{p_{123}\bar{n}}
    \frac{1-\eps}{2z_3}\frac{s_{123}\,t_{12,3}}{s_{12}^2}\\\nonumber
    &\;+\frac{\tilde{p}_{1,23}^\mu\bar{n}^\nu+\tilde{p}_{1,23}^\nu\bar{n}^\mu}{p_{123}\bar{n}}
    \bigg[\frac{s_{123}s_{12}}{s_{23}s_{13}}\bigg(\frac{1-4z_1(1-z_1)}{2(1-z_1)(1-z_2)}
    +\frac{(1-z_1)(1-2z_1)}{2z_1z_2}-\frac{z_3(2-3z_1)}{z_1(1-z_1)}
    +\frac{1-2z_1}{2(1-z_1)}\bigg)\\\nonumber
    &\quad\;+\frac{s_{123}}{s_{23}z_1}\bigg(\frac{1-4z_1(1-z_1)}{2(1-z_1)(1-z_2)}
    +\frac{z_3(1-2z_1)}{z_2}+\frac{z_1(2-3z_1)}{2(1-z_1)}\bigg)-\frac{s_{123}s_{23}}{s_{13}s_{12}}\bigg(
    \frac{(1-z_3+z_2z_3)(1-2z_3)}{2z_2(1-z_2)(1-z_3)}-\frac{2z_1}{1-z_2}\bigg)\\
    &\quad\;+\frac{s_{123}}{s_{13}}\bigg(
    \frac{1-z_2(1-2z_2)}{2z_2(1-z_2)(1-z_3)}
    +\frac{2-5z_2+6z_2^2}{2z_1z_2}
    +\frac{z_2(3-2z_2)}{2(1-z_1)(1-z_2)}
    -\frac{3-5z_2+7z_2^2}{z_2(1-z_2)}
    -\frac{1-2z_2}{z_3}\bigg)\bigg]\\\nonumber
    &\;-\frac{1}{z_1(1-z_1)}\frac{1}{z_3}
    \frac{\bar{n}^\mu\bar{n}^\nu}{(p_{123}\bar{n})^2}\bigg[
    \frac{s_{13}^3}{4s_{12}s_{23}}\frac{z_2+2z_1z_3}{1-z_3}
    +s_{23}\bigg(9(1-z_2)+\frac{1+6z_2 z_3}{1-z_2}
    -\frac{z_2^2(1+8z_2)-18z_2+11}{2(1-z_2)(1-z_3)}\bigg)\\\nonumber
    &\quad\;+\frac{s_{12}s_{13}}{4s_{23}}
    \bigg(8z_2+\frac{z_3(1+z_1)}{(1-z_2)(1-z_3)}\bigg)
    +\frac{s_{13}^2}{2s_{23}}\bigg(\frac{1-z_2}{1-z_3}-1+4z_2\bigg)
    +\frac{s_{12}^2}{s_{23}}(1-z_1)\bigg]\bigg\}
    +(\text{5 permutations})\;.
\end{align}

\bibliography{main}

\end{document}